\documentclass[12pt,preprint]{aastex}

\usepackage{graphics,graphicx}
\usepackage{subfigure}
\usepackage{color}
\usepackage{lscape}

\begin{document}

\renewcommand{\thefootnote}{\fnsymbol{footnote}}

\title{Exploring the Connection Between Star Formation and AGN Activity in the Local Universe}

\author{Stephanie M. LaMassa$^{1}$\footnote{S. M. LaMassa is now at Yale University.}, T. M. Heckman$^{1}$, A. Ptak$^{2}$, D. Schiminovich$^{3}$, M. O'Dowd$^{4}$, B. Bertincourt$^{3}$}

\affil{$^1$The Johns Hopkins University
$^2$NASA/GSFC
$^3$Columbia University
$^4$Lehman College, City University of New York
}

\begin{abstract}

We study a combined sample of 264 star-forming, 51 composite, and 73 active galaxies using optical spectra from SDSS and mid-infrared (mid-IR) spectra from the Spitzer Infrared Spectrograph. We examine optical and mid-IR spectroscopic diagnostics that probe the amount of star formation and  relative energetic contributions from star formation and an active galactic nucleus (AGN). Overall we find good agreement between optical and mid-IR diagnostics. Misclassifications of galaxies based on the SDSS spectra are rare despite the presence of dust obscuration. The luminosity of the [NeII] 12.8 $\mu m$ emission-line is well correlated with the star formation rate (SFR) measured from the SDSS spectra, and this holds for the star forming, composite, and AGN-dominated systems. AGN show a clear excess of [NeIII] 15.6 $\mu m$ emission relative to star forming and composite systems. We find good qualitative agreement between various parameters that probe the relative contributions of the AGN and star formation, including:  the mid-IR spectral slope, the ratio of the [NeV] 14.3 $\mu m$ to [NeII] $\mu m$ 12.8 fluxes, the equivalent widths of the 7.7, 11.3, and 17 $\mu m$ PAH features, and the optical ``D" parameter which measures the distance a source lies from the locus of star forming galaxies in the optical BPT emission-line diagnostic diagram. We also consider the behavior of the three individual PAH features by examining how their flux ratios depend upon the degree of AGN-dominance. We find that the PAH 11.3 $\mu m$ feature is significantly suppressed in the most AGN-dominated systems.

\end{abstract}

\section{Introduction}

It has been established that a strong link exists between supermassive black holes (SMBHs) and the galaxies in which they live \citep{FM, Gebhardt, m-sigma} and between accreting SMBHs (or active galactic nuclei, AGN) and star formation in particular \citep[e.g.][]{Kauffmann, Cid}. Clues to this link can be revealed by detailed analysis of the spectra of galaxies and AGN. Star formation processes imprint their signatures on the spectrum. Stellar populations of various ages produce strong continuum emission in the ultraviolet (UV) through near-IR region, while gas photo-ionized by the hot young stars produces prominent emission-lines in the UV, optical, and IR. The accretion disk surrounding the supermassive black hole in an AGN  produces its own strong UV and optical continuum emission (the ``big blue bump \citep[]{Shields}). This emission can swamp that of the stellar population of the host galaxy when there is a direct view of the accretion disk (i.e. in Type 1 AGN). Dust in the interstellar medium of the host galaxy will absorb optical and UV light from both stars and the AGN and re-radiate this in the IR. In AGN, this reprocessing also occurs in the dusty obscuring torus \citep{Antonucci, Urry}. In type 2 AGN, the direct view of the AGN is blocked by the obscuring torus, and the presence of the AGN can be inferred by prominent optical and infrared emission lines from highly ionized gas located beyond the torus and the strong mid-IR continuum emission from the torus. In such objects the signatures of host galaxys young stars and star formation are also present in optical and mid-IR spectra. These type 2 AGN are therefore ideal laboratories for studying the connection between AGN and star formation, provided that the tracers of this activity can be disentangled. 

Multi-wavelength star formation indicators have been extensively studied in samples of quiescent star forming galaxies \citep[see][for a review]{Ken98}. Star formation rates (SFRs) have been calibrated in the optical based on H$\alpha$ emission, which results from recombination following ionization due to energetic photons from massive O and B stars, and in the UV based on the continuum which provides a window into emission from young stars. Dust in the galaxy reprocesses optical and UV photons and re-radiates in the infrared (IR), thus revealing obscured star formation. Other IR indicators include emission from fine structure lines excited by starburst activity, such as [NeII] 12.81$\mu$m and [NeIII] 15.56$\mu$m \citep{Ho}, as well as spectral emission features associated with polycyclic aromatic hydrocarbons (PAHs), since these macro-molecules/small-grains can be ionized by single optical or UV photons \citep{Li} and are prominent in IR spectra of star-forming galaxies \citep{PAHFIT}. 

Such host galaxy star formation diagnostics can be contaminated by the presence of an AGN. Photons emitted by the accretion disk enhance the optical and UV emission used to derive SFRs. The obscuring medium enshrouding the central engine contributes significantly to the infrared emission, constituting approximately 20\% of the AGN bolometric luminosity \citep{12m}. Disentangling the contribution from the AGN vs. star formation then becomes a necessary intermediate step in using continuum IR emission as a tracer of star formation in active galaxies, and such a process can be plagued with uncertainties. Previous studies have indicated that emission from PAHs tend to be suppressed in galaxies hosting an active nucleus, which is sometimes attributed to the harsh radiation field produced by the AGN accretion disk destroying PAHs \citep[e.g.][]{Voit, O'Dowd, Treyer, me, DS10}. 

A complementary view of the AGN and starburst connection is to study spectroscopic signatures that parameterize the relative importance of these two processes. The optical BPT diagram (a plot of [OIII]$\lambda$5007/H$\beta$ vs. [NII]$\lambda$6583/H$\alpha$) provides a useful diagnostic to differentiate between star-forming galaxies, composite systems (galaxies with comparable amounts of star-formation and AGN activity) and Type 2, or obscured, AGN in the local universe \citep{BPT}. The optical ``D'' parameter is the distance a source lies from the locus of star forming galaxies on the BPT diagram: a higher D value indicates greater AGN dominance \citep{Kauffmann}. Ratios of IR fine structure lines also parameterize the ionization field hardness. For instance, [OIV]26$\mu$m and [NeV]14.32$\mu$m are primarily ionized by AGN  \citep{Rigby, DS09, Armus, Gould10} whereas [NeII]12.81$\mu$m is excited by star formation \citep{Ho}. The ratio of these lines can then indicate the relative importance of these two processes \citep{Genzel, me, PS}. The mid-infrared (MIR) spectral slope, $\alpha_{20-30\mu m}$, steepens as the amount of emission from cold dust heated by stars increases relative to the amount of emission from hotter dust heated by the AGN. It is thus another potentially useful tool to assess the relative amount of star formation in active systems \citep[e.g.][]{Buchanan, Deo}. The equivalent width (EW) of PAH grains is another diagnostic to probe the relative amount of star formation to AGN activity. An anti-correlation between ionization field hardness and the ratio of PAH features at 7.7$\mu$m and 11.3$\mu$m has been reported and is interpreted as due to an increasing contribution to the mid-IR continuum emission by AGN-heated dust. The selective destruction of smaller PAH grains may also play a role \citep[e.g.][]{PAHFIT, O'Dowd, Treyer, Wu10}, and so the flux ratios of the different PAH features can then also potentially trace the relative contribution of AGN and star formation in active galaxies.

In this paper, we combine samples of star-forming galaxies, obscured AGN (Seyfert 2 galaxies, or Sy2s) and composite systems to study the interplay between AGN activity and star formation. Using quiescent star forming galaxies as a base-line, we investigate the effects of AGN activity on the following star formation diagnostics: optically derived SFRs from the Sloan Digital Sky Survey \citep{Brinchmann}, the luminosities of the IR fine structure lines [NeII] 12.81$\mu$m and [NeIII] 15.56$\mu$m \citep{Ho} and the luminosities of the polycyclic aromatic hydrocarbons (PAHs) at 7.7$\mu$m, 11.3$\mu$m and 17$\mu$m \citep{PAHFIT, Farrah, DS10}. Which of these star formation proxies agree the best among star-forming galaxies and AGN and are thus least affected by the presence of an AGN? 

We expand upon the work of \citet{me} to test the agreement among diagnostics that parameterize the relative contributions of AGN activity and star formation, including the equivalent width (EW) of PAHs at 7.7, 11.3 and 17$\mu$m, ratios of PAH fluxes \citep{PAHFIT, O'Dowd, Treyer, DS10}, ratios of IR fine structure lines \citep{Genzel, Armus, PS}, mid-IR spectral slope \citep[$\alpha_{20-30\mu m}$][]{Buchanan, Deo} and the optical D parameter \citep{Kauffmann}. In \citet{me} we found that Sy2s with stronger PAH emission tend to have a softer ionization field and that the MIR spectral slope was well correlated with PAH EW. By expanding this parameter space into the regime of quiescent star forming galaxies, we test if PAH features and the mid-IR spectral slope are dependent on the hardness of the radiation field, regardless of its source, or if ionization by an AGN is necessary to appreciably affected observed features. Finally, we use the results of this analysis to empirically decompose the mid-IR (MIR) emission into a star-forming and AGN component. 

\renewcommand{\thefootnote}{\arabic{footnote}}

\section{Sample Selection}
To assess the efficacy of star formation diagnostics in active galaxies, we have combined samples of AGN, quiescent star forming galaxies, and composite objects. Sources were categorized by their BPT designations \citep{BPT} using the boundaries from \citet{Kewley} and \citet{Kauffmann}: galaxies above the \citet{Kewley} delineation are classified as pure Seyfert 2s, sources in between the \citet{Kewley} and \citet{Kauffmann} demarcations are classified as composites and the sources below the \citet{Kauffmann} boundary are classified as quiescent star-forming galaxies.

As discussed in \citet{me09}, we selected an optical sample of Sy2s from the Sloan Digital Sky Survey (SDSS). These sources were identified as pure AGN on the basis of their [OIII]$\lambda$5007/H$\beta$ and [NII]$\lambda$6583/H$\alpha$ flux ratios which place them above the theoretical starburst demarcation from \citet{Kewley} on the BPT diagram.  This sample, complete to an [OIII] flux limit of  $10^{-14}$ erg/s/cm$^{-2}$ (hereafter ``[OIII] sample''), comprises 20 Sy2s. We complement these Sy2s with a mid-infrared (MIR) selected sample, the original 12$\mu$m sample \citet{12m} from the {\it IRAS} point source catalog, which contains 31 galaxies. Using published or SDSS measured values of [OIII]$\lambda$5007/H$\beta$ and [NII]$\lambda$6583/H$\alpha$, we have classified 27 12$\mu$m sources as Sy2s, 2 as composites (F08572+3915 and NGC 7590) and 2 were dropped due to lack of published [NII]/H$\alpha$ values (F04385-0828 and NGC 3982). As both [OIII] and MIR fluxes are indicators of intrinsic AGN power \citep[e.g][]{12m, H05, me}, these Sy2 samples should to first order represent unbiased views of the AGN population. The [OIII] sample has optical spectroscopy from SDSS whereas the optical data for the majority of the 12$\mu$m sample were based on literature values. Both samples have low and high resolution {\it Spitzer} Infrared Spectrograph \citep[IRS][]{Houck} data. 

Our comparison samples of predominantly normal galaxies have SDSS spectroscopy, {\it Spitzer} IRS data and ultraviolet photometry from the Galaxy Evolution Explorer ({\it Galex}). The {\it Spitzer}-SDSS-{\it GALEX} Spectroscopic Survey (SSGSS, P.I. Schiminovich) contains 101 local (0.03 $< z <$ 0.21) galaxies overlapping the Lockman Hole region \citep{O'Dowd, Treyer, O'Dowd2}. This sample is augmented by the {\it Spitzer} SDSS Spectroscopic Survey (S5, P.I. Schiminovich), increasing the sample size of local (in this case, 0.05 $ < z <$ 0.1) star-forming galaxies to 300 (Bertincourt et al., in prep). Both the SSGSS and S5 samples have IRS Short-Low (SL) and Long-Low (LL) coverage. The S5 sample also has IRS Short-High (SH) for all sources, but only 1/3 of the SSGSS sources (a sub-sample consisting of the brightest galaxies) have high resolution infrared spectra. Based on the BPT designations, there are 26 Sy2s, 49 composites and 264 star-forming galaxies in the SSGSS and S5 surveys.

Combining all four samples, we garner 264 star-forming galaxies, 51 composites, and 73 AGN for this analysis. Throughout this paper, we have adopted a cosmology of $H_o$ = 70 km/s/Mpc, $\Omega_M$ = 0.27, and $\Omega_{\lambda}$ = 0.73.

\section{Data Analysis}

\subsection{Optical Data}
Optical spectra for the [OIII], SSGSS and S5 samples were drawn from SDSS, using the MPA/JHU galaxy value-added catalog.\footnote{http://www.mpa-garching.mpg.de/SDSS/DR7/} This catalog includes ancillary derived data, such as star formation rates \citep[SFRs,][]{Brinchmann}, which we utilize in the subsequent analysis. Only four of the 12$\mu$m selected sources have SDSS spectra (3 Sy2s and 1 composite). Optical data for the remainder of these sources were culled from the literature, as noted in \citet{me}. Hence, these 24 Sy2s and 1 composite are omitted from the SDSS SFR analysis in Section 4. We emphasize here that all SDSS-derived quantities are based on the SDSS spectra alone. 

\subsection{Infrared Data Reduction}
The details of the {\it Spitzer} data reduction for the [OIII] and 12$\mu$m samples were presented in \citet{me}, the SSGSS sample in \citet{O'Dowd, O'Dowd2} and the S5 sample in Bertincourt et al., in prep., which follows the same procedures as \citet{O'Dowd}. In short, the IRS data were run through the standard {\it Spitzer} calibration software. Sky subtraction was performed for all low-resolution data and for the high resolution sources that had dedicated off source observations. All but 68 observations, from the SSGSS sample, have high resolution data, leaving 223 star-forming galaxies, 33 composites and 64 Sy2s with high resolution IRS coverage. Only eight sources, from the 12$\mu$ sample, did not have dedicated high resolution background observations, but since we utilize these data to measure emission line fluxes, the addition of the background to the underlying continuum has negligible effect on our derived results \citep[see][]{me}. Spectra were extracted with SPICE for the [OIII], SSGSS and S5 sources and with SMART \citep{SMART} for the staring mode 12$\mu$m observations. CUBISM \citep{CUBISM} was used to extract spectra from 12$\mu$m spectral mapping observations.

\subsection{Emission Line Measurements}
Emission line fluxes used in this analysis ([NeII] 12.81$\mu$m, [NeV] 14.32$\mu$m and [NeIII] 15.56$\mu$m) were measured from the high resolution spectra, specifically the Short-High (SH) module. The local continuum around the emission feature was fit with a local zero or first order polynomial. This local continuum fit was frozen and the emission feature was then fit with a Gaussian. The error was estimated by smoothing the local continuum by the line width (or the nominal minimum resolvable line width\footnote{width = $\lambda_{nom}$/R, where $\lambda_{nom}$ is the nominal wavelength of the emission line and R is the resolution of the SH module, $\sim$600} in the case of non-detections) and calculating the RMS around this smoothed continuum. In order for a line to be considered detected, it had to be significant at the 5$\sigma$ level or above, have a line width at least on the order of the SH resolution (i.e. full width half max $\geq$ 0.01 $\mu$m) and have a line centroid near the nominal wavelength of the emission line (i.e. $\leq$ 0.03$\mu$m). The [NeII] and [NeIII] fluxes and upper limits for the SSGSS, S5, [OIII] and 12$\mu$m samples are listed in Tables \ref{ssgss_flux} - \ref{12m_flux}. Since we consider [NeV] upper limits in aggregate rather than individually (see Section 7.1), we list only the [NeV] detections in these tables.

\subsection{PAH Measurements}
We measured the fluxes and equivalent widths (EWs) of the polycyclic aromatic hydrocarbons (PAH) at 7.7$\mu$m, 11.3$\mu$m and 17$\mu$m with PAHFIT, a spectral decomposition routine that models dust continuum, starlight continuum, unresolved emission lines, PAH emission features and silicate absorption/emission \citep{PAHFIT}. PAHFIT requires a single stitched low-resolution spectrum. The Long-Low (LL) spectra was therefore shifted to match the Short-Low (SL) spectra in their overlapping regions, with typical adjustments $\sim$20\% or lower \citep[see][]{O'Dowd2, me}. This choice of normalization was made since the SL and SH apertures sizes are well-matched, better enabling us to compare quantities derived from the high and low resolution spectra. 

The routine was run with a mixed extinction model on all sources. We note that in \citet{me}, PAHFIT was run with a screen extinction, which assumes a foreground absorbing slab rather than dust mixed with gas, starlight and grains. We have re-run PAHFIT on the [OIII] and 12$\mu$m sources using the default mixed extinction option to be consistent with the SSGSS and S5 samples, though differences in results were slight. For these two Sy2 samples, we also added a dust component at 1000K since this may better model strong AGN \citep{Deo}, though we note the difference in the fluxes and EWs between the default temperature array and this modified version is negligible for a majority of the sources and a factor of 2 or lower for most of the remaining sources. We consider a PAH feature to be detected if it is present at greater than the 5$\sigma$ level. In the subsequent analysis, we consider PAH upper limits on the flux, but only include EWs for detected sources. PAH measurements are listed in Tables \ref{ssgss_pah} - \ref{12m_pah}.

\section{Parameterizing Star Formation: Effects of AGN Activity}
Optical and IR fluxes have been utilized to calibrate SFRs in quiescent galaxies. Below we investigate if emission from the AGN contaminates these parameters and if composite systems are more similar to star-forming galaxies or Seyfert 2s. Here we consider optical SFRs derived from SDSS using the methodology described in \citet{Brinchmann}. We also explore IR star formation proxies, namely the fine structure [NeII] 12.81$\mu$m and [NeIII] 15.56$\mu$m emission lines as well as the luminosity of the polycyclic aromatic hydrocarbon (PAH) features at 7.7, 11.3 and 17 $\mu$m.

\subsection{SDSS SFR$_{fiber}$: Circumnuclear Star Formation}

For star-forming galaxies, SDSS SFR$_{fiber}$ is calculated from the H$\alpha$ luminosity, corrected for dust extinction, measured through the 3'' SDSS fiber. In these systems, H$\alpha$ emission results from the recombination of ionizing photons from massive O stars, making this flux a useful parameterization of recent star formation. Using a Kroupa IMF \citep{Kroupa}, H$\alpha$ luminosity is then converted into a SFR \citep{Brinchmann}.

Such a straightforward calculation fails for active systems due to emission powered by the AGN contaminating the H$\alpha$ flux. However, a SFR can be estimated in these systems using the strength of the Dn(4000) feature, which is a break in the optical spectrum at 4000\AA\ that indicates the age of the host galaxy stellar population. The SFR$_{fiber}$ for composites and AGN in SDSS is calculated starting with the strength of the Dn(4000) break. This is then converted into a SFR by adopting the relationship between SFR$_{fiber}$/M$_{*,fiber}$ and Dn(4000) measured for quiescent systems and then using measured value of M$_{*,fiber}$ \citep{Brinchmann}.

\subsection{IR SFR diagnostics: Neon Emission Lines and PAH Luminosities}
In the infrared, [NeII] at 12.81$\mu$m has an ionization potential of 21.56 eV and is therefore excited by star formation activity. [NeIII] at 15.56$\mu$m has a higher ionization potential than [NeII] at 41 eV, but it becomes more dominant in lower-metallicity star forming systems \citep[e.g.][]{Wu06}. \citet{Ho} have demonstrated that the sum of [NeII] and [NeIII] correlates better with total infrared luminosity (a reliable proxy of SFR) than just [NeII] as this sum better accommodates the low metallicity systems. They subsequently calibrated a SFR based on [NeII] + [NeIII].

PAH macro-molecules/small-grains are ionized by single ultraviolet or optical photons \citep{Li} and are very prominent in the spectra of star forming galaxies \citep{PAHFIT}. SFRs have consequently been calibrated on PAH luminosities \citep[e.g.][]{Farrah}.  However, the PAH luminosity may be affected by AGN activity, and the effect may be different for the diferent PAH features \citep[e.g.][]{O'Dowd, Treyer, DS10}. In this analysis, we consider the sum of the PAH luminosities at 7.7, 11.3 and 17$\mu$m ($L_{\Sigma PAH 7.7,11.3,17\mu m}$). In Section 5, we examine each PAH feature individually as a function of increasing ionization field strength.

\subsection{Comparison of Optical and IR SFR diagnostics}
\begin{figure}[ht]
\centering
\subfigure[]{\includegraphics[scale=0.35,angle=90]{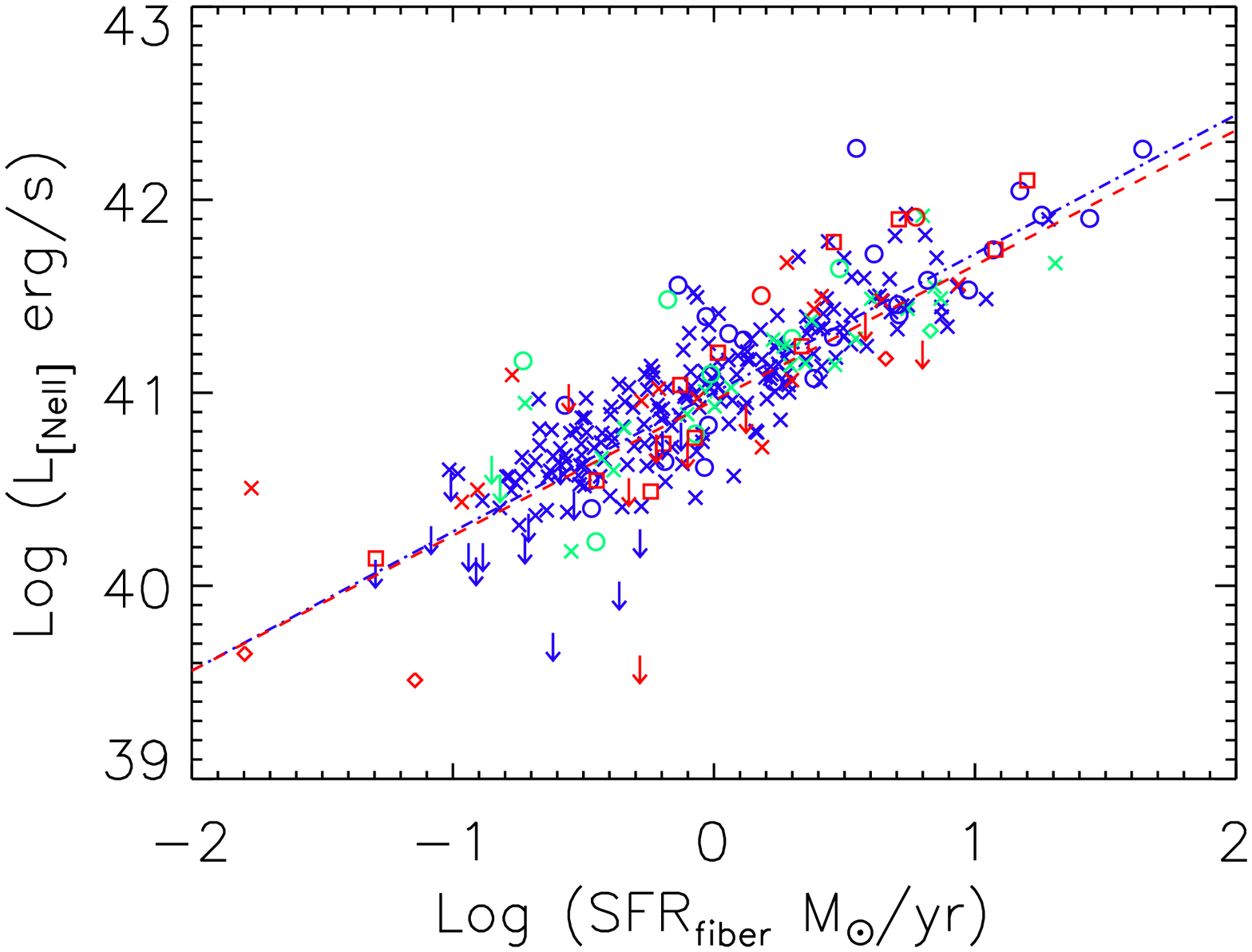}}
\subfigure[]{\includegraphics[scale=0.35,angle=90]{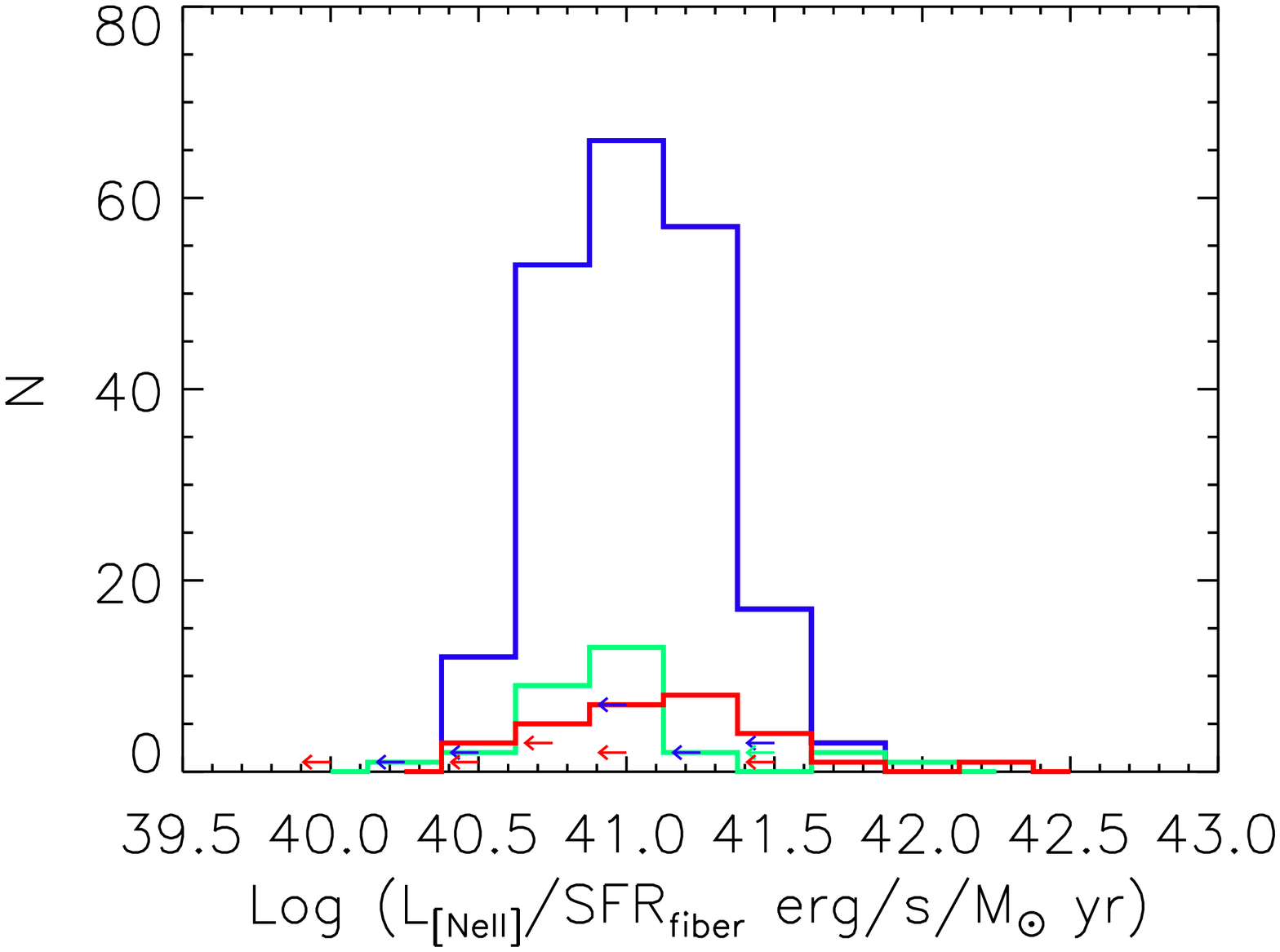}}
\caption[L$_{[NeII]}$ vs. SFRs]{\label{ne2_sdss}Left: L$_{[NeII]}$ vs. SDSS SFR$_{fiber}$. Right: Distribution of log(L$_{[NeII]}$/SDSS SFR$_{fiber}$). The symbol coding is as follows: S5 - crosses, SSGSS - circles, [OIII] sample - squares, 12$\mu$m sample - diamonds. Blue data points represent star-forming galaxies, green indicates composites and red denotes AGN. The dashed-dotted line indicates the best fit to the star-forming galaxy sub-sample and the red dashed line shows the best fit to the AGN population. Active galaxies agree with quiescent in terms of the L$_{[NeII]}$ distribution relative to SDSS SFR$_{fiber}$.}
\end{figure}

\begin{figure}[ht]
\subfigure[]{\includegraphics[scale=0.35,angle=90]{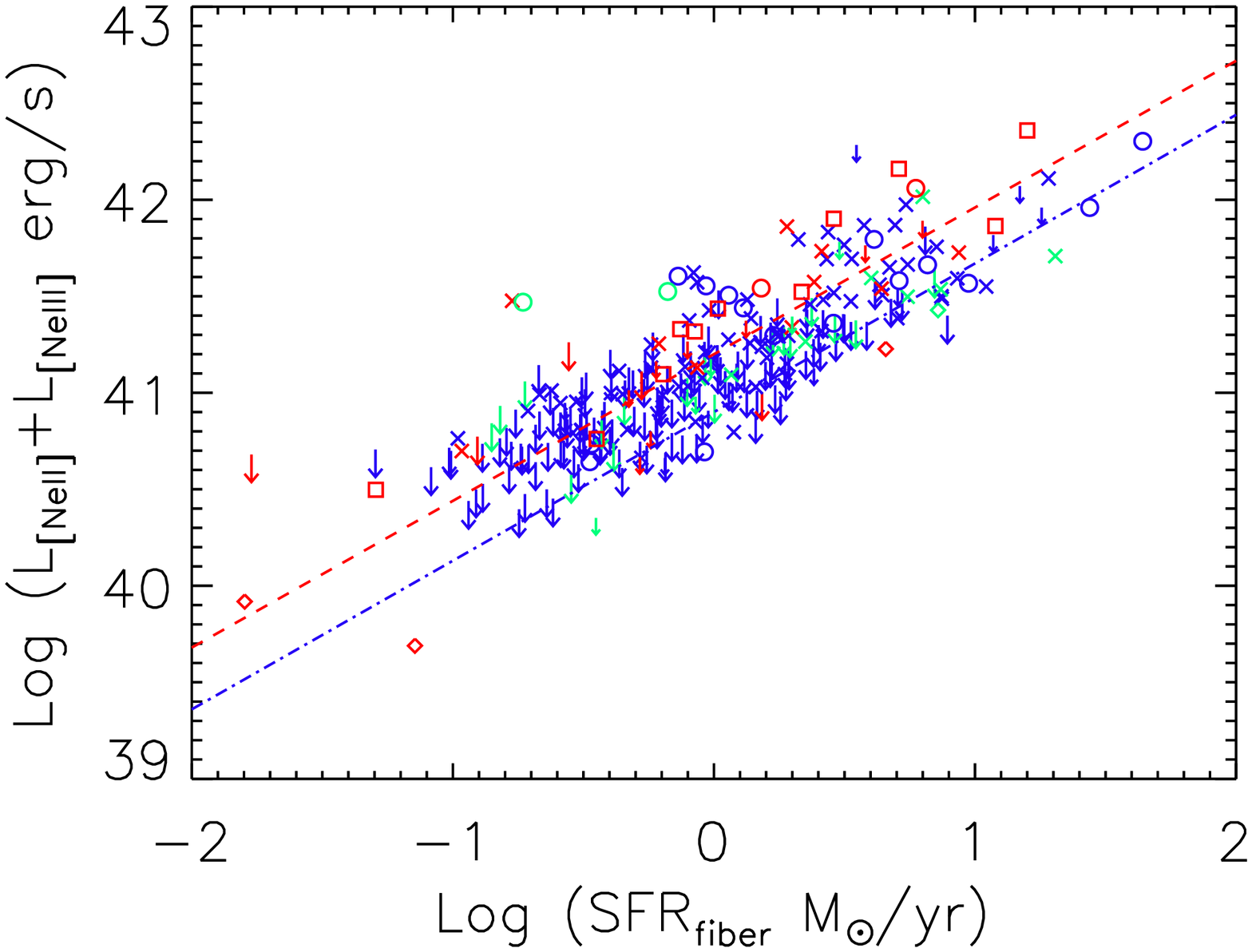}}
\subfigure[]{\includegraphics[scale=0.35,angle=90]{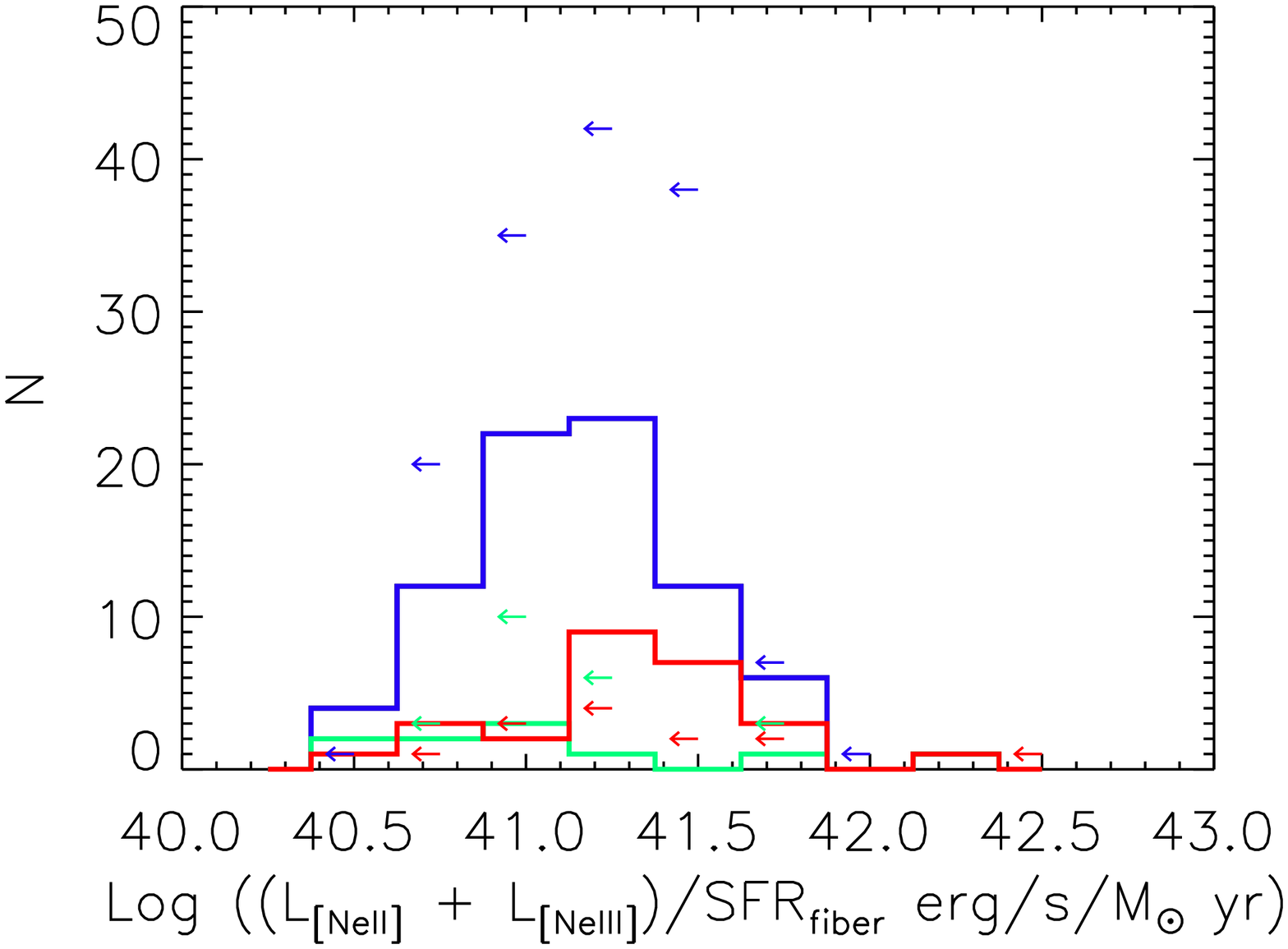}}
\caption[]{\label{nesum_sdss}Left: L$_{[NeII]}$ + L$_{[NeIII]}$ vs. SDSS SFR$_{fiber}$. Right: Distribution of log((L$_{[NeII]}$+L$_{[NeIII]}$/SDSS SFR$_{fiber}$). We note that the relatively large number of non-detections causes the best-fit relation from survival analysis to lie below the majority of plotted upper limits. Nevertheless, the AGN clearly have boosted L$_{[NeII]}$ + L$_{[NeIII]}$ values compared with star-forming and composite galaxies, which can be seen by comparing the best-fit trend to the Sy2s (red dashed line) with the star-forming galaxies (blue dotted-dashed line). Compared with Figure 1, a clear offset is discernable between the best-fit relations between Sy2s and quiescent galaxies, suggesting AGN enhances [NeIII] emission. Color, symbol and line coding same as Figure 1.}
\end{figure}

\begin{figure}[ht]
\centering
\subfigure[]{\includegraphics[scale=0.35,angle=90]{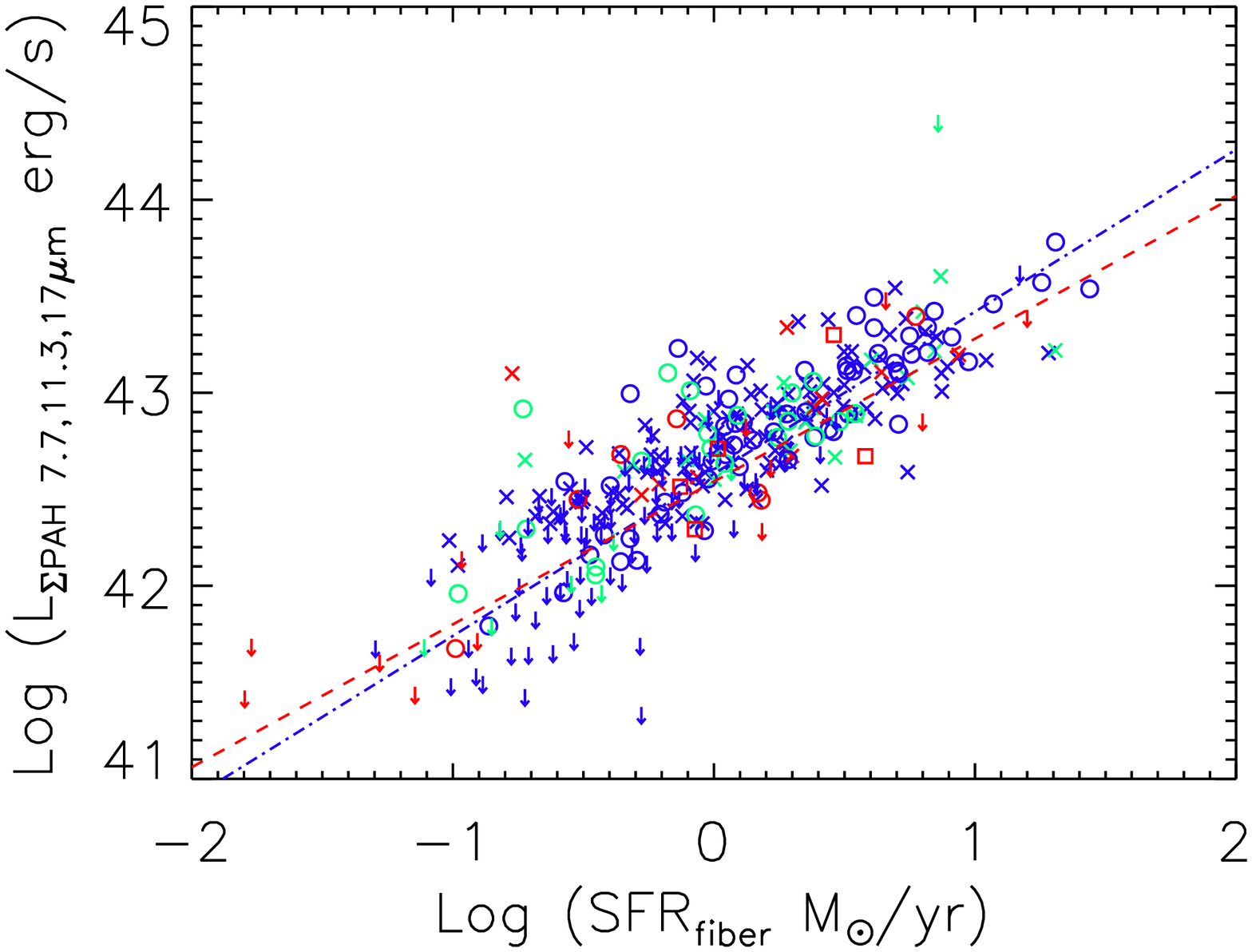}}
\subfigure[]{\includegraphics[scale=0.35,angle=90]{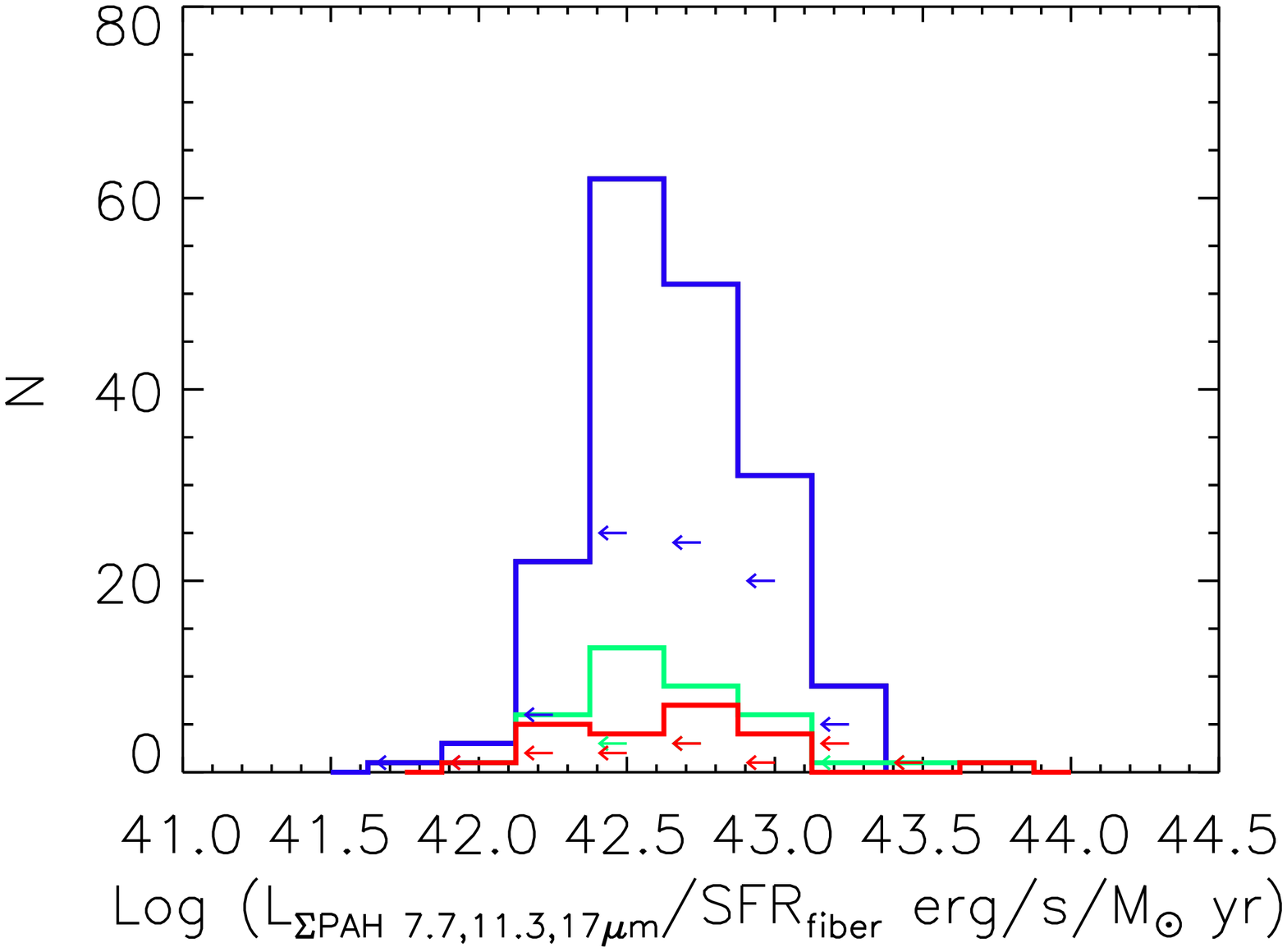}}
\subfigure[]{\includegraphics[scale=0.35,angle=90]{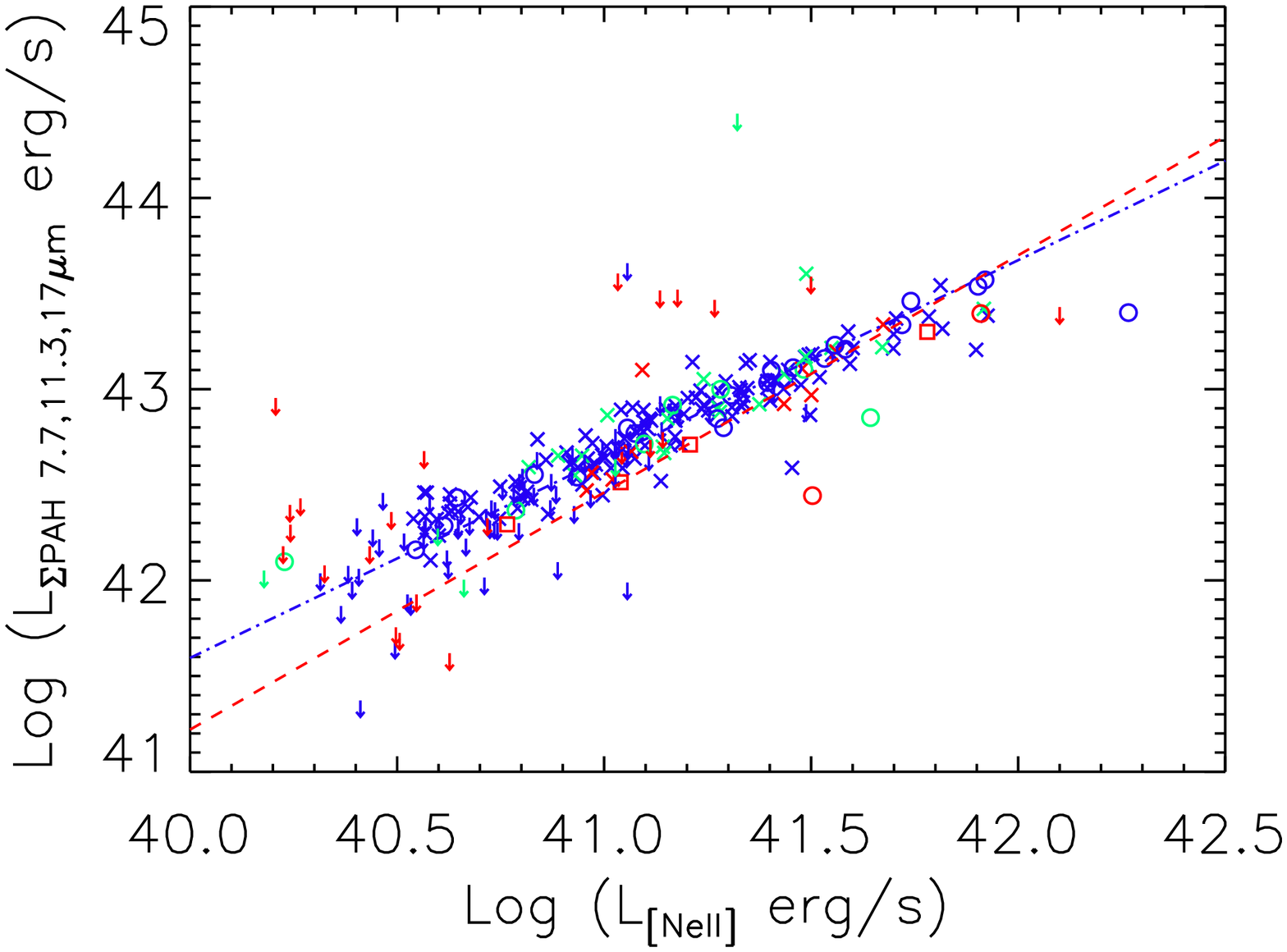}}
\subfigure[]{\includegraphics[scale=0.35,angle=90]{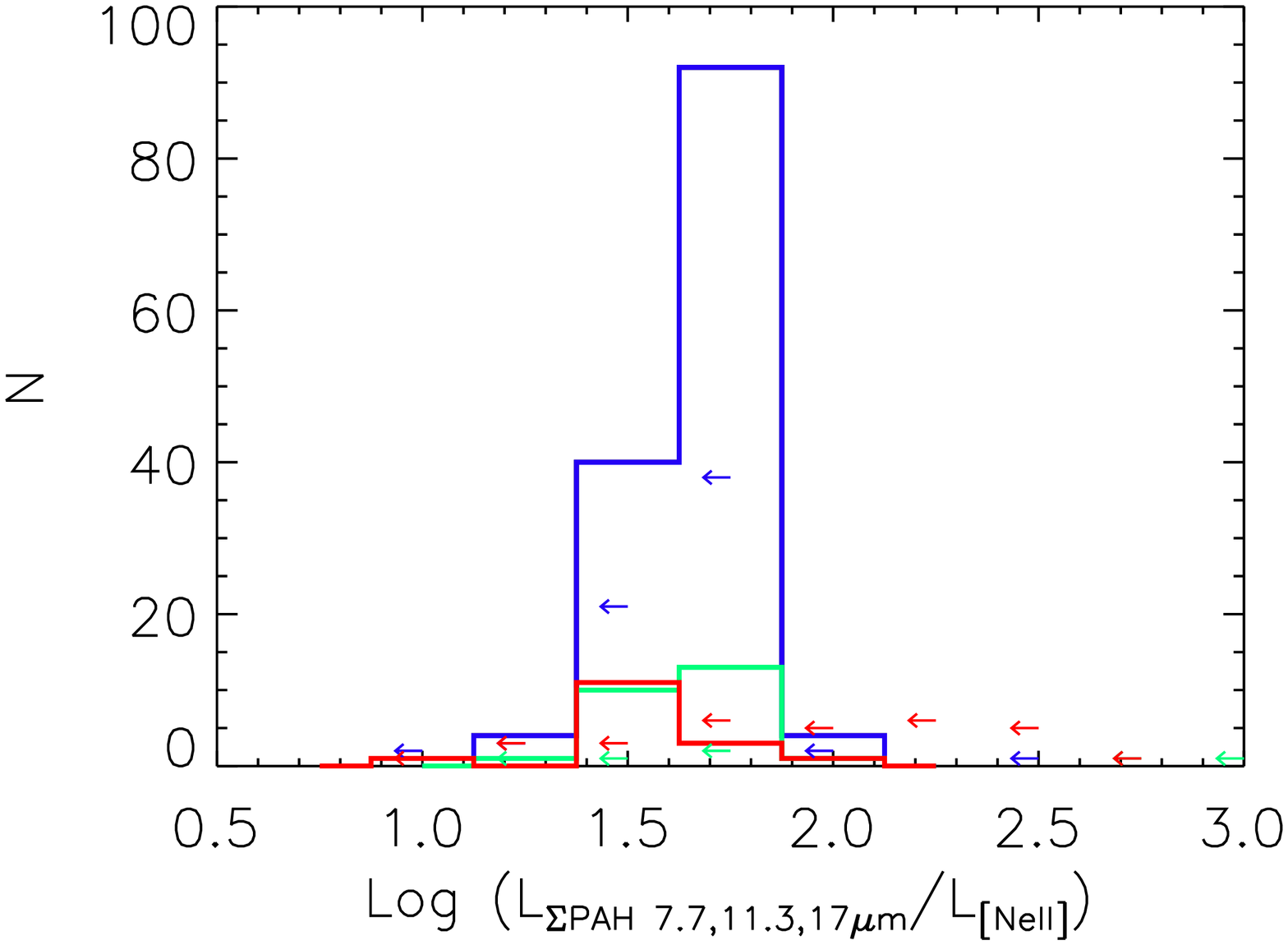}}
\caption[]{\label{pahsum_sfr}Top Left: L$_{\Sigma PAH 7.7,11.3,17\mu m}$ vs SDSS SFR$_{fiber}$. Top Right: Distribution of log(L$_{\Sigma PAH 7.7,11.3,17\mu m}$/SDSS SFR$_{fiber}$). Bottom Left: L$_{\Sigma PAH 7.7,11.3,17\mu m}$ vs L$_{[NeII]}$. Bottom Right: Distribution of log(L$_{\Sigma PAH 7.7,11.3,17\mu m}$/L$_{[NeII]}$). The sum of the PAH luminosities relative the SFR$_{fiber}$ is consistent among star-forming galaxies, composites and AGN. However, when compared with L$_{[NeII]}$, the PAH emission in AGN is suppressed  by about 32\% on-average in relation to the observed PAH emission in star-forming galaxies and composites: the mean log(L$_{\Sigma PAH 7.7,11.3,17\mu m}$)/L$_{[NeII]}$ is 1.50 dex for star-forming galaxies, yet 1.33 dex for Sy2s. The best-fit trend to the Sy2s (red dashed line) also illustrates the different relationship between the PAH luminosities in AGN compared with quiescent galaxies (blue dotted-dashed line). Color, symbol and line coding same as Figure 1. }
\end{figure}

We compare the optical and IR SFR diagnostics pair-wise in Figures \ref{ne2_sdss} - \ref {pahsum_sfr} as both scatter plots (left hand panels) and histograms of the ratios of the two proxies under consideration (right hand panels) to search for systematic offsets among galaxy type.\footnote{In these plots and subsequent plots of PAH fluxes and EWs, the error bars are smaller than or on the order of the symbol size and therefore not plotted. These relatively small errors are due to the stringent 5$\sigma$ threshold we required for detection.} In these plots, blue represents star-forming galaxies, green denotes composites and red indicates AGN. To take into account upper limits, we used survival analysis \citep[ASURV,][]{Feigelson, Isobe, Is2} to determine the best-fit trends to the star-forming galaxies and Sy2s (shown as the blue dotted-dashed lines and red dashed lines, respectively, in the left hand panels of Figures \ref{ne2_sdss} - \ref {pahsum_sfr}), to calculate the mean ratio of SFR proxies for each galaxy type (Table \ref{SFR_ratios}) and to quantify the significance of agreement/disagreement of the distributions of the SFR proxy ratios among star-forming galaxies, composites and AGN (Table \ref{SFR_2samp}). We use the Gehan’s Generalized Wilcoxon Test to calculate the probability that the distributions of SFR proxy ratios among the sub-samples being considered are drawn from the same parent population. A probability value ($P$) $<0.05$ indicates that we can reject the null hypothesis and that the distributions differ at a statistically significant level. We note that not all sources are included in each calculation, either due to absence of data (i.e. 25 12$\mu$m sources have no SDSS coverage) or lack of significant detections in the diagnostics being considered. The number of sources used for each calculation are listed in Tables \ref{SFR_ratios} - \ref{SFR_2samp}.

As shown in Figure \ref{ne2_sdss}  and Table \ref{SFR_2samp}, the [NeII] luminosity in relation to SDSS SFR$_{fiber}$ is consistent among star-forming galaxies, composites and AGN. However, the sum of [NeII] and [NeIII] does show signs of AGN contamination as these sources are systematically offset to higher L$_{[NeII]}$ + L$_{[NeIII]}$ values with respect to star-forming galaxies, as seen in  Figure \ref{nesum_sdss}. Indeed, the distribution of L$_{[NeII]}$ + L$_{[NeIII]}$ normalized by SFR$_{fiber}$ is significantly different between star-forming galaxies and AGN (Table \ref{SFR_2samp}). This result is due to the harder ionization field from the AGN boosting [NeIII] to higher values, biasing the sum as a SF indicator in active systems. This result is consistent with previous works \citep[e.g.][]{Deo07, T08, Mel08, Wu10, PS} that demonstrated that [NeIII]/[NeII] is correlated with AGN activity and thus a tracer of ionization field hardness. Interestingly, the composite galaxies are not affected: their L$_{[NeII]}$ + L$_{[NeIII]}$ sum differs significantly from AGN, but not from star-forming galaxies. Hence, the SFR calibrated on L$_{[NeII]}$ and L$_{[NeIII]}$ is reliable for star-forming and composite galaxies, but not AGN. Consequently, to test the efficacy of PAH luminosities in parameterizing star formation in AGN, we compare PAH luminosities to SDSS SFR$_{fiber}$ and L$_{[NeII]}$ and not the sum of L$_{[NeII]}$ and L$_{[NeIII]}$.

Both [NeII] and PAH luminosities are upper limits for at least a handful of sources. Since ASURV is unable to accommodate doubly censored data points, we calculated the ratio means and performed two sample tests using upper limits on [NeII] and PAH luminosities separately. We report the results of both analyses in Tables \ref{SFR_ratios} and \ref{SFR_2samp}, including the number of sources for each calculation. We find that the distribution of PAH luminosities relative to SFR$_{fiber}$ is consistent among star-forming galaxies, composites and AGN, though the slope of the relation for Sy2s differs markedly compared with star-forming galaxies (Figure \ref{pahsum_sfr}). The difference in PAH emission becomes more pronounced when compared to L$_{[NeII]}$, regardless of whether L$_{\Sigma PAH 7.7,11.3,17\mu m}$ or L$_{[NeII]}$ is treated as the upper limit (see Table \ref{SFR_2samp}). As Table \ref{SFR_ratios} demonstrates, there is a decrease of about 32\% in the mean in the ratio of PAH to [NeII] luminosity in AGN compared to both the quiescent star-forming galaxies and the composites. At face value we can not say whether this is caused by supression of the PAH emission or enhancement in the [NeII] emission. We will return to this issue later, when we explore the effects of ionization field hardness on the relative strengths of PAH features at 7.7, 11.3 and 17 $\mu$m individually.

\section{Relative Importance of AGN to Star Formation Activity}
Expanding on the work of \citet{me}, we now investigate the agreement among multi-wavelength diagnostics of ionization field hardness to determine where ``normal'' star-forming galaxies live in this parameter space. As AGN activity weakens and star formation becomes more dominant, do the trends found for AGN smoothly extend into the regime of quiescent galaxies?

As discussed in \citet{me}, the [OIV]26$\mu$m/[NeII]12.81$\mu$m flux ratio is a reliable tracer of the incident radiation field hardness \citep{Genzel, PS} since [OIV] is primarily ionized by AGN activity \citep{Rigby, DS09, Mel08, me} whereas [NeII]12.81$\mu$m is excited by star formation \citep{Ho}. However, we do not have Long-High (LH) coverage for the SSGSS and S5 samples, and thus lack high resolution data for the [OIV] line at 26$\mu$m. We therefore use [NeV]/[NeII] as a diagnostic of the relative importance of AGN versus starburst activity \citep{Armus, PS}. We compare this ratio with other diagnostics studied in \citet{me}. First we use the optical D parameter ($D=\sqrt{[log([NII]/H\alpha) + 0.45]^2 + [log([OIII]/H\beta) + 0.5]^2}$), which is the distance a source lies on the BPT diagram from the locus of star-forming galaxies, so that a higher D value indicates a greater dominance of AGN activity \citep{Kauffmann}. Second we use $\alpha_{20-30\mu m}$ (MIR spectral index) which increases in the presence of cold dust associated with star formation activity \citep{Buchanan, Deo}. Finally we use the PAH equivalent widths (EWs) which are expected to decrease as AGN activity becomes more dominant \citep[e.g.][]{Genzel}. We also investigate if the ratios of the fluxes of the different PAH features are affected by AGN activity, as found by \citet{O'Dowd}, \citet{Wu10} and \citet{DS10}.

\subsection{[NeV]}
\begin{figure}[ht]
\centering
\subfigure[]{\includegraphics[scale=0.35,angle=90]{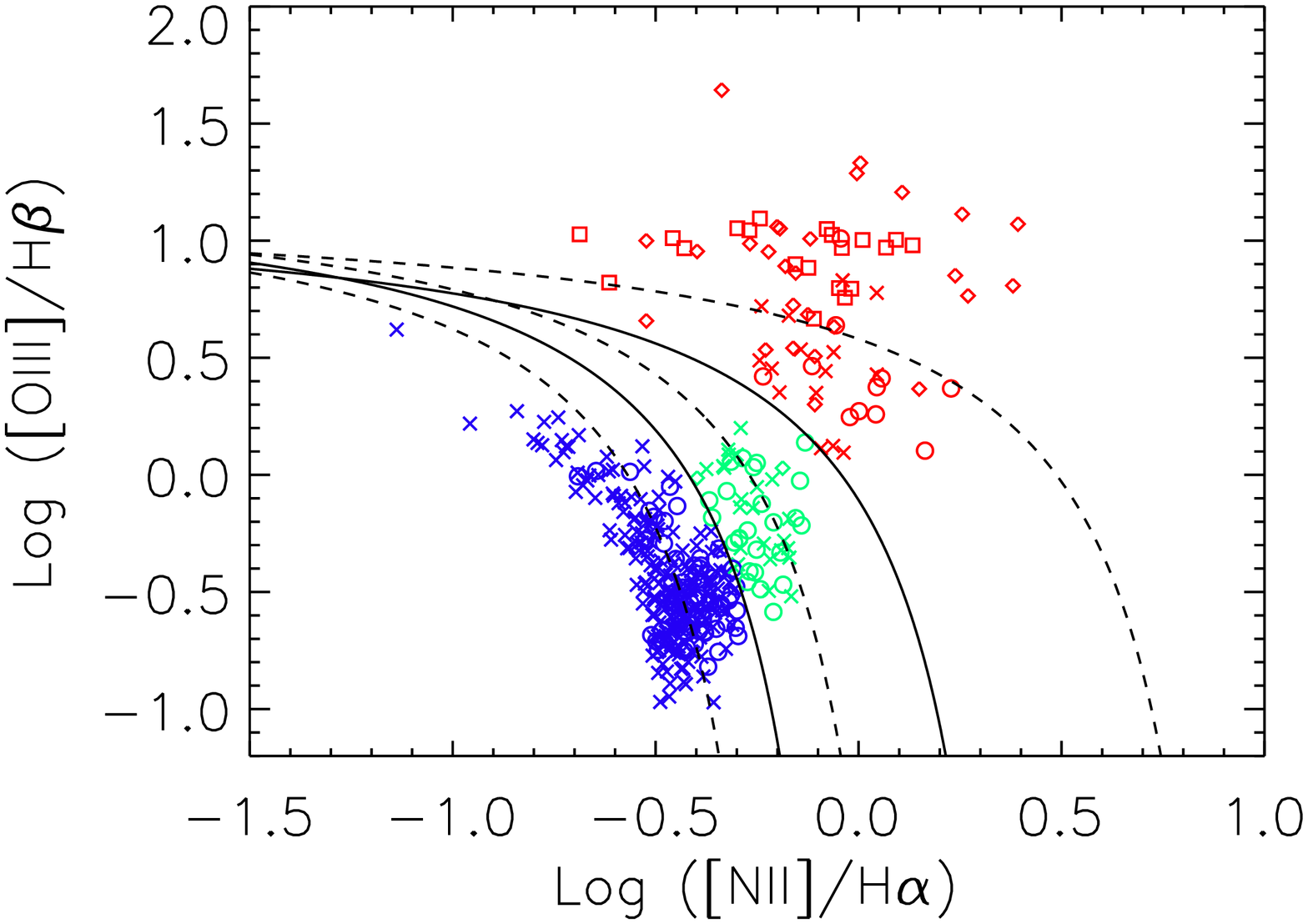}}
\subfigure[]{\includegraphics[scale=0.35,angle=90]{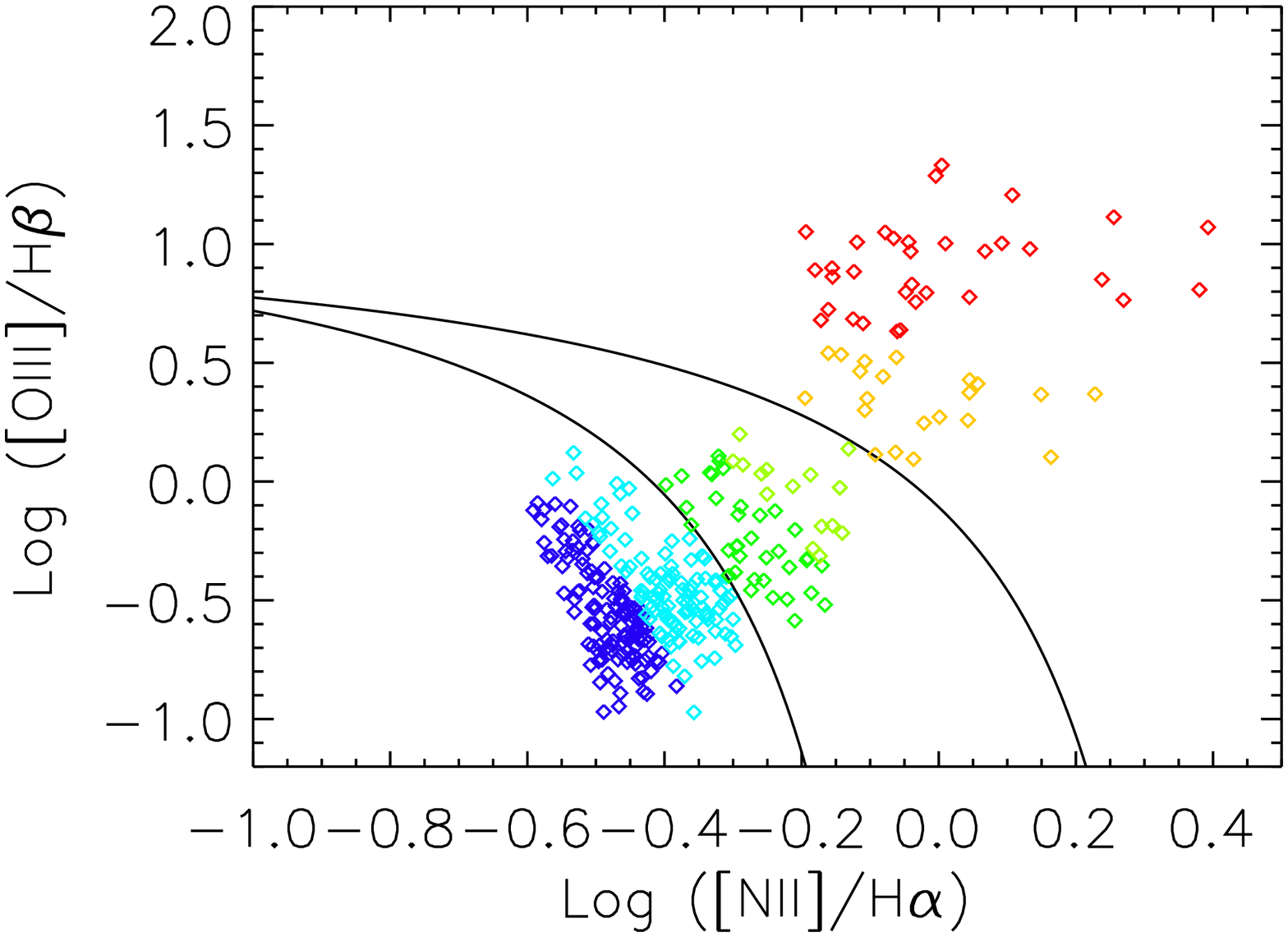}}
\caption[BPT diagrams for stacking SH spectra.]{\label{bpt_nev}Left: BPT diagram for all sources in our sample. The unbroken black lines indicate the \citet{Kewley} (upper) and \citet{Kauffmann} (lower) demarcations between star-forming galaxies, composites and AGN. The dashed black lines illustrate the dividing lines we used to separate the sample to stack the short-high IRS spectra. Symbol and color coding is same as Figure 1.Right: After removing the low-metallicity systems and using the demarcations shown in the left-hand plot, the groups into which our sample were separated our indicated by the colors in this plot: star-forming group 1 - dark blue, star-forming group 2 - light blue, composite group 1 - dark green, composite group 2 - light green, AGN group 1 - gold, AGN group 2 - red.}
\end{figure}

Due to its high ionization potential (97.1 eV), [NeV] has been claimed to cleanly trace SMBH accretion \citep{Genzel, Armus, PS} and to be an unambiguous signature of AGN activity \citep{Gilli, Gould10}. Also, as it is formed in the narrow line region, it should ideally be unaffected by toroidal obscuration and therefore a tracer of intrinsic AGN flux, similar to the [OIII] 5007\AA\  and [OIV] 26$\mu$m lines \citep{H05, Mel08, Rigby, DS09, me}. However, in a majority of our spectra, this feature was not significant above the noise. To use [NeV]/[NeII] as a tracer of ionization field hardness, we consider the non-detections in aggregate, stacking the spectra among sub-groups of galaxy type, rather than employing individual upper limits. For this stacking exercise, we separated star-forming galaxies, composites and AGN into two groups each, based on their location in the BPT diagram, in order of increasing nuclear activity. Figure \ref{bpt_nev} illustrates the method we used. The left hand plot is a BPT diagram for our full sample, with the unbroken black lines showing the \citet{Kewley} (upper) and \citet{Kauffmann} (lower) demarcations between star-forming galaxies, composites and AGN. The dashed lines illustrate the additional empirical separations we used, approximately tracing the halfway point within each region. These lines correspond, in order of lowest to highest, to the following equations:
$$Log([OIII]/H\beta) = \frac{0.61}{Log([NII]/H\alpha)+0.1} + 1.3$$
$$Log([OIII]/H\beta) = \frac{0.61}{Log([NII]/H\alpha)-0.2} + 1.3$$
$$Log([OIII]/H\beta) = \frac{0.61}{Log([NII]/H\alpha)-1} + 1.19$$

We also employed a cut on [NII]/H$\alpha$ to exclude low metallicity systems: Log ([NII]/H$\alpha$) $>$ -0.6 dex for star-forming galaxies and Log ([NII]/H$\alpha$) $>$ -0.2 dex for AGN. The right-hand panel of Figure \ref{bpt_nev} shows our resulting sub-groups depicted in different colors with the \citet{Kewley} (upper) and \citet{Kauffmann} (lower) demarcations overplotted: star-forming group 1 - dark blue, star-forming group 2 - light blue, composite group 1 - dark green, composite group 2 - light green, AGN group 1 - gold, AGN group 2 - red.

\begin{figure}[ht]
\centering
\subfigure[]{\includegraphics[scale=0.35,angle=90]{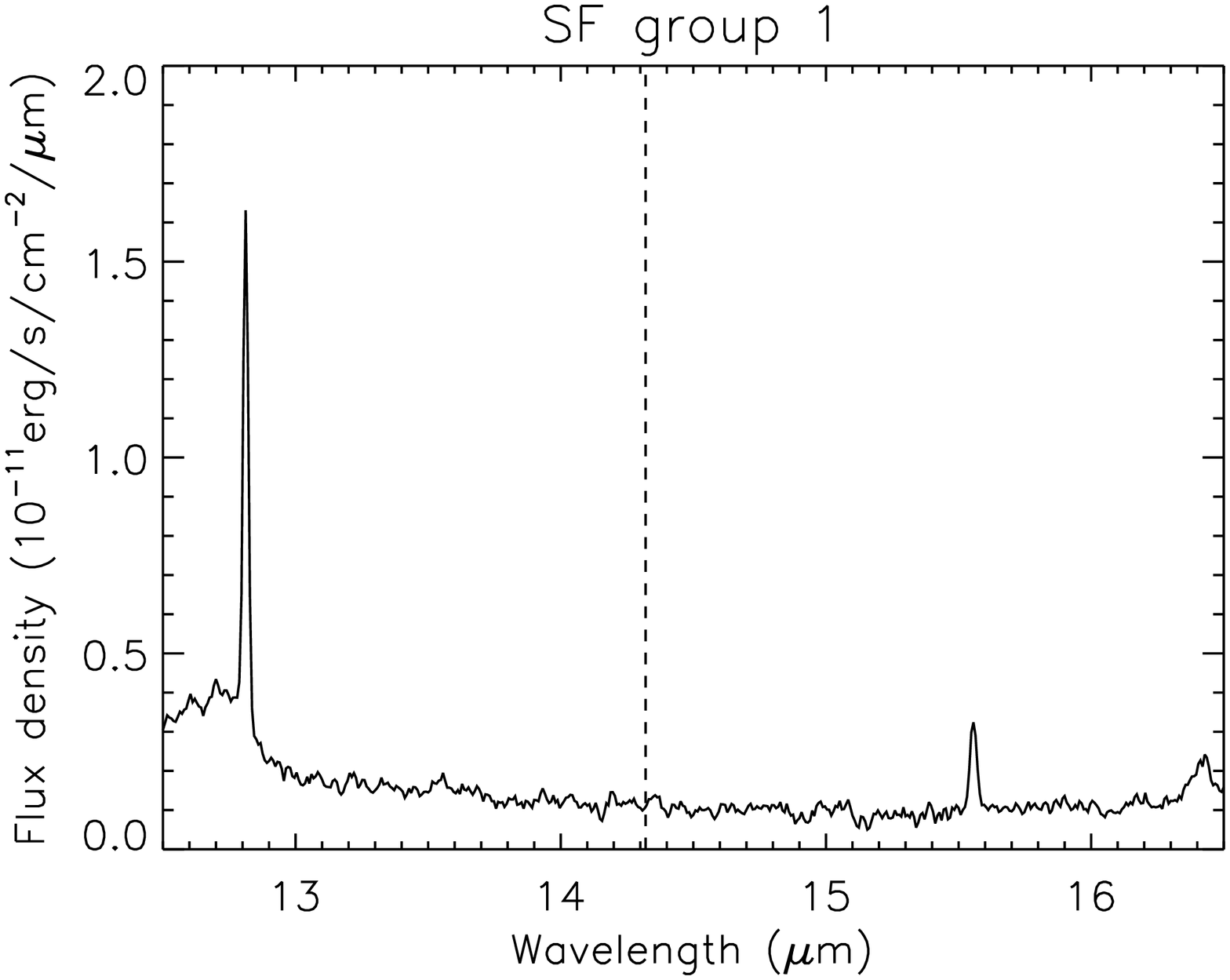}}
\subfigure[]{\includegraphics[scale=0.35,angle=90]{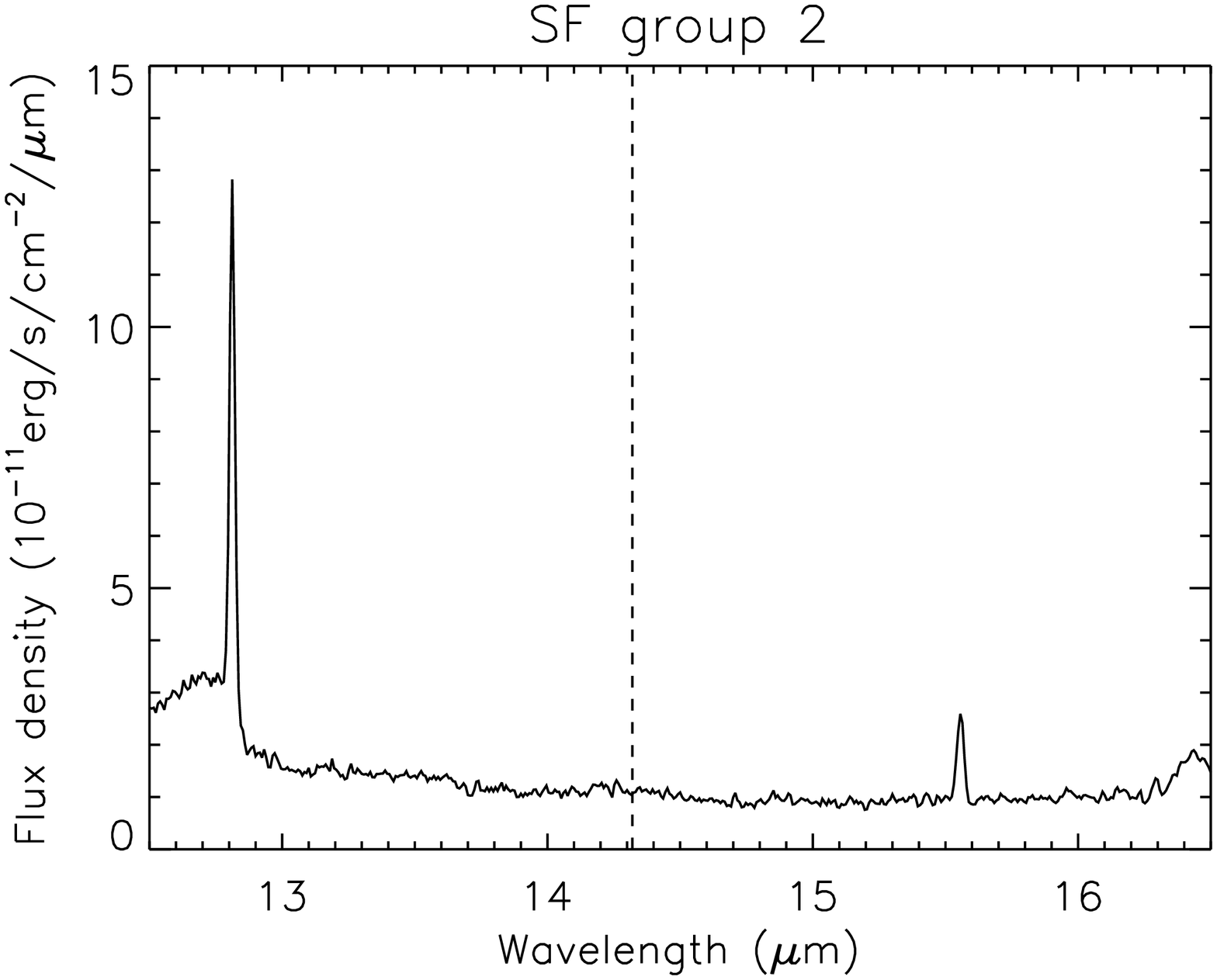}}
\subfigure[]{\includegraphics[scale=0.35,angle=90]{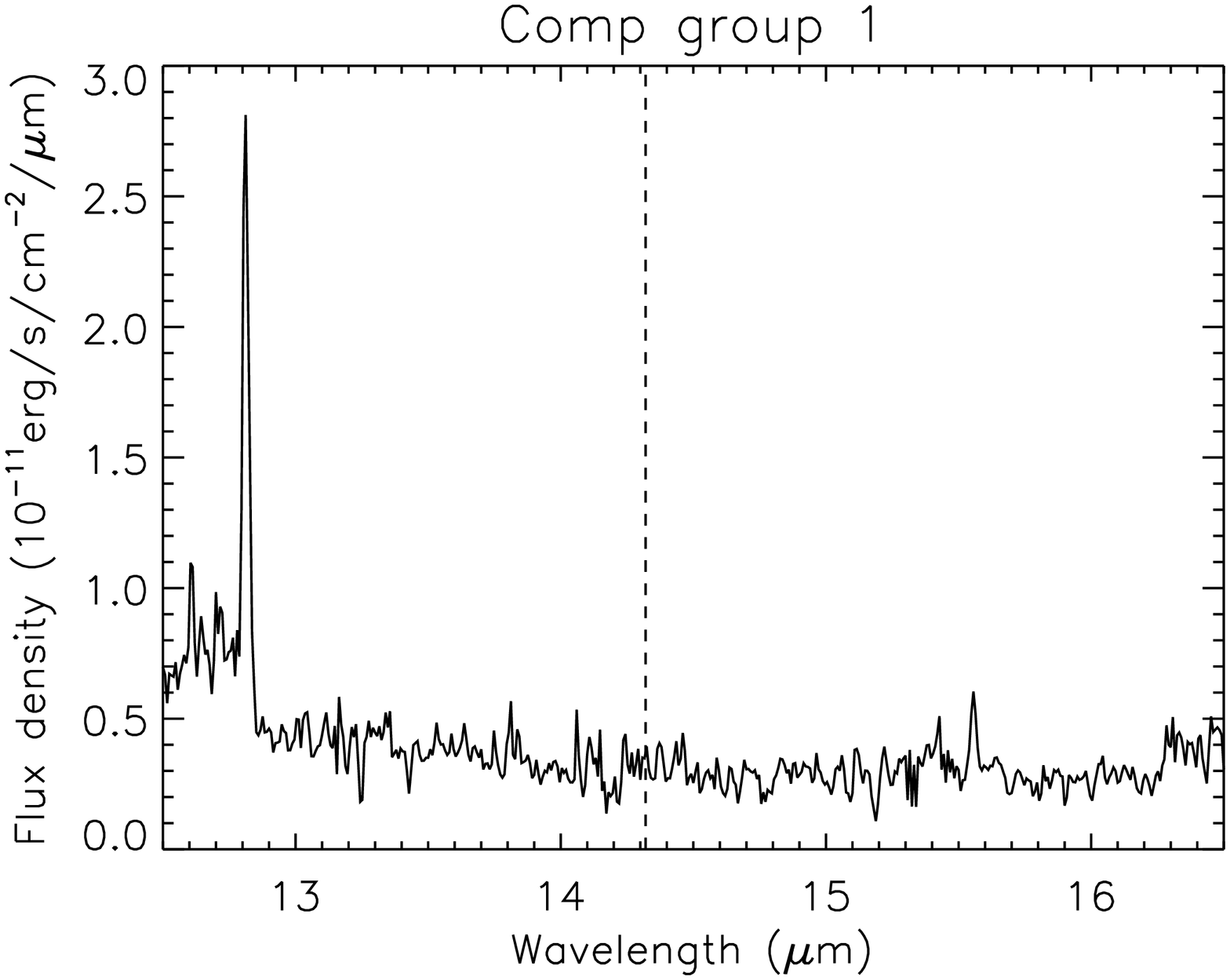}}
\subfigure[]{\includegraphics[scale=0.35,angle=90]{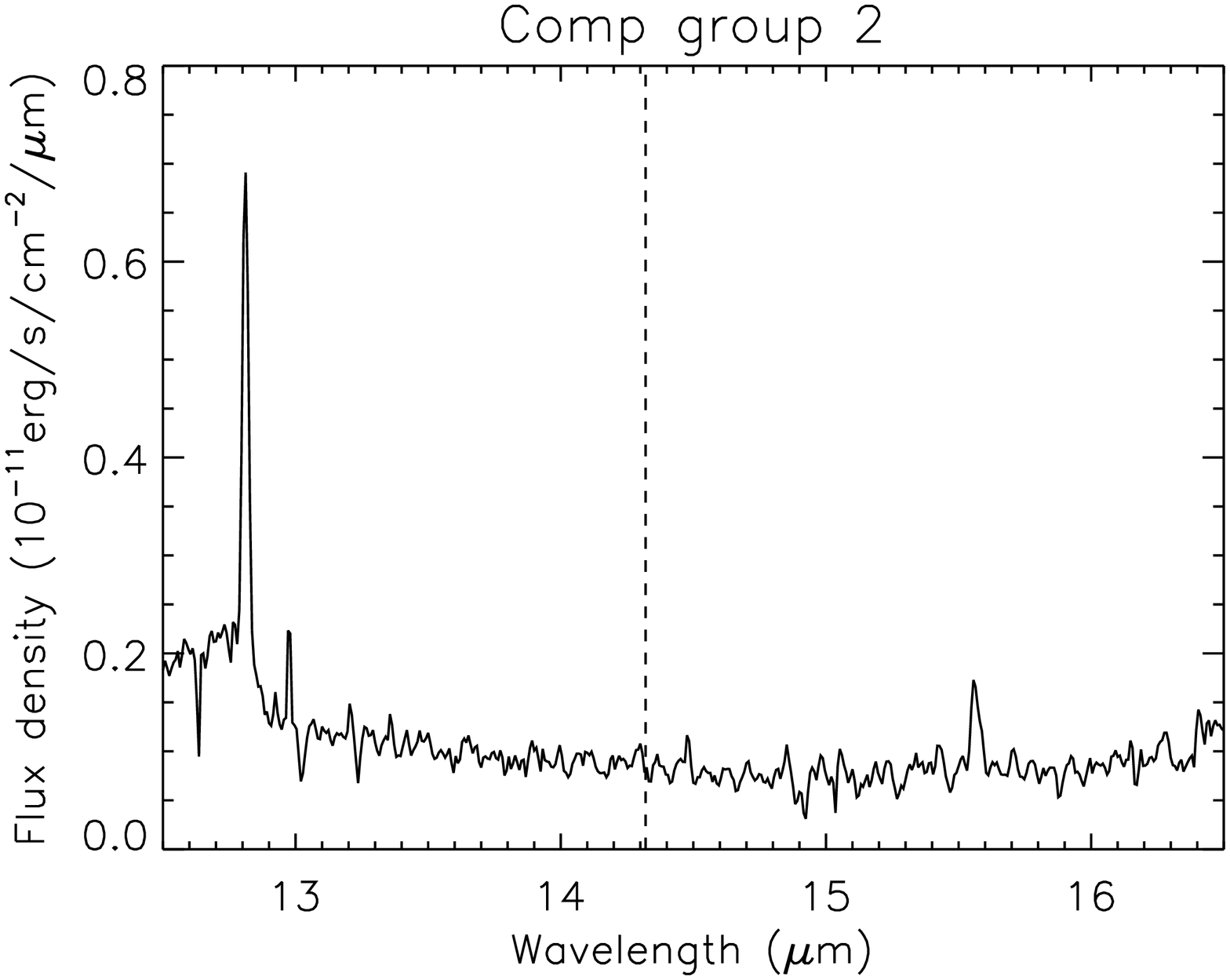}}
\subfigure[]{\includegraphics[scale=0.35,angle=90]{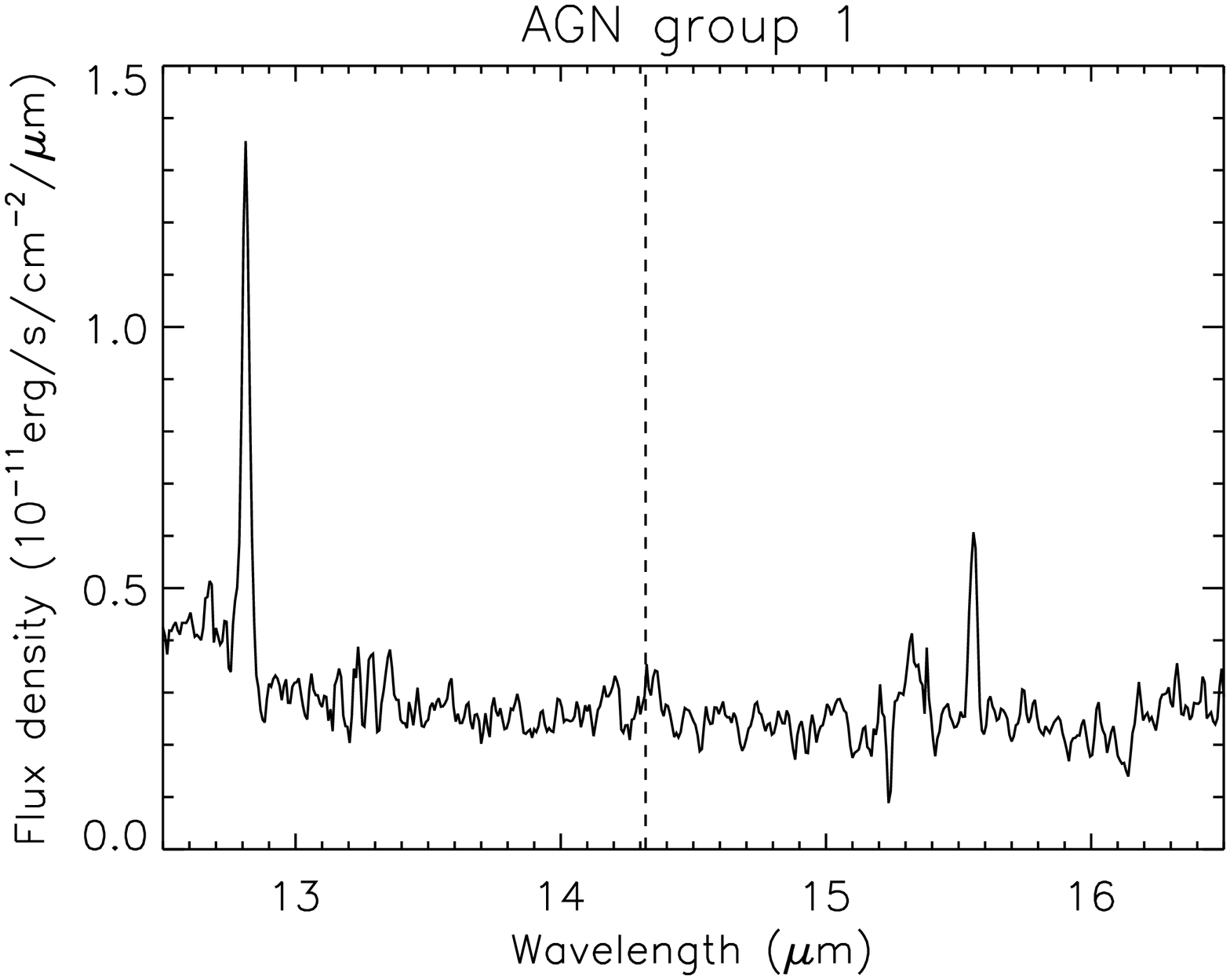}}
\subfigure[]{\includegraphics[scale=0.35,angle=90]{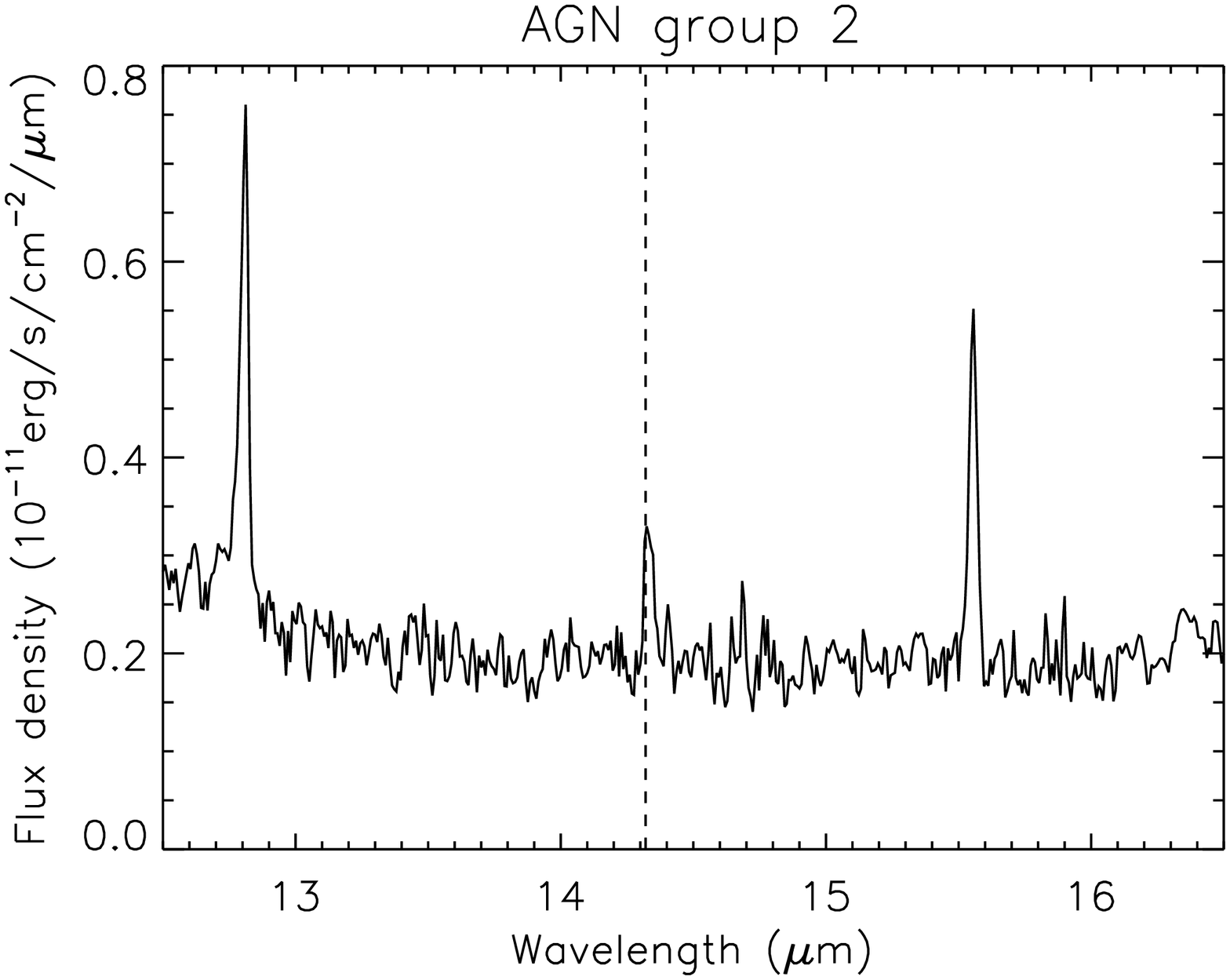}}
\caption[Stacked SH spectra for star-forming galaxies.]{\label{stack_nev}Stacked SH spectra around the [NeV] emission line at 14.32 $\mu$m for each group of galaxies illustrated in Figure \ref{bpt_nev} b). The dashed line indicates the nominal [NeV] wavelength. [NeV] was detected above the 5$\sigma$ level in AGN group 2 only.}
\end{figure}

We stacked together the IRS SH spectra for the sources in each color bin from Figure \ref{bpt_nev} b) not detected in [NeV]. Each spectrum was shifted to the rest-frame, rebinned to a common wavelength grid of equal pixel spacing and normalized by the the average MIR flux between 13.25$\mu$m and 13.75$\mu$m; this window was chosen since it is free of strong emission features. Figure \ref{stack_nev} shows the stacked SH spectra for each galaxy bin. We used a 3$\sigma$ clipping on the average flux in each pixel to reject outliers that would have otherwise introduced erroneous features into our stacked spectra due to noise. [NeV] was detected above the 5$\sigma$ level in just one of these bins, AGN group 2. For the other bins, we obtained [NeV] upper limits using the procedure outlined in Section 3. [NeV]/[NeII] was calculated by normalizing the [NeV] upper limit/detection by the stacked [NeII] emission feature in each group, which was detected at high significance in all cases. In the following analysis, we plot the [NeV]/[NeII] detections and stacked [NeV]/[NeII] upper limits against the optical D value, PAH EW (for the sub-set of sources where the PAH luminosity was detected above the 5$\sigma$ level) and $\alpha_{20-30\mu m}$. For the stacked [NeV]/[NeII] ratios, we calculate the average D, PAH EW and $\alpha{20-30\mu m}$ value in each bin. The [NeV] stacked detection is illustrated by the triangle in the subsequent figures.

\begin{figure}[ht]
\centering
\subfigure[]{\includegraphics[scale=0.35,angle=90]{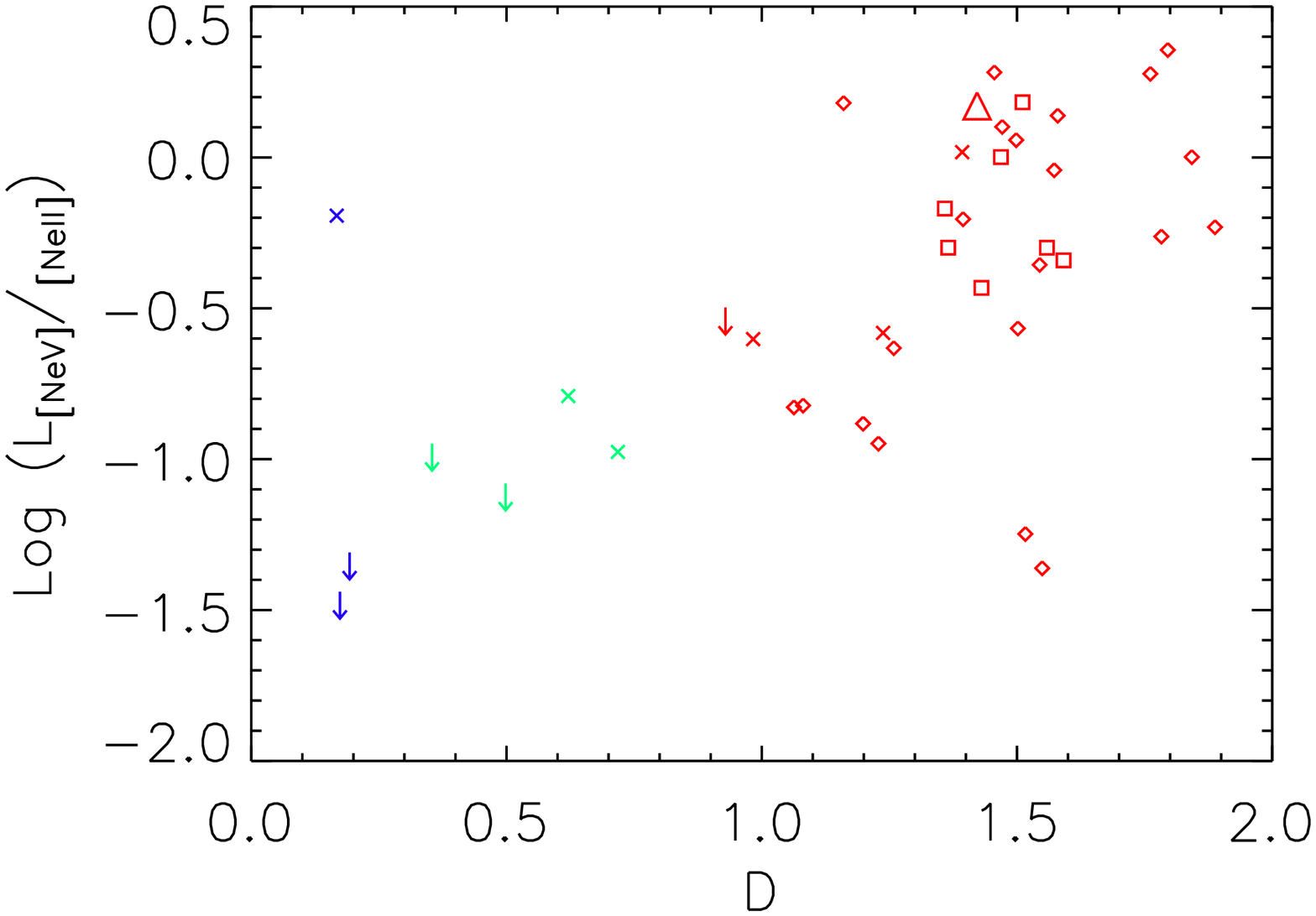}}
\subfigure[]{\includegraphics[scale=0.35,angle=90]{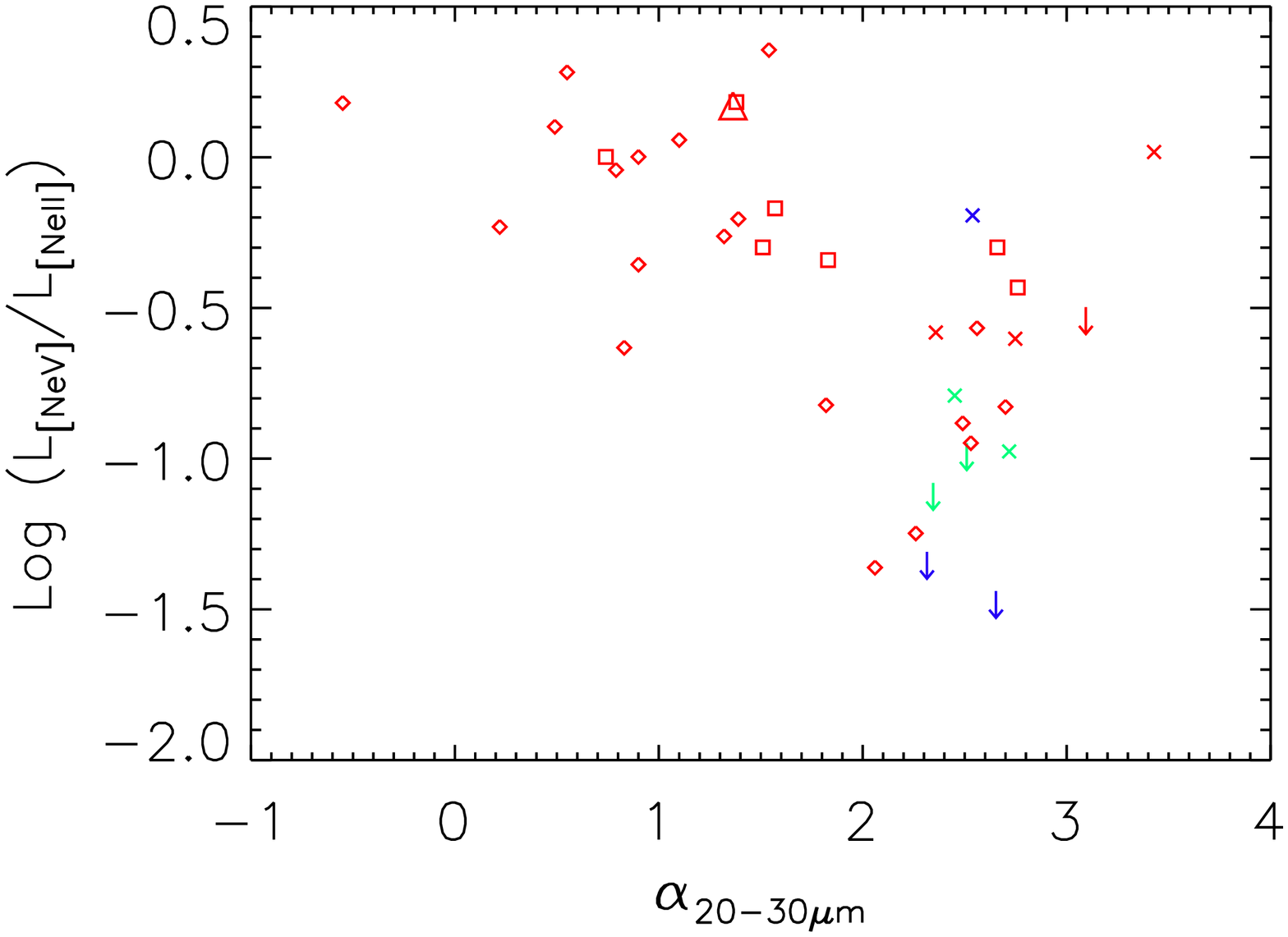}}
\caption[L$_{[NeV]}$/L$_{[NeII]}$ vs. D and $\alpha_{20-30\mu m}$]{\label{nev_ne2_d_alpha}Left: L$_{[NeV]}$/L$_{[NeII]}$ vs. the optical D parameter. Right: L$_{[NeV]}$/L$_{[NeII]}$ vs. $\alpha_{20-30\mu m}$. These quantities are statistically correlated and anti-correlated: $\rho$=0.562 and $\rho$=-0.568, respectively. Color and symbol coding same as Figure \ref{bpt_nev}. The red triangle indicates the [NeV] detection for the stacked spectra in AGN group 2.  For these and subsequent plots with IR emission line and PAH fluxes, error bars are smaller or on the order of the symbol size and therefore not plotted.}
\end{figure}

\begin{figure}[ht]
\centering
\subfigure[]{\includegraphics[scale=0.35,angle=90]{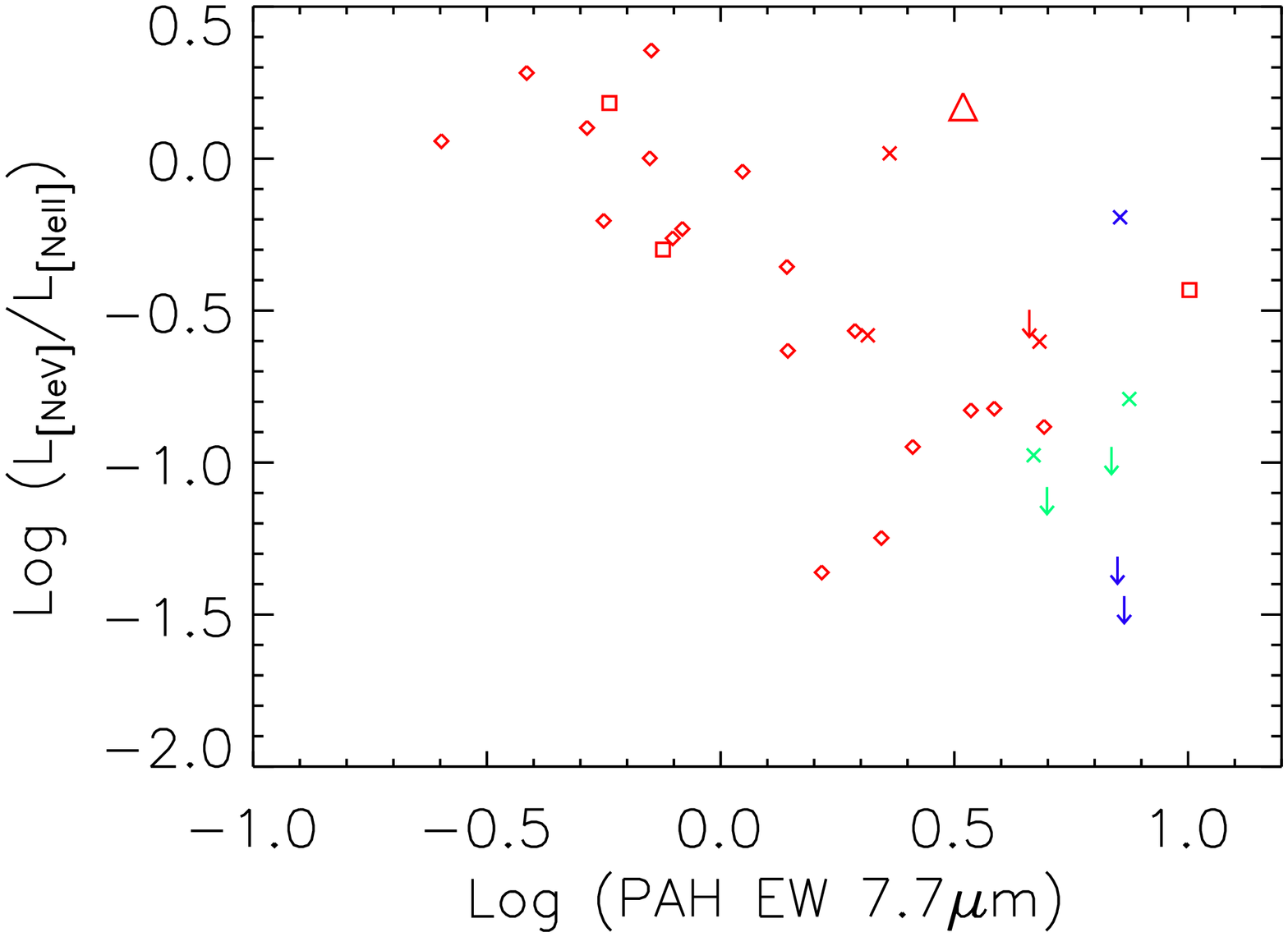}}
\subfigure[]{\includegraphics[scale=0.35,angle=90]{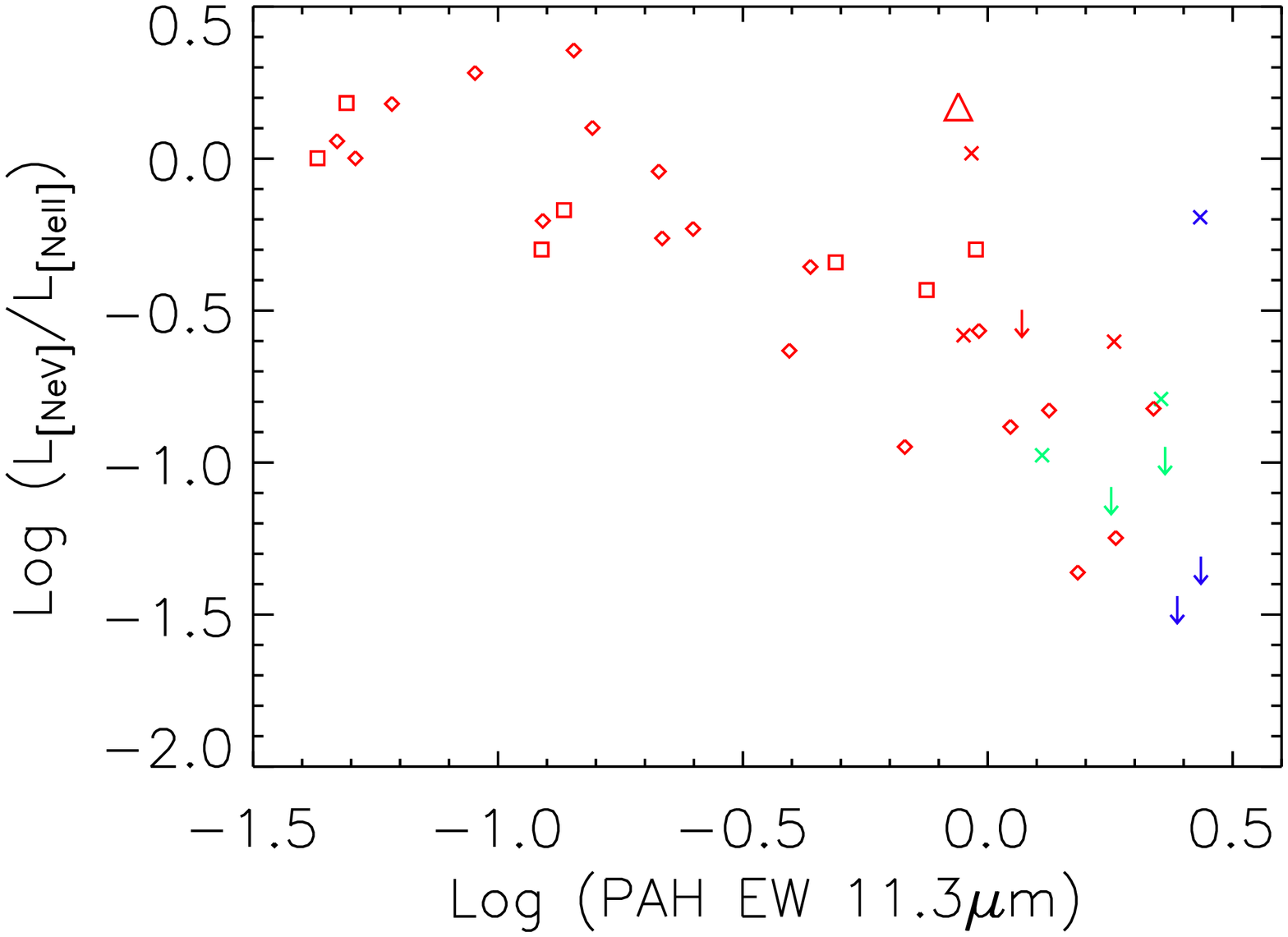}}
\subfigure[]{\includegraphics[scale=0.35,angle=90]{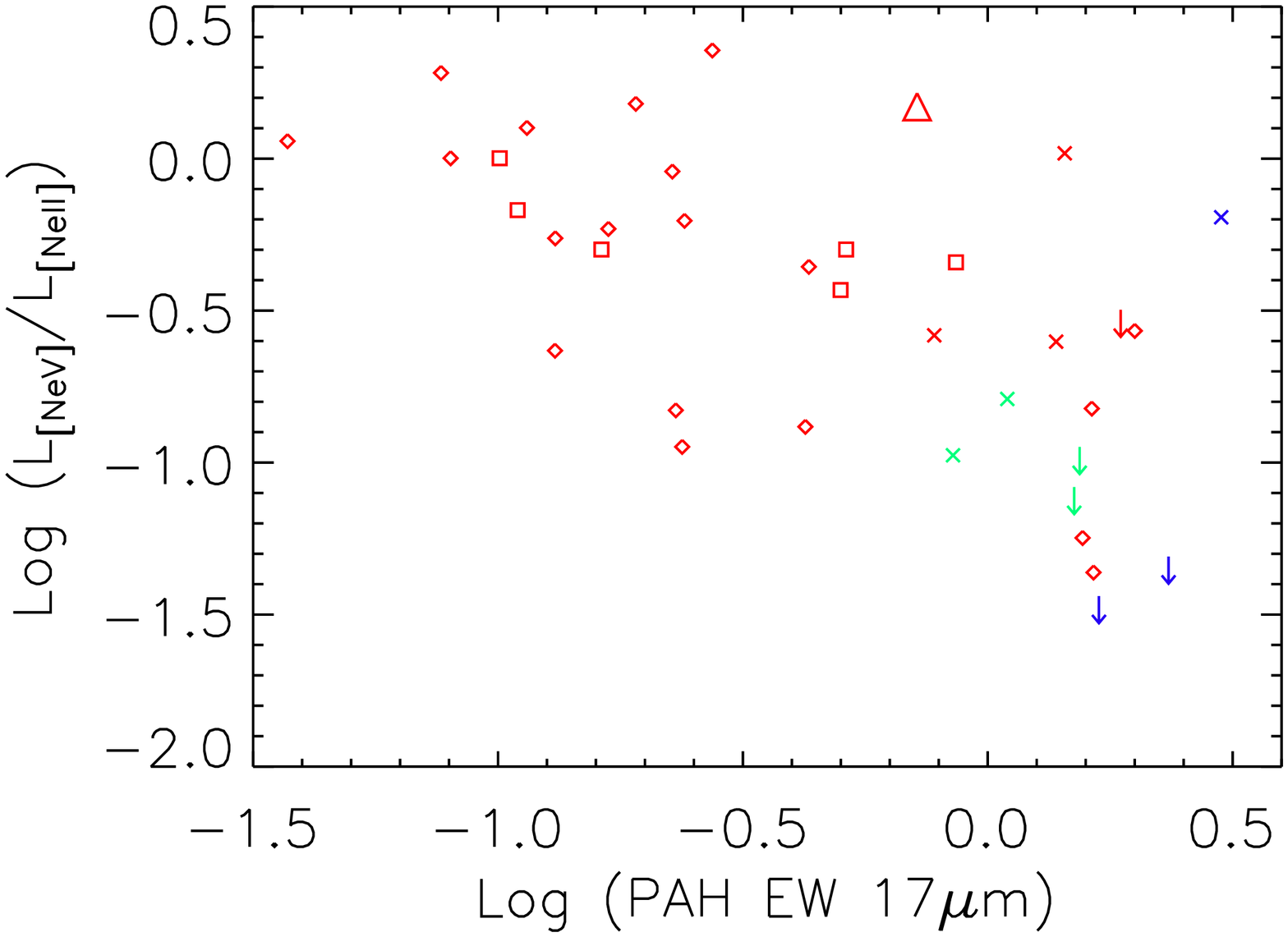}}
\caption[L$_{[NeV]}$/L$_{[NeII]}$ vs. PAH EWs]{\label{nev_ne2_ew}a) L$_{[NeV]}$/L$_{[NeII]}$ vs. PAH EW at 7.7 $\mu$m. b) L$_{[NeV]}$/L$_{[NeII]}$ vs. PAH EW at 11.3 $\mu$m. c) L$_{[NeV]}$/L$_{[NeII]}$ vs. PAH EW at 17 $\mu$m. These quantities are statistically anti-correlated: $\rho$=-0.702, $\rho$=-0.779 and $\rho$=-0.628, respectively. The red triangle indicates the [NeV] detection for the stacked spectra in AGN group 2. Color and symbol coding same as Figure 1. }
\end{figure}

Figures \ref{nev_ne2_d_alpha} - \ref{nev_ne2_ew} illustrate the trends between [NeV]/[NeII] and the other diagnostics of radiation field hardness. As expected, [NeV]/[NeII] is significantly correlated with D and significantly anti-correlated with $\alpha_{20-30\mu m}$ and PAH EWs, as seen in the correlation coefficients ($\rho$) and associated probabilities that the two quantities are uncorrelated ($P_{uncorr}$) reported in Table \ref{rel_agn_sfr}. We note that one galaxy classified as a quiescent star-forming galaxy based on its optical (SDSS) spectrum has a [NeV] detection. This suggests that it is harboring a ``hidden'' AGN, i.e. an AGN embedded in very dusty host galaxy that attenuates the optical signatures used to identify AGN. However, we note that this source has other IR parameters (i.e. $\alpha_{20-30\mu m}$, PAH EWs) more consistent with those of typical star-forming galaxies. This may argue against ``hidden'' AGN interpretation, which may then call into question using solely [NeV] detection as an unambiguous signature of AGN activity.

{\it In any case, we emphasize that the optical and mid-IR classifications of galaxies as being dominated by star formation or by a Type 2 AGN agree in the overwhelming majority of cases.} This is good news, since it shows that SDSS classifications (which are available for roughly one million galaxies) are highly reliable. We do not find a large population of optically hidden AGN, as found by Goulding \& Alexander (2009). Over half of the [NeV] detected sources in their volume limited sample of nearby galaxies are not optically classified as AGN. However, they do not include optically classifed LINERS or composites as ``optically classified AGN,'' while we do consider these to contain AGN. The fraction of [NeV] detected sources in their sample that are classified optically as quiescent star-forming galaxies is 24\% (4 of 17), which is still a substantially larger fraction than what we find.

\subsection{$\alpha_{20-30\mu m}$}
\begin{figure}[ht]
\centering
\subfigure[]{\includegraphics[scale=0.35,angle=90]{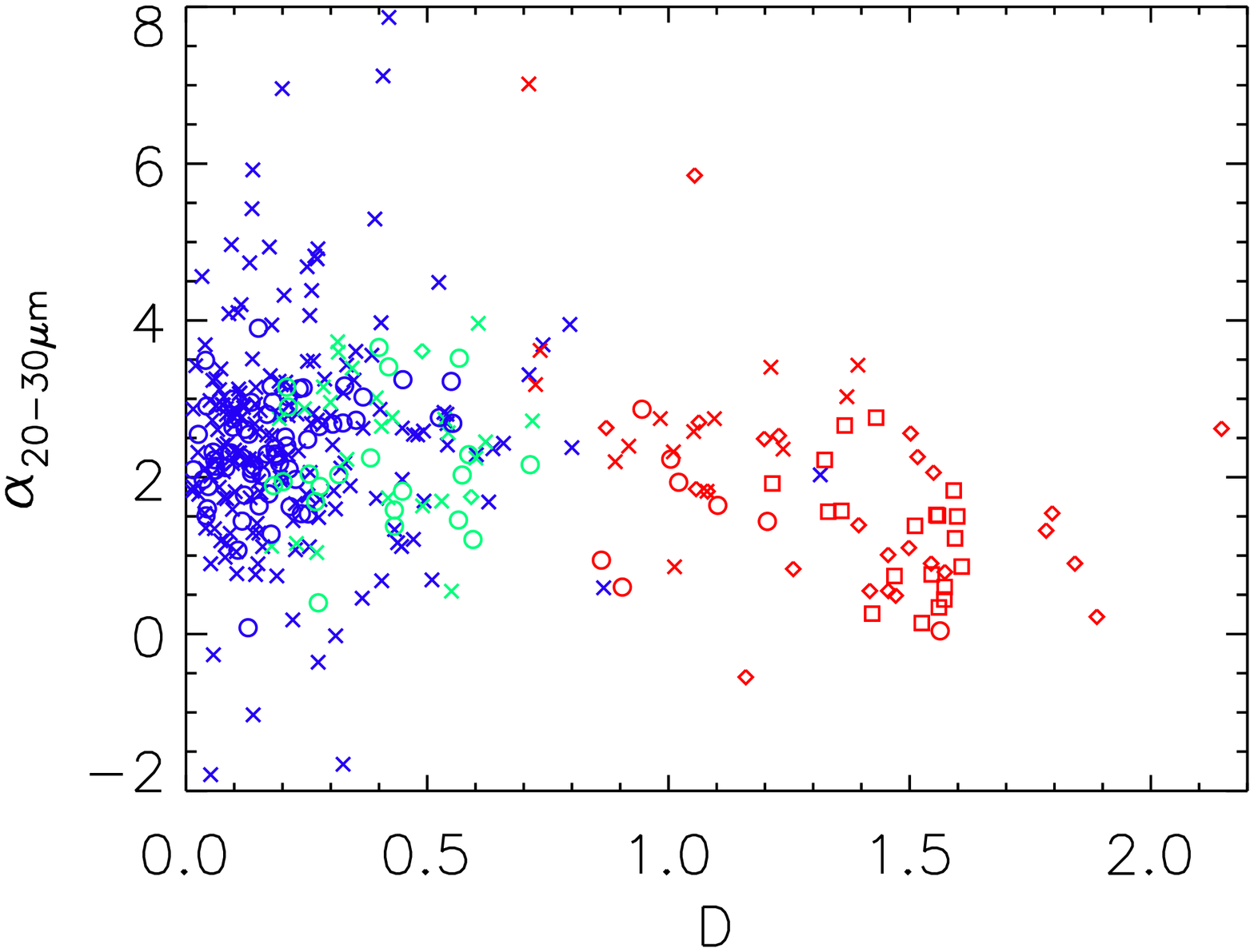}}
\caption[$\alpha_{20-30\mu m}$ vs. D]{\label{alpha_d}$\alpha_{20-30\mu m}$ vs. the optical D parameter. These parameters are correlated for AGN, $\rho$=-0.474, but not for the full sample, $\rho$=-0.195. Color and symbol coding same as Figure \ref{bpt_nev}. }
\end{figure}

The mid-IR spectral slope index, $\alpha_{20-30\mu m}$, increases with the relative amount of star formation \citep{Buchanan, Deo}. We plot $\alpha_{20-30\mu m}$ as a function of the optical D parameter in Figure \ref{alpha_d} which shows that a wide range of $\alpha_{20-30\mu m}$ values are present for the star-forming galaxies. Consequently, there is no significant anti-correlation present in the full sample, with correlation coefficient ($\rho$) of -0.195. However, if we consider just the AGN, a significant trend is apparent: $\rho$=-0.474. This correspondence likely drives the significant anti-correlation found for [NeV]/[NeII] vs $\alpha_{20-30\mu m}$. The IR spectral index is thus useful in parameterizing the relative contribution of star formation in active galaxies but this appears to break down in the quiescent galaxy regime. Consequently, we do not utilize $\alpha_{20-30\mu m}$ as a tracer of ionization field hardness in the subsequent analysis.

\subsection{PAH EW and Flux Ratios}
Previous studies have indicated that the PAH equivalent widths (EWs) are anti-correlated with the presence of increasing AGN activity. This can be due to two effects. First, as the AGN becomes increasingly dominant energetically, the emission due to warm dust heated by the AGN will rise. Even if the intrinsic PAH emission is unaffected, this increase in the mid-IR continuum flux will cause the PAH EWs to decrease. A second possible effect would be due to the actual destruction of dust grains by the hard ionization field from the AGN central engine \citep{Voit}. Our new expanded data set that spans the range from quiescent systems to strong AGN provides the opportunity to test these processes over a full range of ionization field hardness and to test which features may be most sensitive to AGN activity. 

\begin{figure}[ht]
\centering
\subfigure[]{\includegraphics[scale=0.35,angle=90]{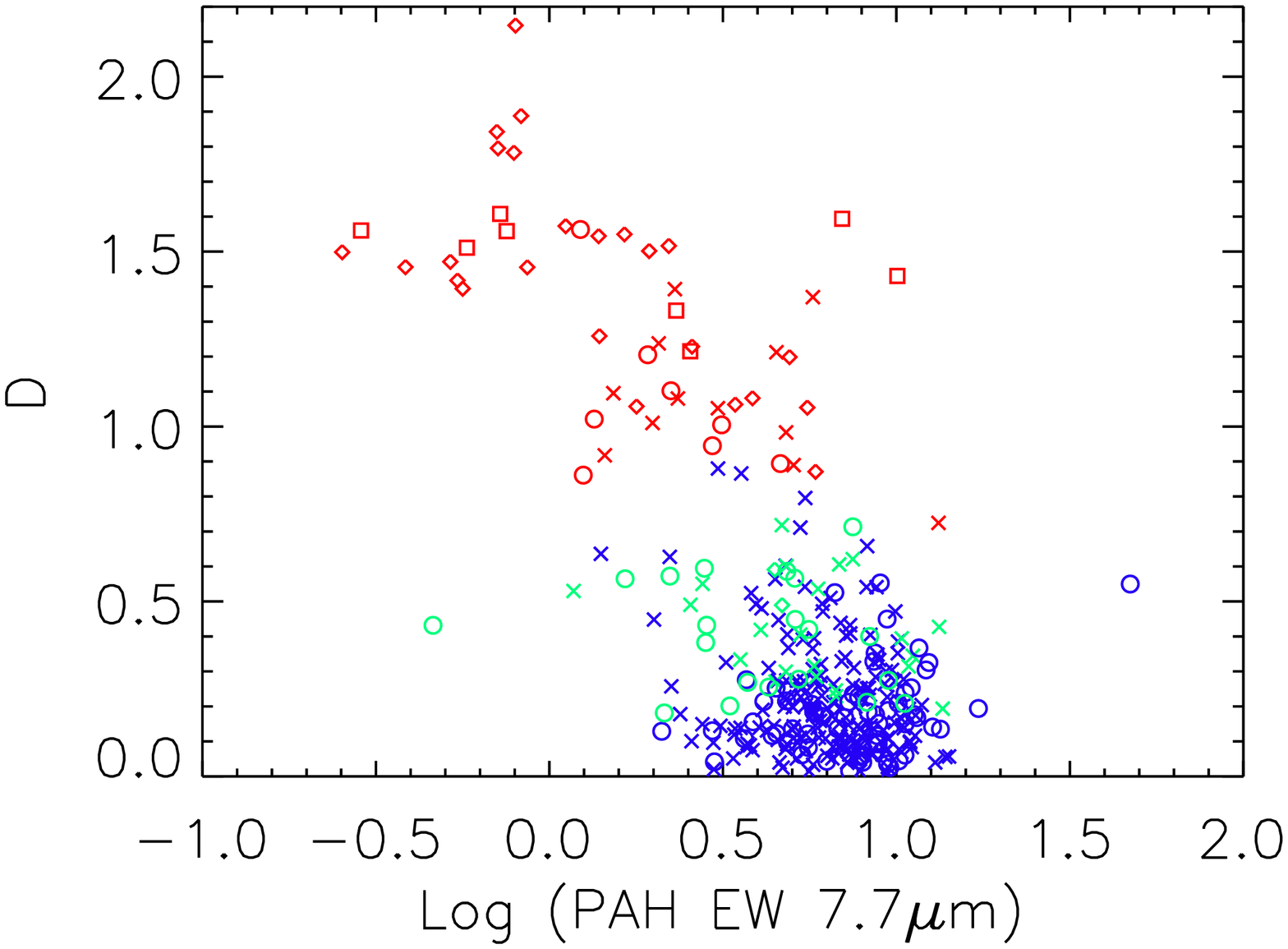}}
\subfigure[]{\includegraphics[scale=0.35,angle=90]{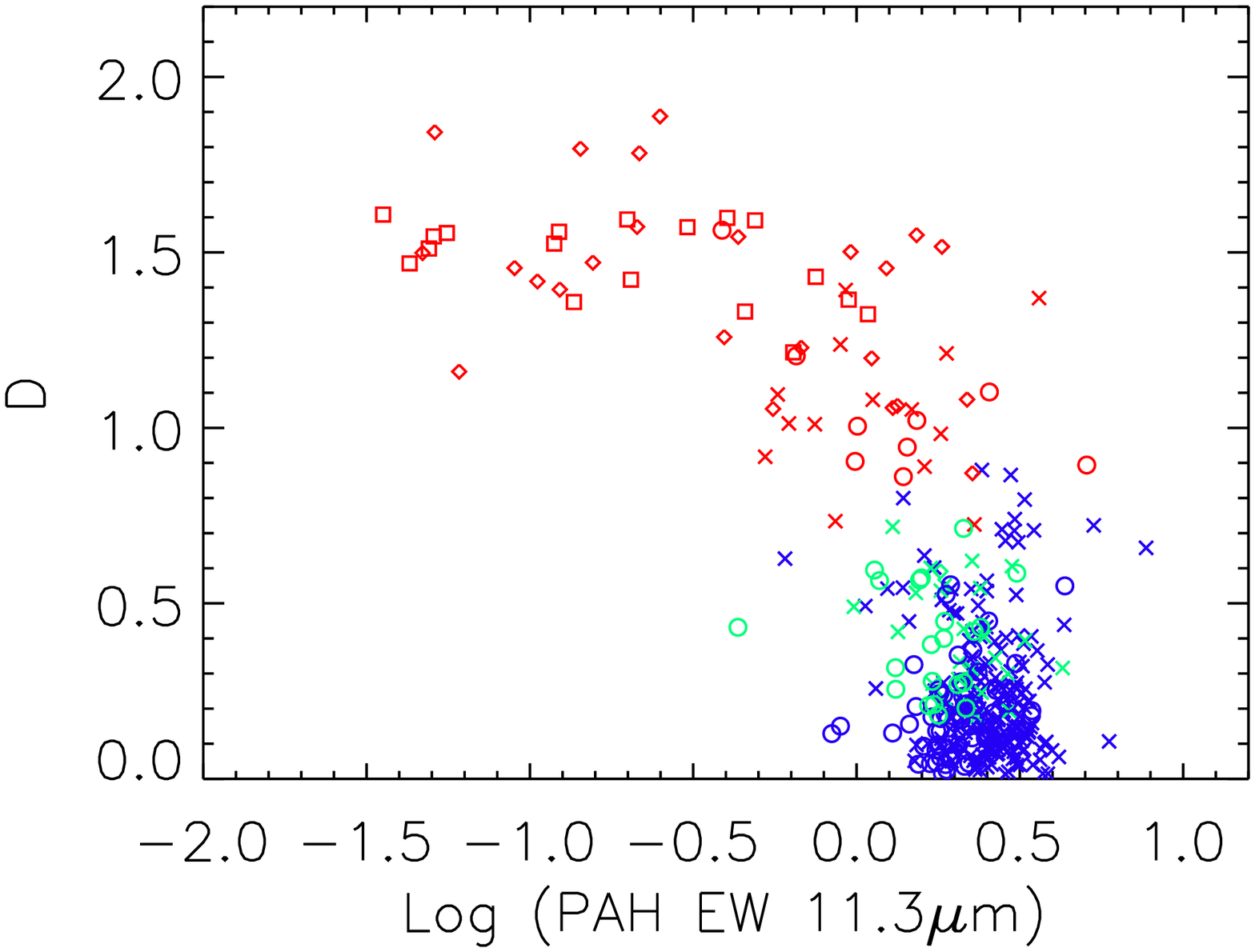}}
\subfigure[]{\includegraphics[scale=0.35,angle=90]{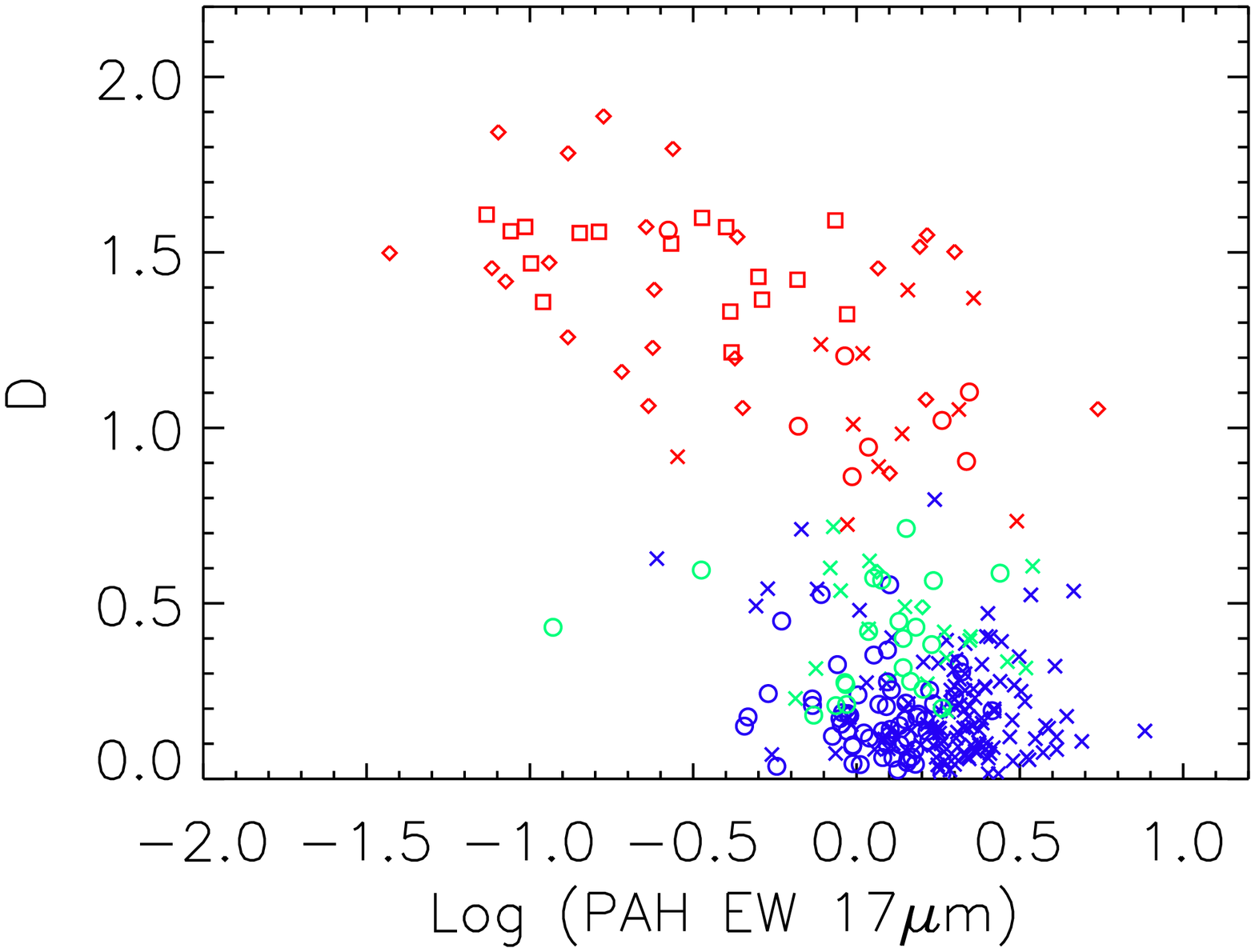}}
\caption[D vs. PAH EWs]{\label{ew_d}The optical D parameter vs a) PAH EW at 7.7 $\mu$m, b) PAH EW at 11.3 $\mu$m  and c) PAH EW at 17 $\mu$m. These quantities are statistically anti-correlated: $\rho$=-0.679, $\rho$=-0.747 and $\rho$=-0.658, respectively. Color and symbol coding same as Figure 1. }
\end{figure}

Figure \ref{ew_d} plots the PAH EW at 7.7 $\mu$m, 11.3 $\mu$m and 17$\mu$m as a function of the optical D parameter. The relationship between D and the PAH EWs is significant, with $\rho <$-0.65 (Table \ref{ew}). However, if we focus on just the star-forming galaxies and composites, this significant anti-correlation disappears: $\rho$=-0.227, -0.125 and -0.230 for the 7.7$\mu$m, 11.3$\mu$m and 17 $\mu$m features, respectively. Hence, it appears that only in the objects that are optically classified as AGN is the PAH EW appreciably affected, suggesting the AGN contributes a significant fraction of the mid-IR continuum or perhaps destroys PAHs.

\begin{figure}[ht]
\centering
\subfigure[]{\includegraphics[scale=0.35,angle=90]{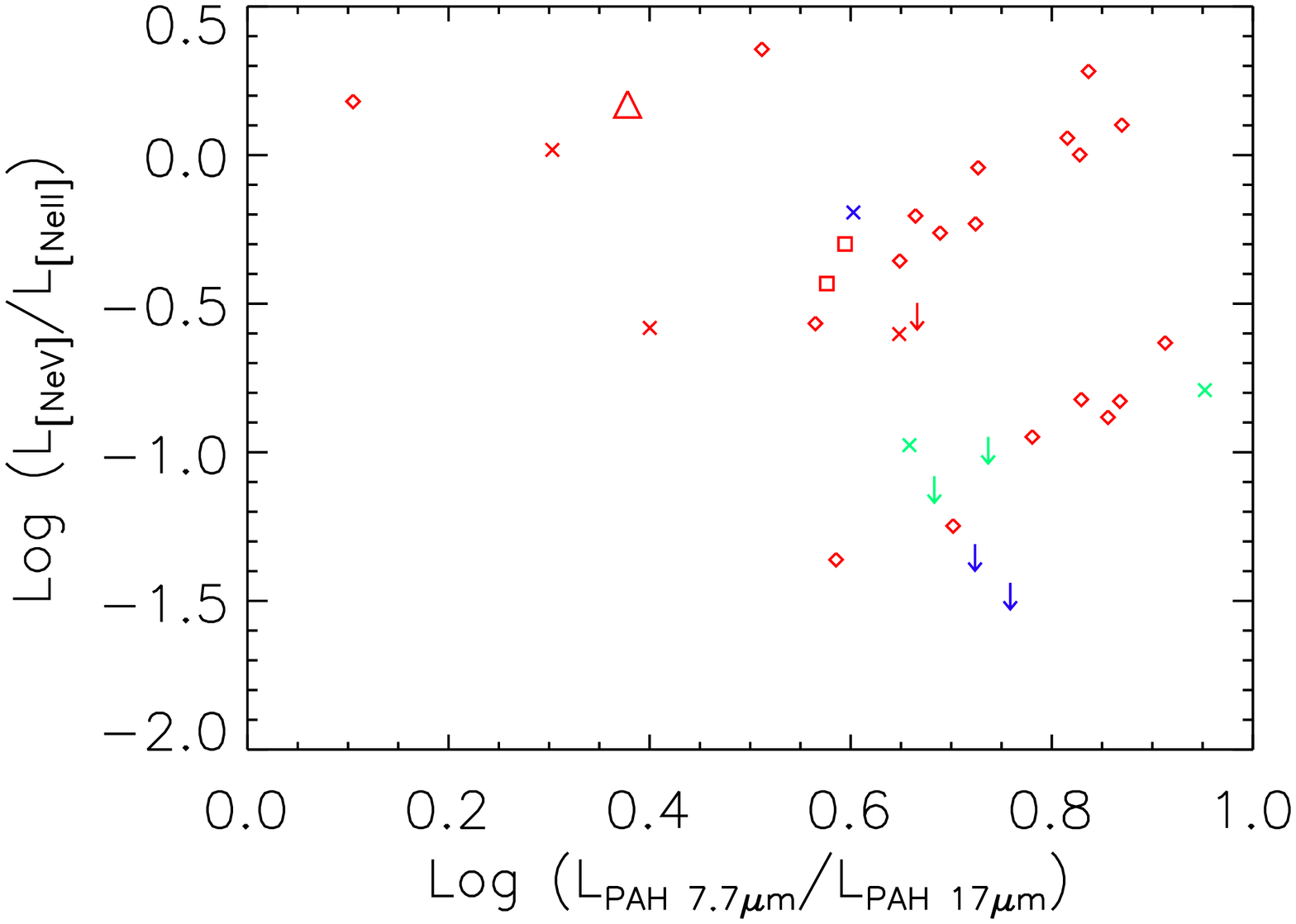}}
\subfigure[]{\includegraphics[scale=0.35,angle=90]{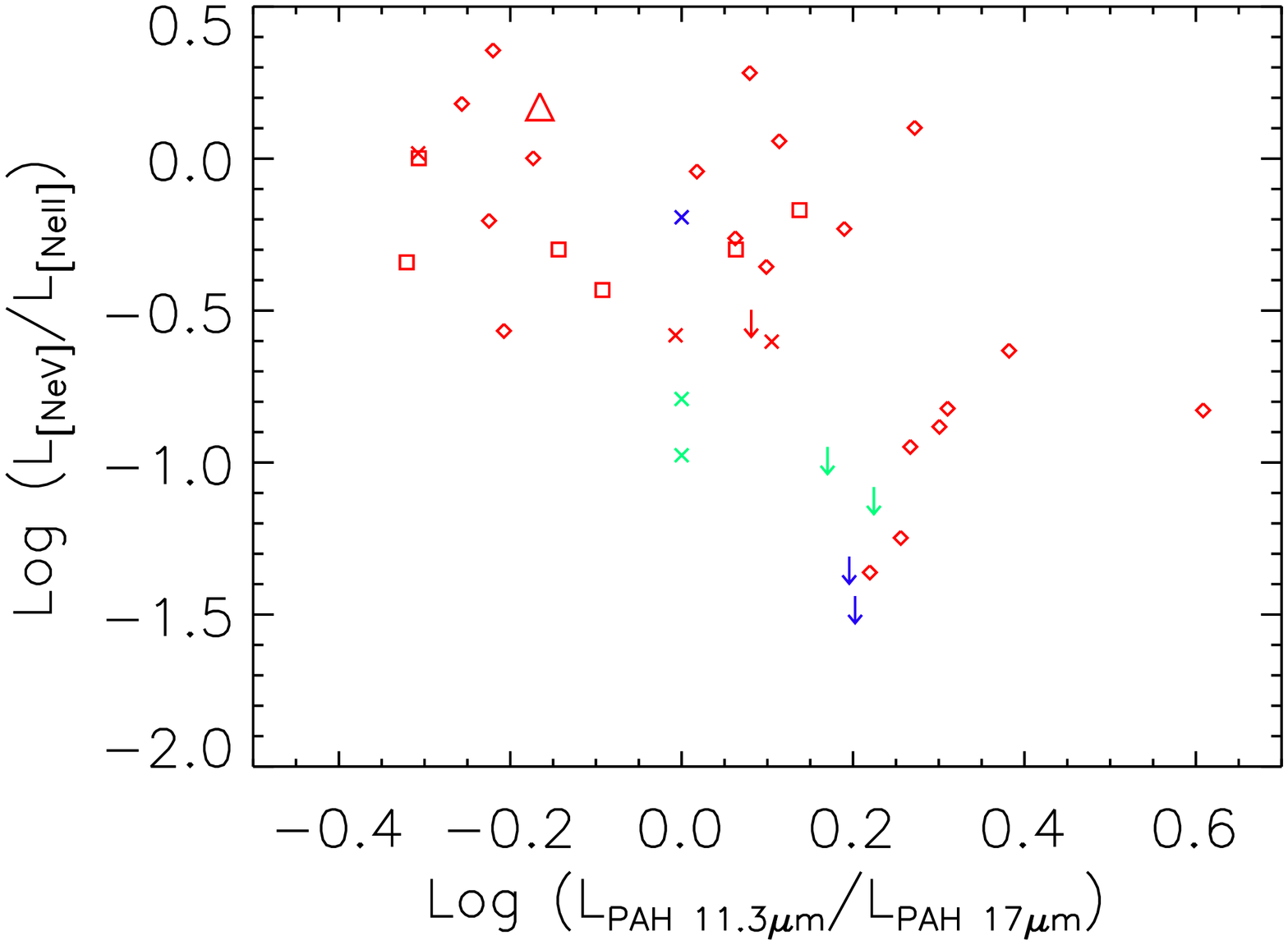}}
\subfigure[]{\includegraphics[scale=0.35,angle=90]{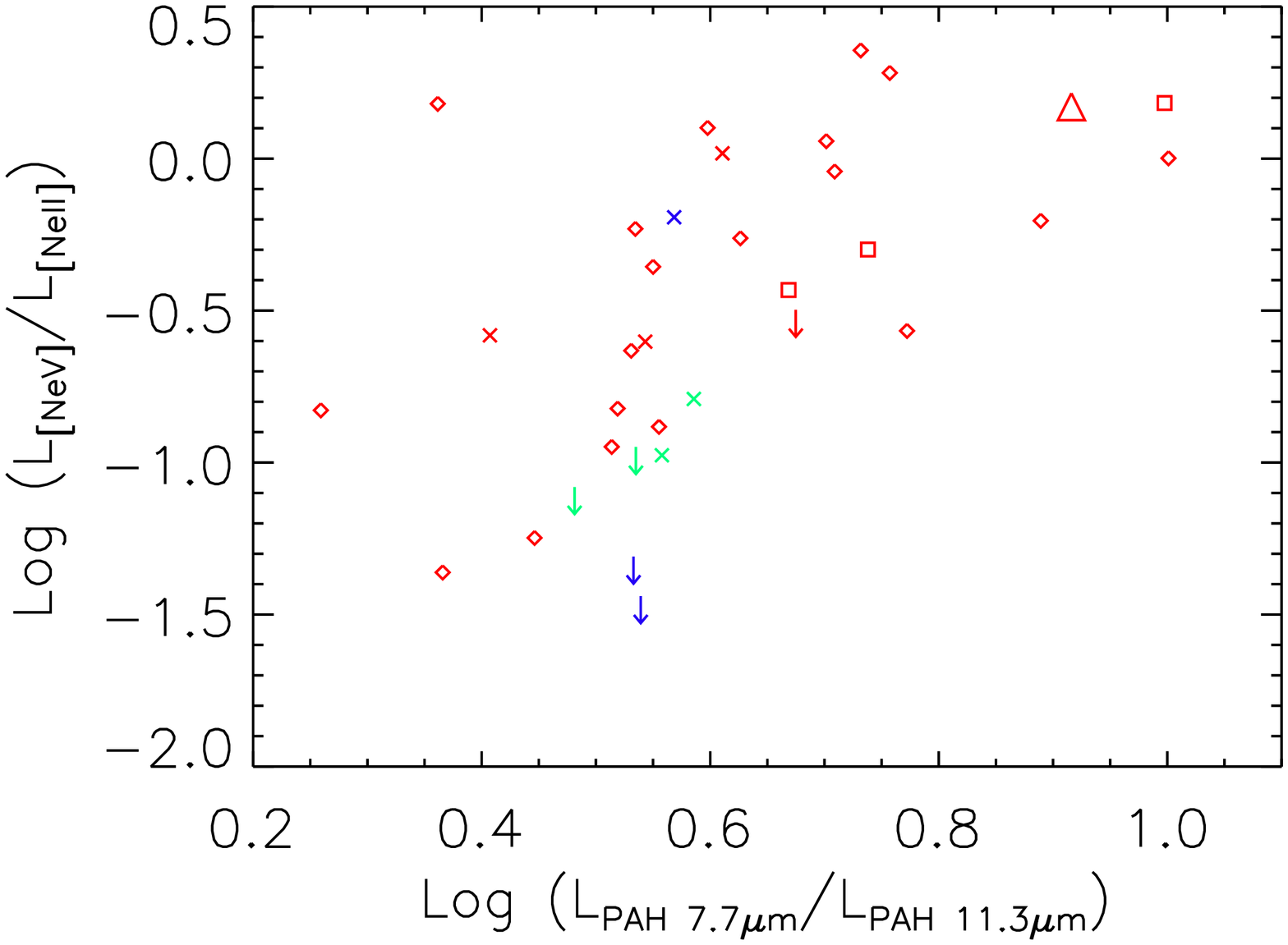}}
\caption[L$_{[NeV]}$/L$_{[NeII]}$ vs. PAH ratios]{\label{ratios_nev_ne2}L$_{[NeV]}$/L$_{[NeII]}$ vs. a) L$_{PAH 7.7\mu m}$/L$_{PAH 17\mu m}$ (no significant trend, $\rho$=-0.241), b) L$_{PAH 11.3\mu m}$/L$_{PAH 17\mu m}$ (significant anti-correlation, $\rho$=-0.564) and c) L$_{PAH 7.7\mu m}$/L$_{PAH 11.3\mu m}$ (significant correlation, $\rho$=0.611). The red triangle indicates the [NeV] detection for the stacked spectra in AGN group 2. Color and symbol coding same as Figure 1. }
\end{figure}

\begin{figure}[ht]
\centering
\subfigure[]{\includegraphics[scale=0.35,angle=90]{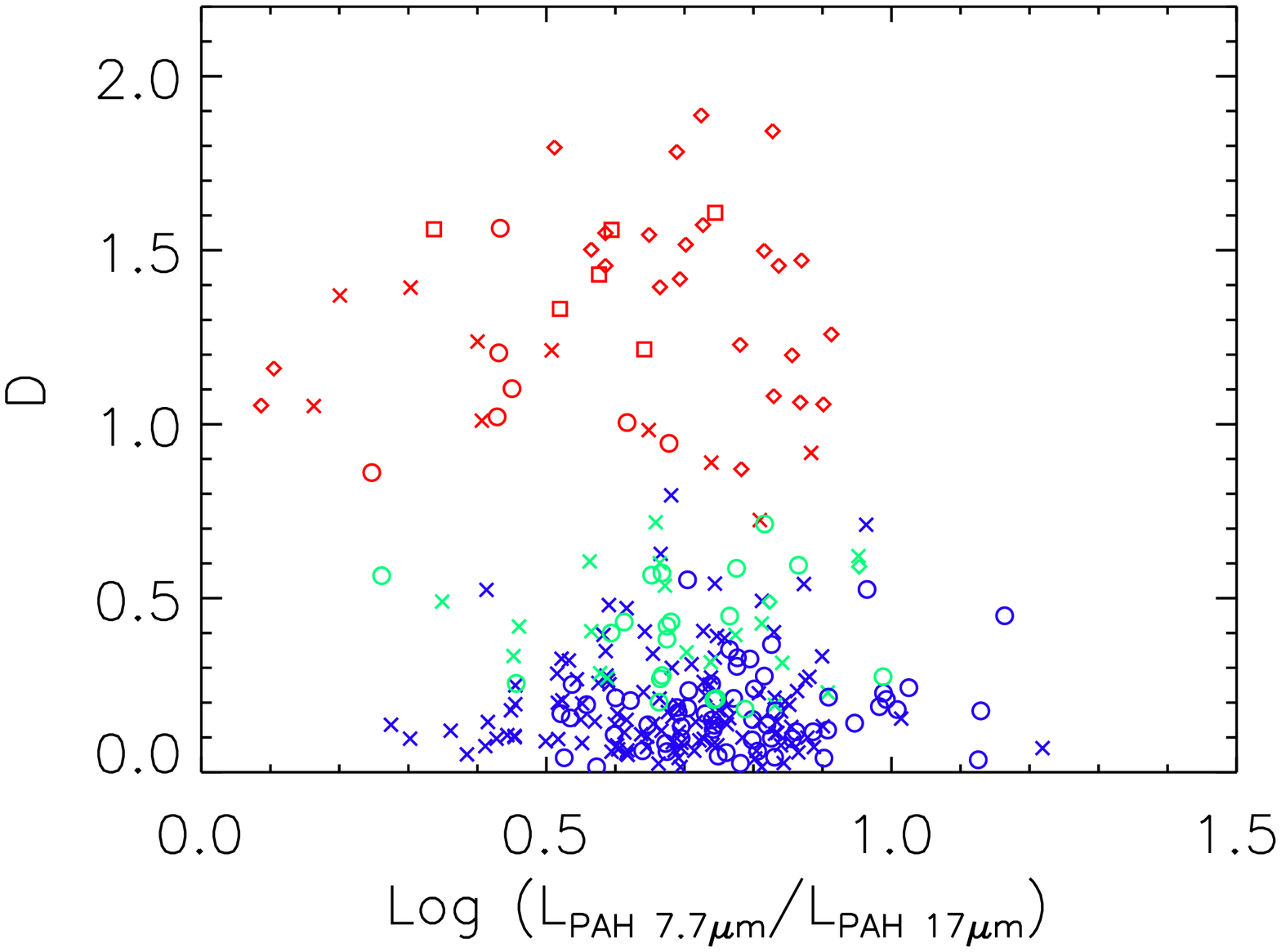}}
\subfigure[]{\includegraphics[scale=0.35,angle=90]{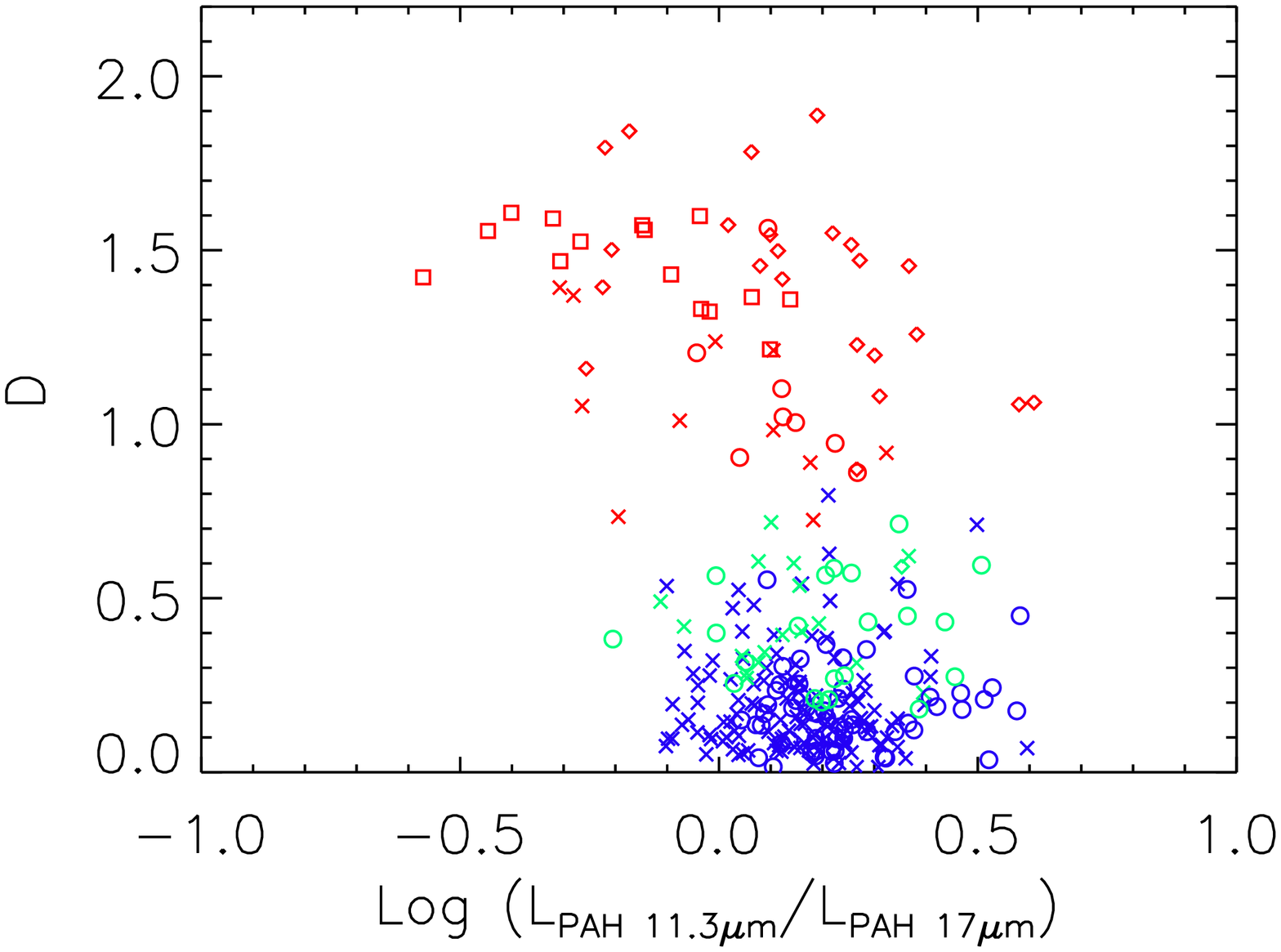}}
\subfigure[]{\includegraphics[scale=0.35,angle=90]{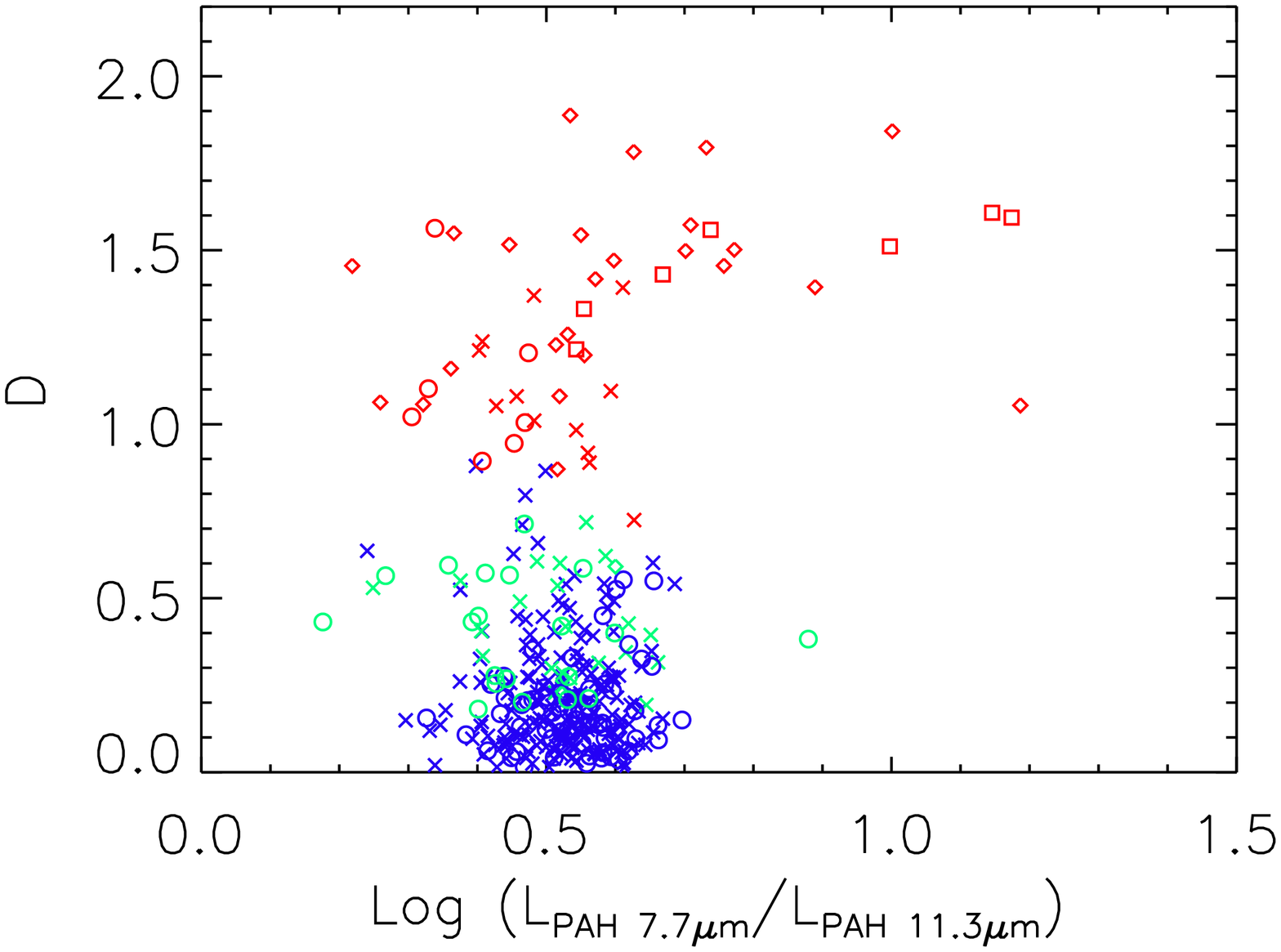}}
\caption[Optical D parameter vs. PAH ratios]{\label{ratios_d}Optical D parameter vs. a) L$_{PAH 7.7\mu m}$/L$_{PAH 17\mu m}$, b) L$_{PAH 11.3\mu m}$/L$_{PAH 17\mu m}$, $\rho$=-0.364) and  c) L$_{PAH 7.7\mu m}$/L$_{PAH 11.3\mu m}$. Color and symbol coding same as Figure 1. }
\end{figure}

As noted above, the PAH EW are sensitive to both the relative amount of mid-IR continuum from the AGN as well as effects on the PAH emission itself. A complementary approach that isolates the effect of the AGN directly on the PAH emission is to consider changes in the ratios of the different PAH features as the AGN becomes more dominant. Thus, similar to the work of \citet{O'Dowd} and \citet{DS10}, we investigate the relationship between radiation field hardness (parameterized by [NeV]/[NeII] and D) and the ratios of the PAH fluxes in Figures \ref{ratios_nev_ne2} and \ref{ratios_d}. Since the [NeV] line is only detected in the optically-classifed AGN (with few exceptions), we compare D to the PAH ratios for the AGN subset alone, as well as for the full sample (including star-forming and composite systems). Qualitatively, it appears that the distribution of the AGN are different in these figures from the star forming and composite objects. For the most AGN-dominated sources (D $>$ 1.2 and log(L$_{[NeV]}$/L$_{[NeII]}$) $> -0.5$), there is a subset of the population that has abnormally low ratios of PAH 11.3 $\mu m$ to 17 $\mu m$ and abormally high ratios of PAH 7.7 $\mu m$ to 11.3 $\mu m$. A similar but weaker effect is seen in the PAH 7.7 $\mu m$ to 17 $\mu m$ ratios. These distributions do not appear to take the form of a gradual correlation across the whole galaxy population, but rather appear as a threshold effect at high values of D and log (L$_{[NeV]}$/L$_{[NeII]}$).

Despite this caveat, we have quantified the statistical significances of overall trends that may be present. We emphasize that the correlations may be weak, but the trends found can be quantified as statistically significant as indicated by the correlation coefficients ($\rho$) listed in Table \ref{pah_ratio}. L$_{PAH 7.7\mu m}$/L$_{PAH 17\mu m}$ displays no significant trend with ionization hardness, regardless of the radiation field parameterization used and whether the full sample or AGN only are considered. Using [NeV]/[NeII] as a probe of the ionization field, a significant trend becomes apparent with L$_{PAH 11.3\mu m}$/L$_{PAH 17\mu m}$ ($\rho$=-0.564) and L$_{PAH 7.7\mu m}$/L$_{PAH 11.3\mu m}$ ($\rho$=0.611). However, with D as a diagnostic of the incident radiation, the L$_{PAH 11.3\mu m}$/L$_{PAH 17\mu m}$  relationship becomes less significant in the full sample ($\rho$=-0.364) and even less correlated for the AGN sub-sampple ($\rho$=-0.247). L$_{PAH 7.7\mu m}$/L$_{PAH 11.3\mu m}$ is not significantly correlated in the full sample ($\rho$=0.179), but becomes more signficant in the AGN sample ($\rho$=0.431). Taken together, these results suggest that the PAH feature at 11.3$\mu$m is suppressed in AGN-dominated systems relative to the 7.7 and 17 $\mu m$ features. This disagrees with the results of \citet{PAHFIT}, \citet{O'Dowd}, \citet{Wu10} and \citet{DS10} who find that ratio of L$_{PAH 7.7\mu m}$/L$_{PAH 11.3\mu m}$ is smaller in AGN than in star-forming galaxies. 

The disagreement with the first three studies listed above could result from the inclusion of a greater number of high-luminosity AGN-dominated in our analysis. Clearly, AGN affect PAH macro-molecules and this effect could be more pronounced in more luminous AGN. The inconsistency with \citet{DS10} could be due to their Seyfert sample being systematically closer. Perhaps the AGN effect on PAHs are different in the region hundreds of parsecs around the central engine, as probed by the \citet{DS10} analysis, rather than the kiloparsec to galaxy-wide scale that we are investigating where there is a greater interplay between host galaxy star formation and AGN activity. Different screening effects at different physical scales may play an important role. More advanced theoretical work on the effect of the AGN ionization field on PAH macro-molecules would help to understand the physical causes of these inconsistencies.

\begin{figure}[ht]
\centering
\subfigure[]{\includegraphics[scale=0.35,angle=90]{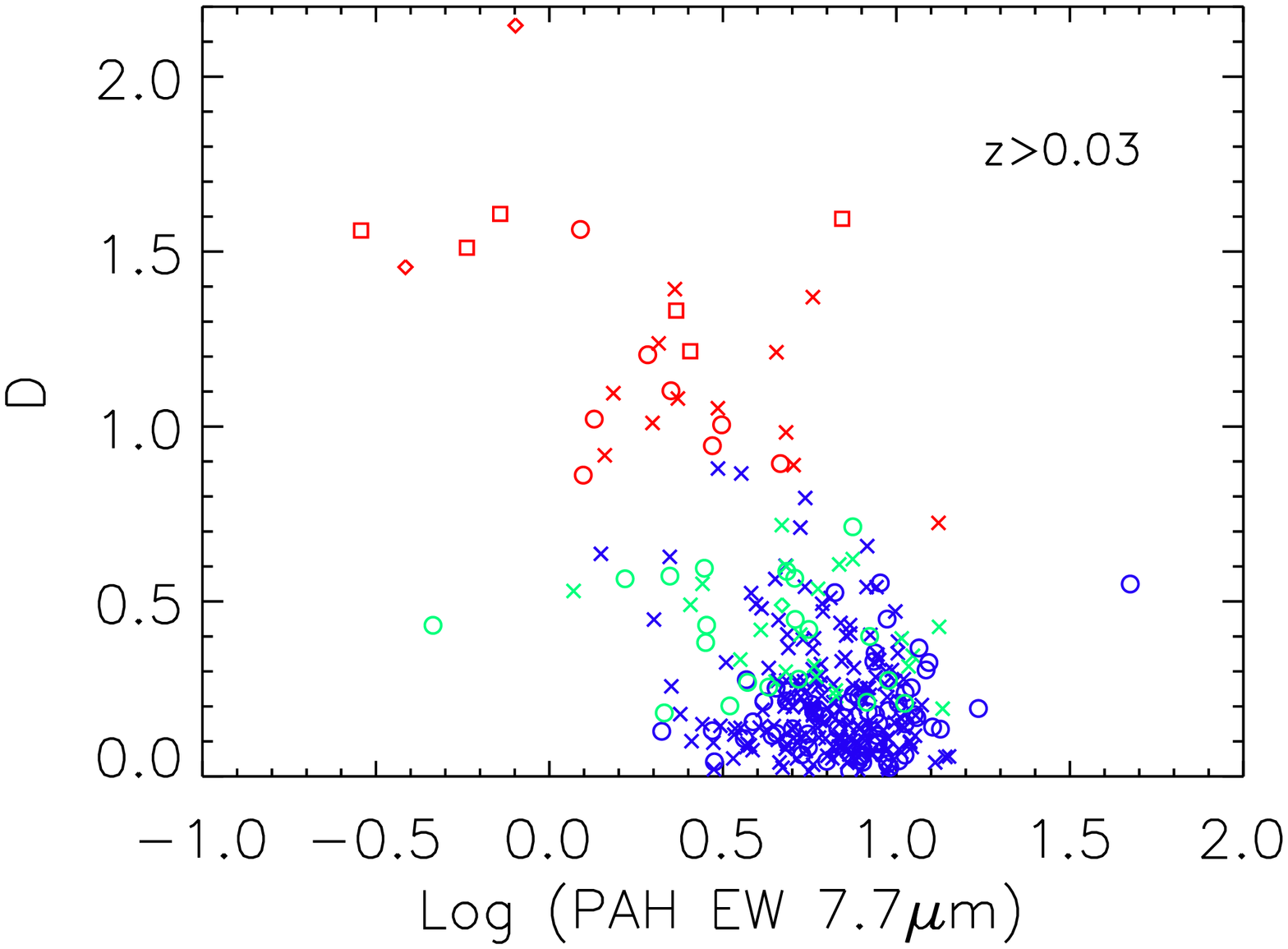}}
\subfigure[]{\includegraphics[scale=0.35,angle=90]{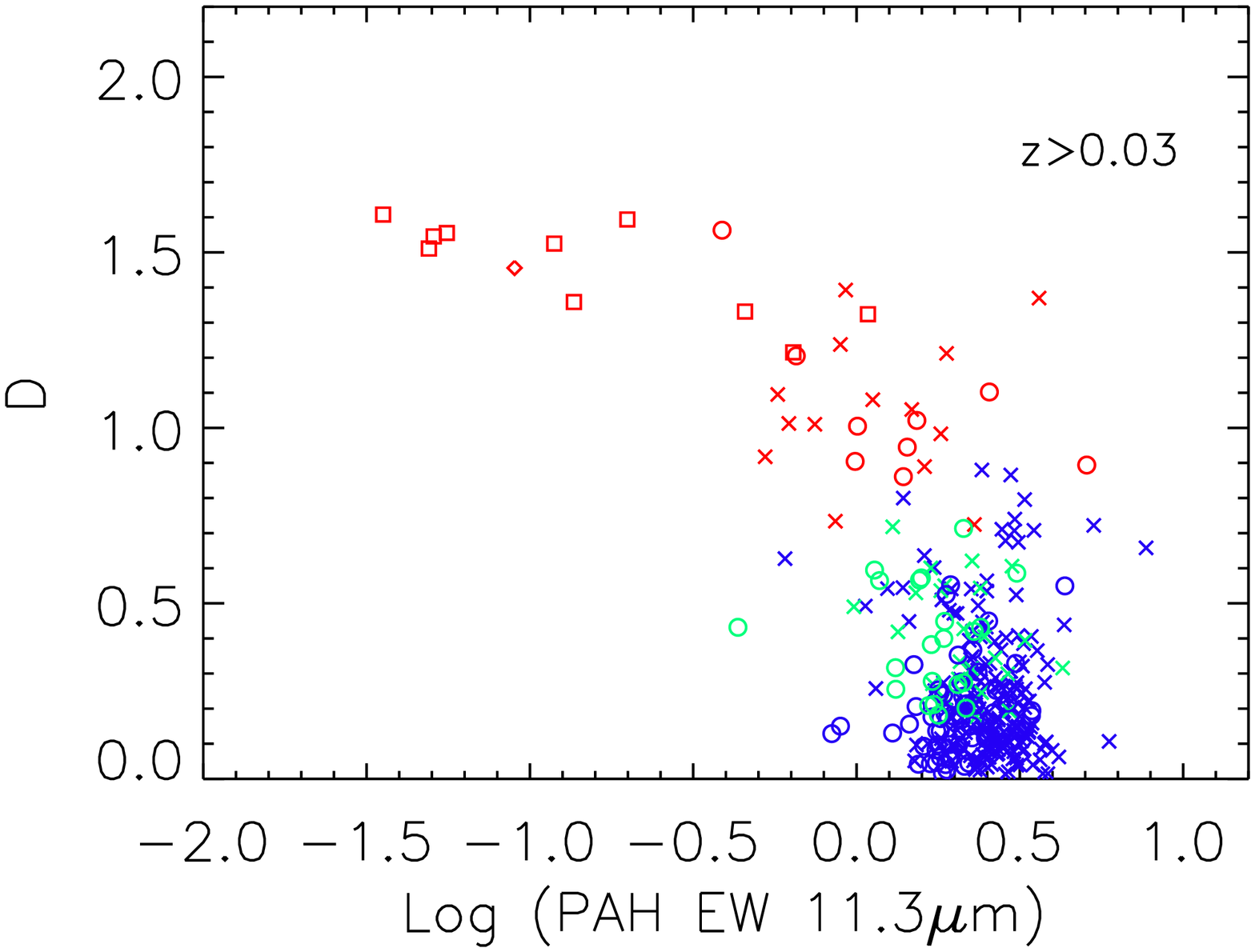}}
\subfigure[]{\includegraphics[scale=0.35,angle=90]{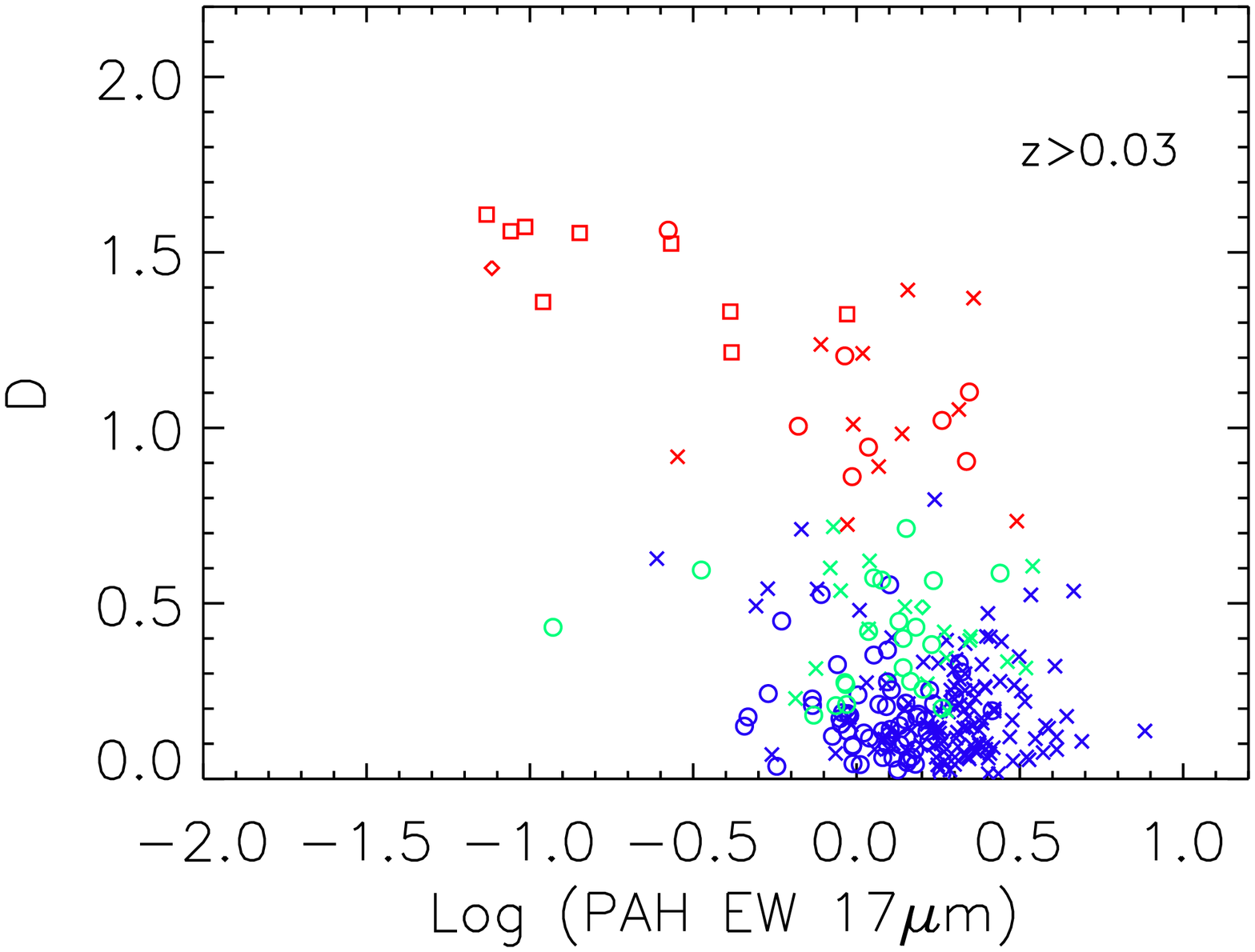}}
\caption[PAH EWs vs. D for $z \geq$ 0.03]{\label{ew_d_z_cut}The optical D parameter vs. PAH EWs. Imposing a redshift cut, to test for possible systematic differences due to physical sizes probed between closeby and far away sources, does not alter the trends seen in the full sample. Color and symbol coding same as Figure 1. }
\end{figure}

\begin{figure}[ht]
\centering
\subfigure[]{\includegraphics[scale=0.35,angle=90]{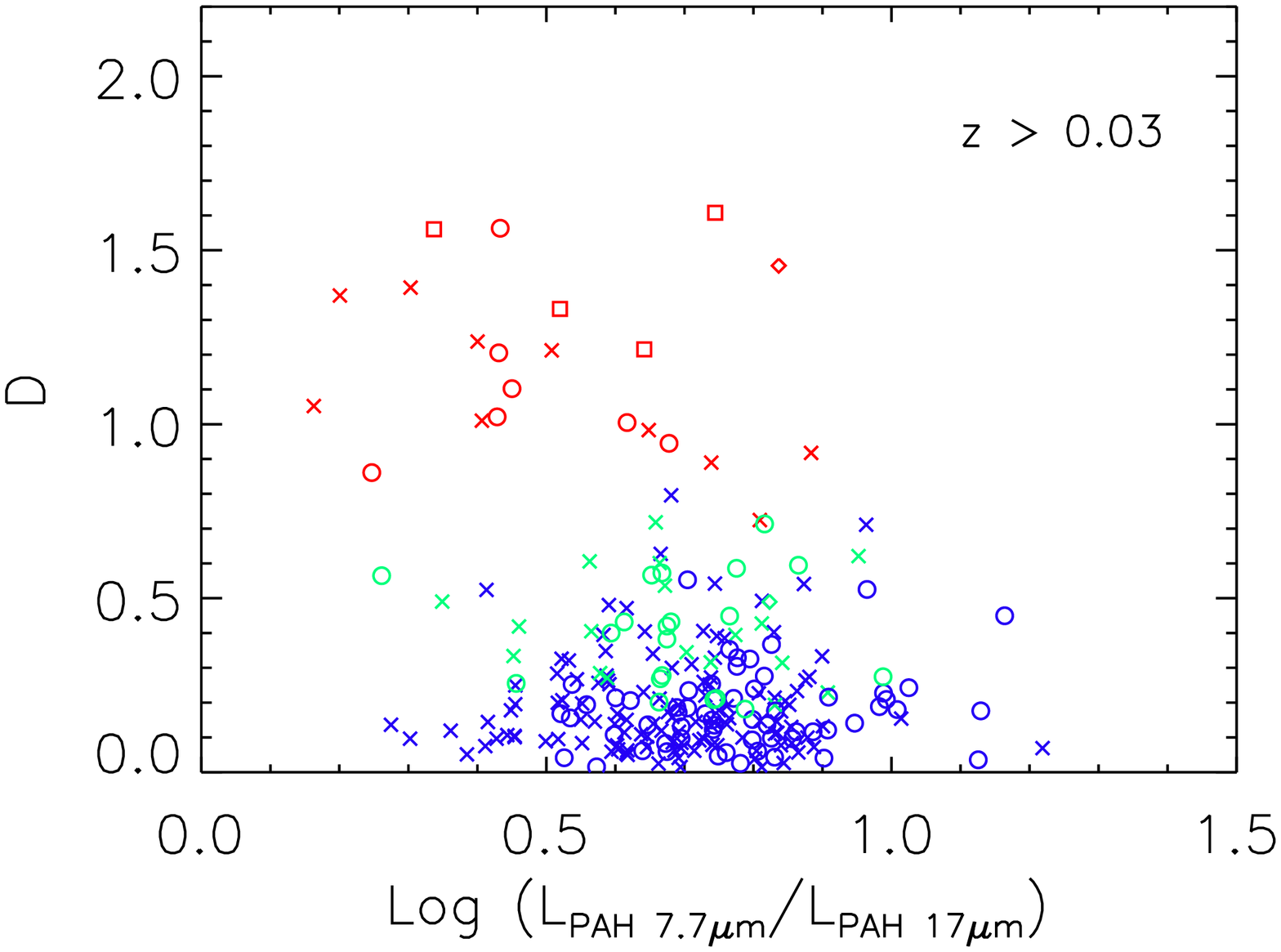}}
\subfigure[]{\includegraphics[scale=0.35,angle=90]{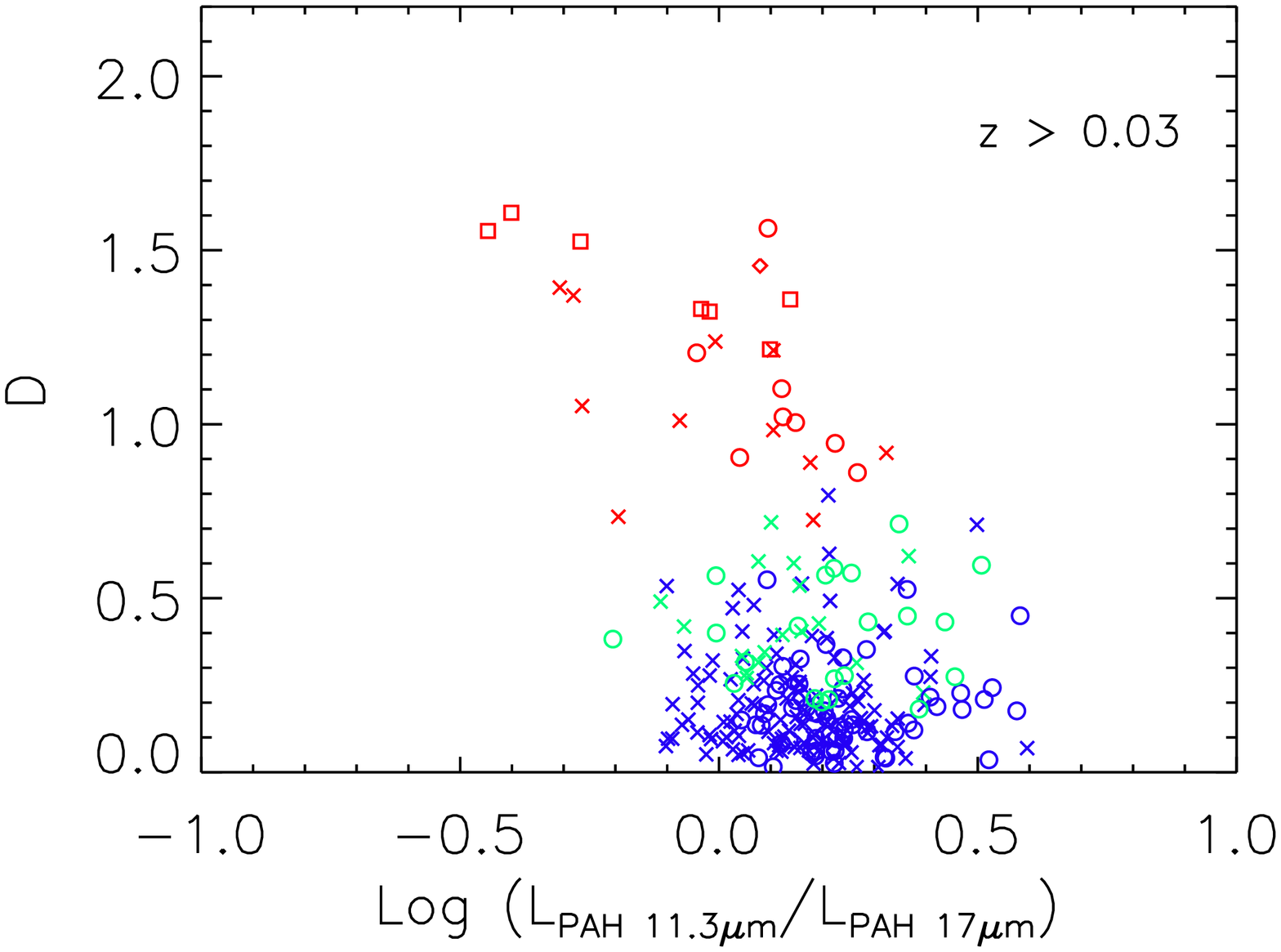}}
\subfigure[]{\includegraphics[scale=0.35,angle=90]{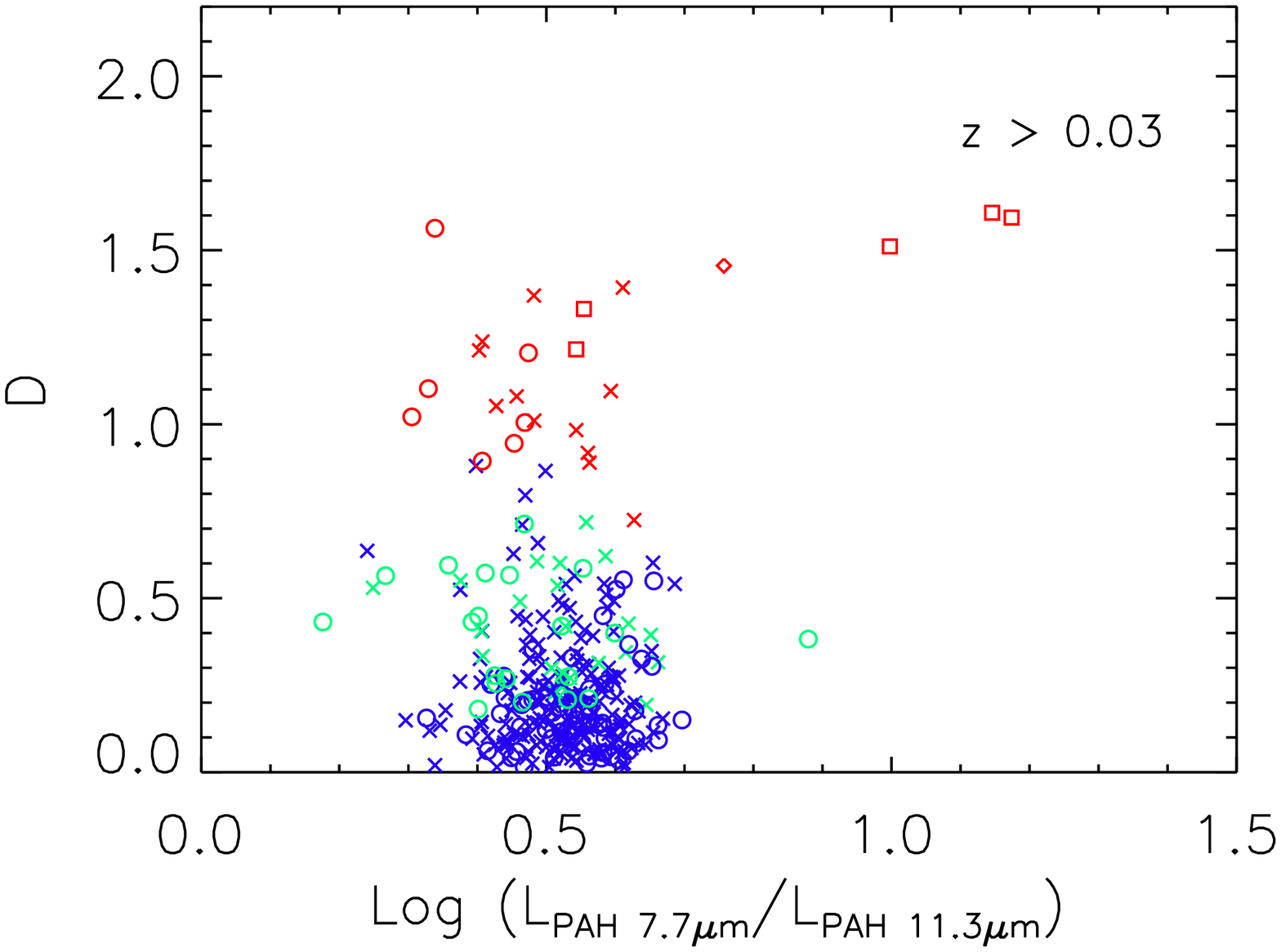}}
\caption[Optical D parameter vs. PAH ratios for z$\geq$0.03]{\label{ratios_d_z_cut}For sources at z$\geq$ 0.03, optical D parameter vs. PAH ratios shown in Figure \ref{ratios_d_z_cut}. Imposing a redshift cut does not alter the trends seen in the full sample. Color and symbol coding same as Figure 1. }
\end{figure}

Because we have combined a number of different samples selected in different ways, our AGN systems (mostly drawn from the SDSS [OIII] and 12$\mu$m samples) are typically at lower redshift than most of the star-forming and composite systems (drawn from SSGSS and S5). Could any of the results above be influenced by the systematic differences in the physical sizes being probed (i.e. smaller scale nuclear emission in closeby AGN versus larger scale host galaxy emission for sources further away)? To test this, we impose a redshift cut of z$\geq$0.03, which corresponds to the lowest redshift source in the SSGSS sample. This cut eliminates most of the 12$\mu$m sources and almost half of the [OIII] sample. However, since we do retain a handful of [OIII]-selected AGN which are strong AGN, we are able to test the relationship between PAH EWs and ionization field hardness over a wide range of nuclear galactic activity. Figure \ref{ew_d_z_cut} and \ref{ratios_d_z_cut} illustrate that the trends found for the full sample still hold for this pared down sample. The AGN at higher D values are associated with suppressed PAH EWs at a statistically significant level even when the host galaxy is more fully sampled (see Table \ref{ew}). The only change in the PAH ratio analysis is the significant anti-correlation of L$_{PAH 11.3\mu m}$/L$_{PAH 17\mu m}$ with respect to D in the AGN sub-sample (see Table \ref{pah_ratio}). We note that though the L$_{PAH 7.7\mu m}$/L$_{PAH 11.3\mu m}$ trend is driven by a handful of AGN, the relationship is similar to the parent dataset where the parameter space of high L$_{PAH 7.7\mu m}$/L$_{PAH 11.3\mu m}$ values is more fully sampled. The thresholds in D and [NeV]/[NeII] for changes in the PAH ratios noted above are still present. 

Taken together, these results suggest that some care be used when using the PAH luminosities to estimate star formation rates in galaxies with dominant AGN. The 11.3 $\mu m$ feature appears to be rather strongly suppressed in the most AGN-dominated systems.

\section{Disentangling Starburst and AGN Contributions to MIR Emission}
Mid-infrared emission in active galaxies results from both the toroidal obscuring medium, accounting for $\sim$20\% of the bolometric AGN flux \citep{12m}, and host galaxy star formation. We define L$_{MIR}$ as the rest-frame luminosity at 13.5$\mu$m averaged over a 3$\mu$m window: this region is free from strong PAH features and provides a reliable estimate of the MIR continuum emission. Using quiescent galaxies as a baseline to estimate the amount of MIR emission due to star formation, we can empircally separate the contributions of starburst activity and toroidal emission. To quantify the relative importance of these two effects, we solve: $$L_{MIR} = \alpha L_{SFR} + \beta L_{AGN} $$

We set L$_{SFR}$ = L$_{[NeII]}$ and L$_{AGN}$ = L$_{[OIII],AGN}$, the AGN contribution to the observed [OIII] emission. To calculate L$_{[OIII],AGN}$, we use Equation 3 in \citet{WHC}, which scales the observed [OIII] flux based on the source's position in the BPT diagram. For the AGN where this correction failed (i.e. L$_{[OIII],AGN} >$ L$_{[OIII],obs}$), we set L$_{[OIII],AGN}$ = L$_{[OIII],obs}$. For the composite and star-forming galaxies where this correction failed (i.e. L$_{[OIII],AGN} < 0$), we set L$_{[OIII],AGN}$ = 0. A vast majority of the star-forming galaxies had L$_{[OIII],AGN}$ set to 0.

\begin{figure}[ht]
\centering
{\includegraphics[scale=0.50,angle=90]{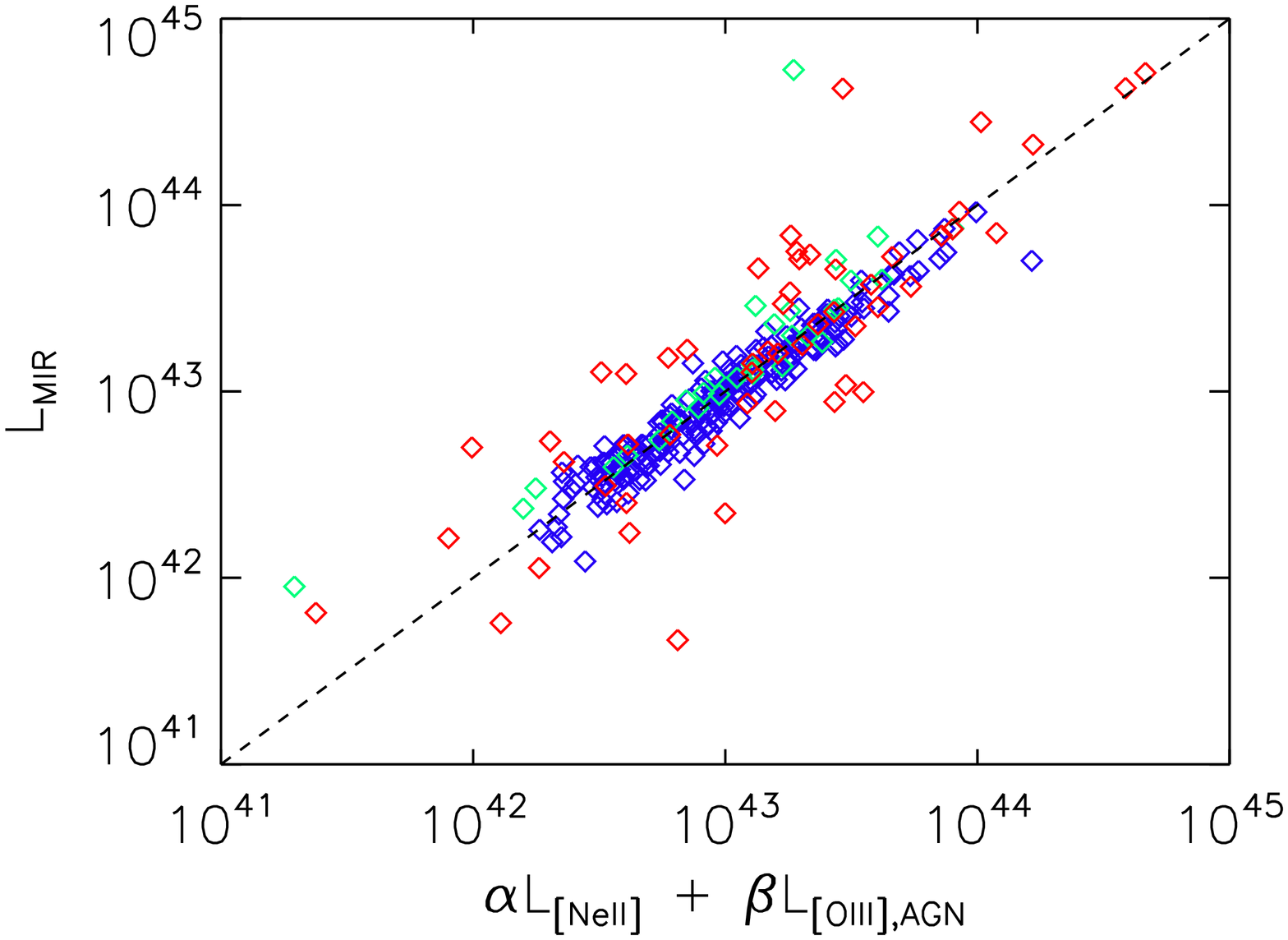}}
\caption[L$_{MIR}$ vs. $\alpha$L$_{SFR}$ + $\beta$L$_{AGN}$]{\label{mir_agn_sfr}L$_{MIR}$ vs. $\alpha$L$_{SFR}$ + $\beta$L$_{AGN}$, where L$_{SFR}$ = L$_{[NeII]}$ and L$_{AGN}$ = L$_{[OIII],AGN}$. $\alpha$ and $\beta$ were found by using a least trimmed squares regression: $\alpha$=89$\pm$1, $\beta$=111$\pm$7. The dashed line indicates where the two quantities are equal. Color coding same as Figure 1. }
\end{figure}

Using the sources that had [NeII] detections, we calculated $\alpha$ and $\beta$ by using a least squares trimmed (LTS) regression, a robust regression technique that removes outliers, on the above equation. As each execution of the LTS regression results in different values for $\alpha$ and $\beta$, we performed this calculation 1000 times to determine the average values and spreads on the distributions. The resulting values, using a 3$\sigma$ clipping mean, are $\alpha$=89$\pm$1 and $\beta$=111$\pm$7. We note that the difference between these values and using a straight average are negligible for $\alpha$ and small for $\beta$ (c.f. 109$\pm$15).  In Figure \ref{mir_agn_sfr}, we plot L$_{MIR}$ vs. $\alpha$L$_{SFR}$ + $\beta$L$_{AGN}$ (i.e. 89 $\times$ L$_{[NeII]}$ + 124 $\times$ L$_{[OIII],AGN}$), with the overplotted line indicating where the two quantities are equal. No systematic offsets among galaxy sub-type are present. This relationship can therefore be valuable for estimating the MIR flux due to star-formation (or conversely from the torus) in active galaxies.

\begin{figure}[ht]
\centering
{\includegraphics[scale=0.50,angle=90]{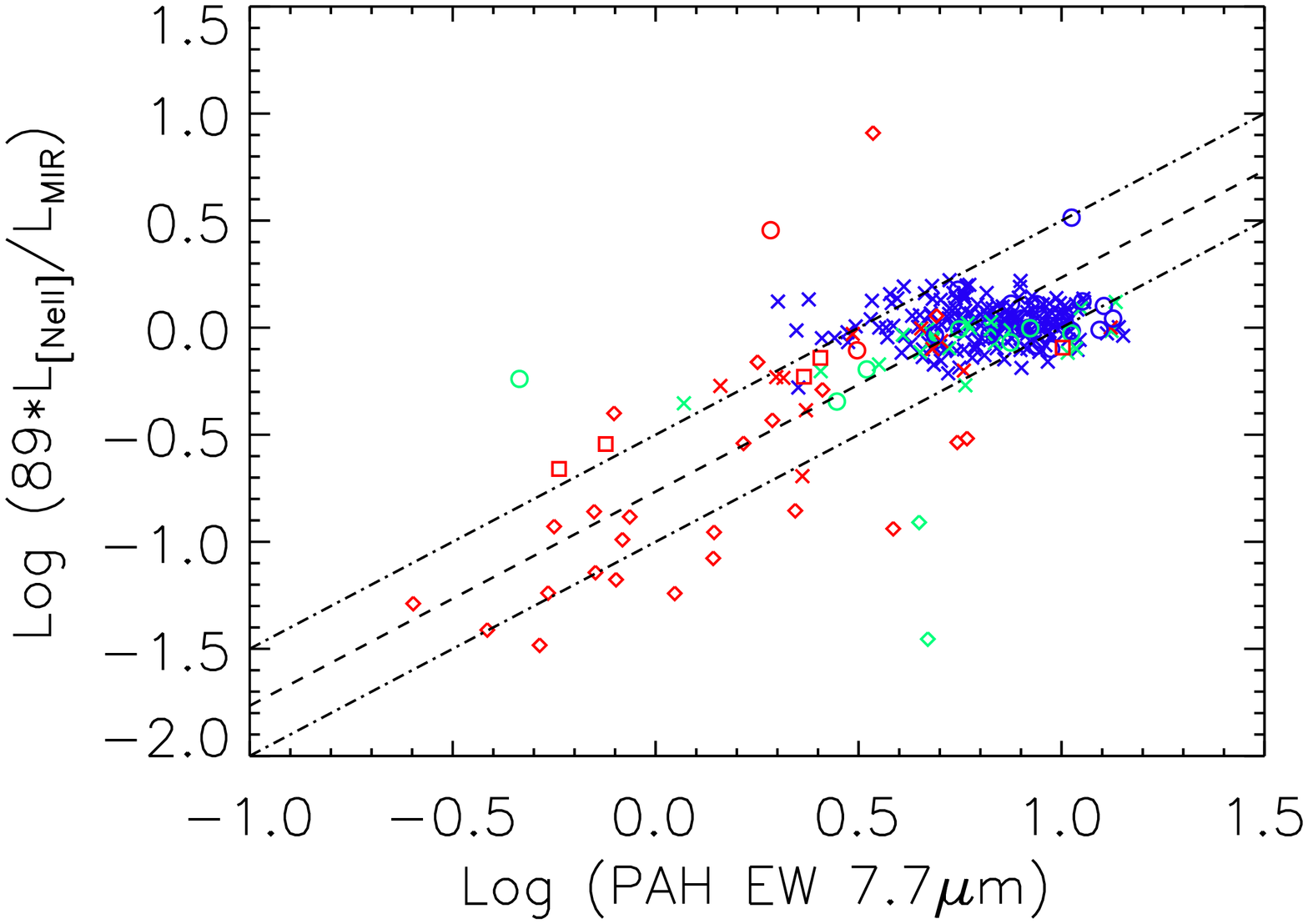}}
\caption[]{\label{aneii_mir_ew77}Log ($\alpha \times$L$_{[NeII]}$/L$_{MIR}$) vs. Log (PAH EW) at 7.7$\mu$m. The two dotted-dashed lines have unity slope and intercept points where the MIR emission is due purely to star formation according to our decomposition (i.e. $\alpha \times$ L$_{[NeII]}$ = L$_{MIR}$) at two fiducial values of PAH 7.7$\mu$m EW (1.0 dex and 0.5 dex) chosen to bracket the star-forming galaxies. The dashed line indicates the best-fit through the AGN, with the slope held constant at unity. The AGN generally fall within the region indicated by the lines, which is consistent with a monotonic decrease in PAH EW, denoting the ``extra'' MIR emission due to AGN, correlating with the empirial MIR AGN parameterization (i.e. decreasing $\alpha \times$L$_{[NeII]}$/L$_{MIR}$ values indicate greater AGN contribution to the MIR luminosity). Color and symbol coding same as Figure 1. }
\end{figure}

To test this decomposition, we have plotted $\alpha \times$L$_{[NeII]}$/L$_{MIR}$ vs. the PAH EW at 7.7$\mu$m in Figure \ref{aneii_mir_ew77}. As we have demonstrated above, this PAH macro-molecule/small-grain is the least affected by AGN activity (i.e. not suppressed like the 11.3$\mu$m feature nor mildly boosted like the 17$\mu$m feature). Hence, to first order the EW of the 7.7$\mu$m PAH feature is tracing the ``extra'' MIR emission from AGN heated dust. If our decomposition is correct, the relative amount of MIR emission due to star formation activity, i.e. $\alpha \times$L$_{[NeII]}$/L$_{MIR}$, should increase monotonically with increasing EW of the PAH 7.7$\mu$m feature. The dotted dashed lines have unity slope and were chosen to intersect the points where the MIR emission is solely due to star formation (e.g. log ($\alpha \times$ L$_{[NeII]}$/L$_{MIR}$) = 0 dex) and fiducial PAH EW values that approximately enclose the star-forming galaxy locus. Many of the AGN fall within the region demarcated by these two lines. The dashed line is the best-fit line to the AGN sources where the slope is held constant at unity. This fit neatly bisects the above region. These results suggest that our empirical decomposition works reasonably well in describing the relative contributions of star formation and AGN activity to the MIR emission.

\section{Conclusions}

We have explored the infrared and optical parameter space where star-forming galaxies, composites and AGN live to analyze diagnostics that parameterize host galaxy star formation. Using star-forming galaxies as a control sample, we have investigated which star formation diagnostics are least affected by AGN activity.  We have studied diagnostics that trace the interplay between AGN and starburst activity and tested whether a smooth transition exists over a range of radiation field hardness. Finally, using the results of this analysis, we present an empirical decomposition of the MIR flux into a star-forming and an AGN component. {\it Our overall result is that the optical and mid-IR diagnostics of star formation and of the relative importance of young stars and the AGN generally agree well.} Our more specific results are summarized as follows:

\begin{description}

\item{\bf{SFR diagnostics}} The SDSS derived SFR$_{fiber}$ and the [NeII] 12.81$\mu$m luminosity agree well, and are the most reliable SF proxies we have considered for use in star-forming galaxies, composites and AGN. The sum of [NeII] and [NeIII] 15.56$\mu$m is systematically offset to higher values for AGN due to the active nucleus boosting the [NeIII] flux. The aggregate PAH luminosity (L$_{\Sigma PAH 7.7,11.3,17\mu m}$) relative to L$_{[NeII]}$ is suppressed in AGN compared with both quiescent and composite galaxies. Though the (L$_{\Sigma PAH 7.7,11.3,17\mu m}$)/SFR$_{fiber}$ ratio agrees between star-forming galaxies and Sy2s, analysis of PAH flux ratios indicate abnormal behavior in some AGN dominated systems, such as suppression of the 11.3$\mu$m feature, suggesting diminished PAH emission in strong AGN rather than [NeII] enhancement. The disparate slopes of (L$_{\Sigma PAH 7.7,11.3,17\mu m}$)/SFR$_{fiber}$ between Sy2s and star-forming galaxies also hint that AGN may contaminate the emission from PAH macro-molecules.

\item{\bf{Ionization Field Hardness}} We have used the optical D parameter, which indicates the distance a source lies from the star-forming galaxy locus on the BPT diagram \citep{Kauffmann}, and L$_{[NeV]}$/L$_{[NeII]}$ \citep{Armus, PS} as probes of the incident radiation field hardness. The infrared diagnostics that parameterize the relative contributions of AGN to star-forming activity that we have considered ($\alpha_{20-30\mu m}$, PAH EWs and PAH flux ratios) only show significant trends with ionization field hardness in the AGN population. This result suggests that rather than parameterizing incident radiation hardness in general, these IR diagnostics are only effective descriptors of ionization field hardness in AGN.

\begin{description}

\item{\boldmath{$\alpha_{20-30\mu m}$}} In AGN, the spectral slope between 20-30 $\mu$m (measured by the spectral index $\alpha_{20-30\mu m}$) has been shown to steepen in the presence of cold dust due to star formation \citep{Buchanan, Deo}. For the AGN in our sample, $\alpha_{20-30\mu m}$ is significantly anti-correlated with  proxies of ionization field hardness, albeit with wide scatter, however this result does not hold when expanded into the regime of quiescent star forming galaxies. 

\item{\bf{PAH EW}} PAH EWs are significantly anti-correlated with the increasing hardness of radiation field, but this is only true when considering the AGN. When considering just the star-forming galaxies, no trend exists between incident radiation hardness and the strength of the resulting PAH feature, which agrees with the findings of \citet{O'Dowd} and \citet{Treyer}. This result is consistent with the hypothesis that the mid-IR continuum has a significant contribution from dust heated by the AGN only in objects where the ionizing radiation field is dominated by the AGN. For the 11.3$\mu$m PAH feature, our analysis of the PAH flux ratios suggests that destruction of the PAH macro-molecule in strong AGN could also contribute to suppression of the PAH 11.3$\mu$m EW.

\item{\bf{PAH flux ratios}} We find that the luminosity of PAH 11.3 $\mu m$ feature is signifcantly suppressed relative to the luminosities of the PAH 7.7 $\mu m$ and 17 $\mu m$ features and relative to the optical and mid-IR derived star formation rates in the most AGN-dominated systems. Thus, some care is required in using the PAH 11.3$\mu$m luminosity as a proxy for the SFR in AGN-dominated systems.

\end{description}

\item{{\bf Composite Systems}} Galaxies that are optically classified as composite systems are more akin to quiescent star forming galaxies in terms of the mid-IR parameters that trace star formation: the smaller relative energetic importance of the AGN in composites does not seem to affect [NeIII] and PAH luminosities.

\item{\bf{[NeV] as unambiguous AGN signature}} Due to the high ionization potential of [NeV] (97.1 eV), it is potentially an unequivocal signature of AGN activity \citep{Gilli, Gould10}. One galaxy that is optically classied as a quiescent star-forming system in our sample has a [NeV] detection, yet other IR parameters (i.e. PAH EW values) are normal for a pure star forming galaxy. The nature of this galaxy as a strong ``hidden'' AGN is not clear.

\item{\bf{Disentangling MIR Emission}} Using multiple linear regression, we fit the relation L$_{MIR}$ = $\alpha$L$_{SFR}$ + $\beta$L$_{AGN}$ to our full sample to determine $\alpha$ and $\beta$. We set L$_{SFR}$ = L$_{[NeII]}$ and L$_{AGN}$=L$_{[OIII],AGN}$ where in L$_{[OIII],AGN}$, we have subtracted out the estimated starburst contribution to the [OIII] flux using Eq. 3 in \citet{WHC}. We find $\alpha$=89$\pm$1 and $\beta$=111$\pm$7. This decomposition can be useful in estimating the MIR emission due to the circumnuclear AGN obscuration or conversely host galaxy star formation in AGN.

\end{description}


\clearpage

\clearpage

\renewcommand*{\thesubfigure}{}

\renewcommand*{\thesubfigure}{(\alph{subfigure})} 

\clearpage


\begin{landscape}
\begin{deluxetable}{lccccccc}
\small
\tablewidth{0pt}
\tablecaption{\label{ssgss_flux}Emission Line Luminosities for SSGSS Sample}
\tablehead{\colhead{Source Number} & \colhead{RA} & \colhead{Dec} & \colhead{{\it z}} & \colhead{Type} & \colhead{[NeII] 12.81$\mu$m} &  \colhead{[NeV] 14.32$\mu$m} & \colhead{[NeIII] 15.56$\mu$m} \\
 & & & & & \colhead{10$^{41}$ erg/s} & \colhead{10$^{41}$ erg/s} & \colhead{10$^{41}$ erg/s}}
\startdata
     33 & 160.291 & 56.932 & 0.185 &AGN &  8.12$\pm$ 0.70 &  ...  &  3.34$\pm$ 0.57\\
     61 & 160.952 & 58.197 & 0.073 &AGN &  3.19$\pm$ 0.08 &  ...  &  0.30$\pm$ 0.05\\
     11 & 161.720 & 56.252 & 0.047 &Comp &  0.17$\pm$ 0.01 &  ...  & $<$ 0.06\\
     12 & 162.268 & 56.224 & 0.072 &Comp &  3.04$\pm$ 0.09 &  ...  &  0.31$\pm$ 0.04\\
     15 & 161.758 & 56.307 & 0.153 &Comp &  4.40$\pm$ 0.30 &  ...  & $<$ 1.64\\
\enddata
\tablenotetext{1}{This table is available in its entirety in machine readable format on-line. We list several row here as a guide.}
\end{deluxetable}
\end{landscape}

\begin{landscape}
\begin{deluxetable}{lcccccc}
\small
\tablewidth{0pt}
\tablecaption{\label{s5_flux}Emission Line Luminosities for S5 Sample}
\tablehead{\colhead{RA} & \colhead{Dec} & \colhead{{\it z}} & \colhead{Type} & \colhead{[NeII] 12.81$\mu$m} &  \colhead{[NeV] 14.32$\mu$m} & \colhead{[NeIII] 15.56$\mu$m} \\
 & & & & \colhead{10$^{41}$ erg/s} & \colhead{10$^{41}$ erg/s} & \colhead{10$^{41}$ erg/s}}
\startdata
 36.380 & -8.419 & 0.055 & AGN &  1.16$\pm$ 0.05 &  ... & 1.02$\pm$ 0.07\\
 36.435 & -7.868 & 0.077 & AGN &  2.72$\pm$ 0.12 &  ... & 1.03$\pm$ 0.08\\
132.186 & 53.746 & 0.083 &AGN &  4.72$\pm$ 0.14 &  1.18$\pm$ 0.09 &  2.54$\pm$ 0.08\\
132.714 & 54.823 & 0.081 & AGN &  3.17$\pm$ 0.11 &  ... & 2.25$\pm$ 0.07\\
132.815 & 54.986 & 0.083 & AGN &  3.64$\pm$ 0.16 &  ... & 1.68$\pm$ 0.09\\
\enddata
\tablenotetext{1}{This table is available in its entirety in machine readable format on-line. We list several row here as a guide.}
\end{deluxetable}
\end{landscape}

\begin{landscape}
\begin{deluxetable}{lccccccc}
\small
\tablewidth{0pt}
\tablecaption{\label{o3_flux}Emission Line Luminosities for [OIII] Sample}
\tablehead{\colhead{Galaxy} & \colhead{RA} & \colhead{Dec} & \colhead{{\it z}} & \colhead{[NeII] 12.81$\mu$m} &  \colhead{[NeV] 14.32$\mu$m} & \colhead{[NeIII] 15.56$\mu$m} \\
 & & & & \colhead{10$^{41}$ erg/s} & \colhead{10$^{41}$ erg/s} & \colhead{10$^{41}$ erg/s}}
\startdata

NGC 291 & 13.375 & -8.768 & 0.019 & 1.61$\pm$ 0.04 &  0.60$\pm$ 0.02 &  1.12$\pm$ 0.02\\

Mrk 609 &  51.356 & -6.144 & 0.034 &  5.51$\pm$ 0.12 & ... &  1.80$\pm$ 0.06\\

IC 0486 & 120.087 & 26.614 & 0.027 & 1.10$\pm$ 0.10 &  0.55$\pm$ 0.06 &  1.04$\pm$ 0.05\\

2MASX J08035923+2345201 & 120.997 & 23.756 & 0.029 & ... & $<$ 0.14 & $<$ 0.42\\

2MASX J08244333+2959238 & 126.180 & 29.990 & 0.025 & 0.54$\pm$ 0.06 &  0.55$\pm$ 0.07 &  0.70$\pm$ 0.06\\

\enddata
\tablenotetext{1}{This table is available in its entirety in machine readable format on-line. We list several row here as a guide.}
\end{deluxetable}
\end{landscape}

\begin{landscape}
\begin{deluxetable}{lccccccc}
\small
\tablewidth{0pt}
\tablecaption{\label{12m_flux}Emission Line Luminosities for 12$\mu$m Sample}
\tablehead{\colhead{Galaxy} & \colhead{RA} & \colhead{Dec} & \colhead{{\it z}} & \colhead{Type} & \colhead{[NeII] 12.81$\mu$m} &  \colhead{[NeV] 14.32$\mu$m} & \colhead{[NeIII] 15.56$\mu$m} \\
 & & & & & \colhead{10$^{40}$ erg/s} & \colhead{10$^{40}$ erg/s} & \colhead{10$^{40}$ erg/s}}
\startdata

 IRAS 00198-7926 & 5.473  & -79.169 &0.073 & AGN &35.65$\pm$ 1.45 & 49.05$\pm$ 1.53 & 60.06$\pm$ 1.31\\

   NGC 0424 & 17.865 & -38.083 &0.012 & AGN & 3.06$\pm$ 0.19 &  4.64$\pm$ 0.27 &  5.67$\pm$ 0.17\\

   NGC 1068 & 40.670 & -0.013 &0.004 & AGN & 5.34$\pm$ 0.28 & 10.11$\pm$ 0.48 & 15.02$\pm$ 0.32\\

   NGC 1144 & 43.790 & -0.178 &0.028 & AGN &10.79$\pm$ 0.26 &  0.61$\pm$ 0.10 &  2.88$\pm$ 0.05\\

   NGC 1320 & 51.203 & -3.042 &0.009 & AGN & 0.47$\pm$ 0.02 &  0.59$\pm$ 0.02 &  0.78$\pm$ 0.02\\

\enddata
\tablenotetext{1}{This table is available in its entirety in machine readable format on-line. We list several row here as a guide.}
\end{deluxetable}
\end{landscape}

\clearpage


\begin{landscape}
\begin{deluxetable}{lccccccc}
\small
\tablewidth{0pt}
\tablecaption{\label{ssgss_pah}PAH Luminosities and EWs for SSGSS Sample}
\tablehead{\colhead{Src Num}  & \colhead{Type} & \colhead{L$_{7.7\mu m}$} &   \colhead{EW 7.7$\mu$m} & \colhead{L$_{11.3\mu m}$} &   \colhead{EW 11.3$\mu$m} & \colhead{L$_{17\mu m}$} &   \colhead{EW 17$\mu$m} \\ 
 & &  \colhead{10$^{41}$ erg/s} & \colhead{$\mu$m} & \colhead{10$^{41}$ erg/s} & \colhead{$\mu$m} & \colhead{10$^{41}$ erg/s} & \colhead{$\mu$m} }
\startdata
     13 & AGN & 16.59$\pm$ 0.82 &  1.23 &   7.61$\pm$ 0.32 &  0.39 &   6.12$\pm$ 1.09 &  0.27 \\
     21 & AGN & 30.66$\pm$ 1.73 &  2.95 &  10.80$\pm$ 0.60 &  1.43 &   6.44$\pm$ 0.49 &  1.09 \\
     23 & AGN & ...  & ... &  ...  & ... &  ...  & ... \\
     33 & AGN &157.18$\pm$ 6.51 &  3.14 &  53.40$\pm$ 1.08 &  1.01 &  37.98$\pm$ 3.62 &  0.66 \\
     50 & AGN & ...  & ... &  ...  & ... &  ...  & ... \\
\enddata
\tablenotetext{1}{This table is available in its entirety in machine readable format on-line. We list several row here as a guide.}
\end{deluxetable}
\end{landscape}

\begin{landscape}
\begin{deluxetable}{lcccccccc}
\small
\tablewidth{0pt}
\tablecaption{\label{s5_pah}PAH Luminosities and EWs for S5 Sample}
\tablehead{\colhead{RA}  & \colhead{Dec} & \colhead{Type} & \colhead{L$_{7.7\mu m}$} &   \colhead{EW 7.7$\mu$m} & \colhead{L$_{11.3\mu m}$} &   \colhead{EW 11.3$\mu$m} & \colhead{L$_{17\mu m}$} &   \colhead{EW 17$\mu$m} \\ 
 & & &\colhead{10$^{41}$ erg/s} & \colhead{$\mu$m} & \colhead{10$^{41}$ erg/s} & \colhead{$\mu$m} & \colhead{10$^{41}$ erg/s} & \colhead{$\mu$m} }
\startdata
 36.380 & -8.419 & AGN & 24.01$\pm$ 2.34 &  5.74 &   7.92$\pm$ 0.71 &  3.62 &  15.13$\pm$ 2.76 &  2.28 \\
 36.435 & -7.868 & AGN & 59.59$\pm$ 4.64 &  1.44 &  16.41$\pm$ 0.62 &  0.53 &   7.79$\pm$ 1.54 &  0.28 \\
132.186 & 53.746 & AGN &144.14$\pm$ 5.69 &  4.81 &  41.27$\pm$ 0.81 &  1.81 &  32.40$\pm$ 1.82 &  1.38 \\
132.714 & 54.823 & AGN & 54.59$\pm$ 5.05 &  4.51 &  21.61$\pm$ 0.71 &  1.89 &  16.96$\pm$ 1.69 &  1.04 \\
132.815 & 54.986 & AGN &107.85$\pm$ 3.30 &  5.05 &  29.54$\pm$ 0.49 &  1.62 &  19.68$\pm$ 1.23 &  1.17 \\
\enddata
\tablenotetext{1}{This table is available in its entirety in machine readable format on-line. We list several row here as a guide.}
\end{deluxetable}
\end{landscape}

\begin{landscape}
\begin{deluxetable}{lcccccc}
\small
\tablewidth{0pt}
\tablecaption{\label{o3_pah}PAH Luminosities and EWs for [OIII] Sample}
\tablehead{\colhead{Galaxy}  &  \colhead{L$_{7.7\mu m}$} &   \colhead{EW 7.7$\mu$m} & \colhead{L$_{11.3\mu m}$} &   \colhead{EW 11.3$\mu$m} & \colhead{L$_{17\mu m}$} &   \colhead{EW 17$\mu$m} \\ 
 &  \colhead{10$^{41}$ erg/s} & \colhead{$\mu$m} & \colhead{10$^{41}$ erg/s} & \colhead{$\mu$m} & \colhead{10$^{41}$ erg/s} & \colhead{$\mu$m} }
\startdata
NGC 0291   & 34.69$\pm$ 0.95 & 10.07 &   7.44$\pm$ 0.15 &  0.75 &   9.20$\pm$ 0.15 &  0.50 \\
Mrk 0609   & ...  & ... &  52.35$\pm$ 0.78 &  1.08 &  54.58$\pm$ 0.54 &  0.94 \\
IC 0486   & 22.69$\pm$ 0.47 &  0.75 &   4.15$\pm$ 0.19 &  0.12 &   5.77$\pm$ 0.12 &  0.16 \\
2MASX J08035923+2345201   & ...  & ... &   1.49$\pm$ 0.05 &  0.30 &   2.10$\pm$ 0.13 &  0.40 \\
2MASX J08244333+2959238   & ...  & ... &   2.80$\pm$ 0.16 &  0.04 &   5.67$\pm$ 0.29 &  0.10 \\
\enddata
\tablenotetext{1}{This table is available in its entirety in machine readable format on-line. We list several row here as a guide.}
\end{deluxetable}
\end{landscape}

\begin{landscape}
\begin{deluxetable}{lccccccc}
\small
\tablewidth{0pt}
\tablecaption{\label{12m_pah}PAH Luminosities and EWs for 12 $\mu$m Sample}
\tablehead{\colhead{Galaxy}  & \colhead{Type} & \colhead{L$_{7.7\mu m}$} &   \colhead{EW 7.7$\mu$m} & \colhead{L$_{11.3\mu m}$} &   \colhead{EW 11.3$\mu$m} & \colhead{L$_{17\mu m}$} &   \colhead{EW 17$\mu$m} \\ 
 & & \colhead{10$^{41}$ erg/s} & \colhead{$\mu$m} & \colhead{10$^{41}$ erg/s} & \colhead{$\mu$m} & \colhead{10$^{41}$ erg/s} & \colhead{$\mu$m} }
\startdata

   NGC 0424 & AGN &   10.25$\pm$ 0.65 &  0.09 &   4.46$\pm$ 0.35 &  0.06 &   8.05$\pm$ 0.61 &  0.19 \\
   NGC 1144 & AGN &  258.85$\pm$ 1.37 &  2.21 &  92.63$\pm$ 0.99 &  1.83 &  51.43$\pm$ 0.73 &  1.56 \\
   NGC 1320 & AGN &    9.69$\pm$ 0.38 &  0.52 &   2.45$\pm$ 0.08 &  0.16 &   1.31$\pm$ 0.12 &  0.11 \\
   NGC 1386 & AGN &    0.84$\pm$ 0.02 &  0.71 &   0.16$\pm$ 0.00 &  0.14 &   0.26$\pm$ 0.01 &  0.27 \\
   NGC 1667 & AGN &   61.97$\pm$ 1.14 &  3.85 &  18.76$\pm$ 0.22 &  2.18 &   9.18$\pm$ 0.13 &  1.63 \\

\enddata
\tablenotetext{1}{This table is available in its entirety in machine readable format on-line. We list several row here as a guide.}
\end{deluxetable}
\end{landscape}


\clearpage

\begin{landscape}
\begin{deluxetable}{lrcrcrcl}
\small
\tablewidth{0pt}
\tablecaption{\label{SFR_ratios}Ratios\tablenotemark{1} of SFR proxies}
\tablehead{\colhead{Ratio} & \colhead{SF\tablenotemark{2}} & \colhead{N\tablenotemark{2}} & \colhead{Composites} & \colhead{N}
& \colhead{AGN} & \colhead{N} & \colhead{Comments}}

\startdata

L$_{[NeII]}$/SFR$_{fiber}$                      & 40.97$^{+0.26}_{-0.26}$ & 223\tablenotemark{3} & 40.91$^{+0.08}_{-0.26}$ & 32 & 40.92$^{+0.26}_{-0.34}$ & 39\tablenotemark{3} \\             

(L$_{[NeII]}$+L$_{[NeIII]}$)/SFR$_{fiber}$      & 40.84$^{+0.17}_{-0.31}$ & 223\tablenotemark{3} & 40.74$^{+0.05}_{-0.30}$ & 32 & 41.11$^{+0.21}_{-0.40}$ & 39 \\

L$_{\Sigma PAH 7.7,11.3,17\mu m}$/SFR$_{fiber}$ & 42.50$^{+0.15}_{-0.26}$ & 261\tablenotemark{3} & 42.59$^{+0.10}_{-0.30}$ & 46 & 42.49$^{+0.16}_{-0.36}$ & 35\tablenotemark{3} \\

L$_{\Sigma PAH 7.7,11.3,17\mu m}$/L$_{[NeII]}$  & 1.50\tablenotemark{4}   & 207\tablenotemark{3} & 1.59$^{+0.05}_{-0.22}$  & 31 & 1.33$^{+0.15}_{-0.59}$  & 21 & PAH ULs \\

L$_{[NeII]}$/L$_{\Sigma PAH 7.7,11.3,17\mu m}$  & -1.61\tablenotemark{4}  & 141\tablenotemark{3} & -1.62$^{+0.04}_{-0.13}$ & 25 & -1.50$^{+0.02}_{-0.21}$ & 17 & [NeII] ULs \\

\enddata

\tablenotetext{1}{{\bf Ratios in log space.} Ratios were calculated using survival analysis. Error bars represent the deviation from the values at the 75th and 25th percentiles from the mean.}
\tablenotetext{2}{``SF'' refers to star-forming galaxies.}
\tablenotetext{3}{Number of sources in each galaxy sub-group used in calculation of mean values.}
\tablenotetext{4}{Upper limit data point changed to a detection for Kaplan-Meier computation, which biases the mean estimate.}
\tablenotetext{4}{Error in percentile calculation, so error bars are not reported.}

\end{deluxetable}
\end{landscape}

\clearpage

\begin{landscape}
\begin{deluxetable}{lccccccl}
\tablewidth{0pt}
\tablecaption{\label{SFR_2samp}Two-Sample Tests on SFR Proxies between Galaxy Sub-Samples\tablenotemark{1}}
\tablehead{\colhead{Ratio} & \colhead{SF vs. Composites} & \colhead{N\tablenotemark{2}} & \colhead{Composites vs. AGN} & \colhead{N} & 
\colhead{SF vs. AGN} & \colhead{N} & \colhead{Comments\tablenotemark{3}} \\
 & \colhead{{\it P}} &  & \colhead{{\it P}}  &  & \colhead{{\it P}} }

\startdata

L$_{[NeII]}$/SFR$_{fiber}$                       & 0.127 & 255 & 0.598  & 71 & 0.624               & 223 \\

(L$_{[NeII]}$+L$_{[NeIII]}$)/SFR$_{fiber}$       & 0.175 & 255 & 0.0005 & 71 & $<$1$\times10^{-4}$ & 262 \\

L$_{\Sigma PAH 7.7,11.3,17\mu m}$/SFR$_{fiber}$  & 0.331 & 307 & 0.365  & 81 & 0.758               & 296 \\

L$_{\Sigma PAH 7.7,11.3,17\mu m}$/L$_{[NeII]}$   & 0.213 & 238 & 0.0002 & 77 & 0.003               & 253 & PAH ULs \\

L$_{[NeII]}$/L$_{\Sigma PAH 7.7,11.3,17\mu m}$   & 0.945 & 166 & 0.006  & 42 & 0.0004              & 158 & [NeII] ULs \\

\enddata

\tablenotetext{1}{``SF'' refers to star-forming galaxies.}
\tablenotetext{2}{Number of sources in two-sample tests.}
\tablenotetext{3}{We report the survival analysis probability from Gehan's Generalized Wilcoxon Test, permutation variance. For the PAH and [NeII] ratios, survival analysis was performed twice: once to account for [NeII] upper limits and once to accommodate the PAH luminosity upper limits.  A {\it P} value $\leq$0.05 indicates that the two sub-populations differ at a statistically significant level.}

\end{deluxetable}
\end{landscape}

\clearpage


\begin{deluxetable}{lcc}
\small
\tablewidth{0pt}
\tablecaption{\label{rel_agn_sfr}Correlation between Diagnostics of Relative Importance of AGN Activity to Star formation\tablenotemark{1}}
\tablehead{\colhead{Diagnostic} & \colhead{$\rho$\tablenotemark{2}} & \colhead{$P_{uncorr}$\tablenotemark{3}}}

\startdata

L$_{[NeV]}$/L$_{[NeII]}$ vs. D                      & 0.562  & 4$\times10^{-4}$  \\
L$_{[NeV]}$/L$_{[NeII]}$ vs. $\alpha_{20-30\mu m}$  & -0.568 & 6$\times10^{-4}$  \\
L$_{[NeV]}$/L$_{[NeII]}$ vs. PAH EW 7.7 $\mu$m      & -0.702 & 1$\times10^{-4}$  \\
L$_{[NeV]}$/L$_{[NeII]}$ vs. PAH EW 11.3 $\mu$m     & -0.779 & $<1\times10^{-4}$ \\
L$_{[NeV]}$/L$_{[NeII]}$ vs. PAH EW 17 $\mu$m       & -0.628 & 2$\times10^{-4}$  \\

\enddata

\tablenotetext{1}{Used ASURV to account for upper limits in L$_{[NeV]}$.}
\tablenotetext{2}{$\rho$ is the correlation coefficient from which P$_{uncorr}$ is derived.}
\tablenotetext{3}{Probability the two quantities are not correlated. A $P_{uncorr}$ value $\leq$0.05 indicates
a significant correlation.}

\end{deluxetable}

\begin{deluxetable}{lcc}
\small
\tablewidth{0pt}
\tablecaption{\label{ew}Relationship between PAH EWs and Ionization Field Hardness}
\tablehead{\colhead{Diagnostic} & \colhead{$\rho$\tablenotemark{1}} & \colhead{N\tablenotemark{2}}}

\startdata

 D vs. PAH EW 7.7 $\mu$m  & -0.679  & 344 \\
 D vs. PAH EW 11.3 $\mu$m & -0.747  & 368 \\
 D vs. PAH EW 17 $\mu$m   & -0.658  & 281 \\

\cline{1-3}\\
\multicolumn{3}{l}{Star-forming Galaxies and Composites Sub-sample}\\
\cline{1-3}\\

 D vs. PAH EW 7.7 $\mu$m  & -0.227 & 291 \\
 D vs. PAH EW 11.3 $\mu$m & -0.125 & 303 \\
 D vs. PAH EW 17 $\mu$m   & -0.230 & 221 \\

\cline{1-3}\\
\multicolumn{3}{l}{Sources with $z \geq$ 0.03}\\
\cline{1-3}\\
 D vs. PAH EW 7.7 $\mu$m  & -0.566  & 318 \\
 D vs. PAH EW 11.3 $\mu$m & -0.653  & 336 \\
 D vs. PAH EW 17 $\mu$m   & -0.549  & 249 \\

\enddata
\tablenotetext{1}{$\rho$ is the correlation coefficient. Higher $|\rho|$ values indicate more significant (anti-)correlations.}
\tablenotetext{2}{Number of sources.}

\end{deluxetable}

\begin{deluxetable}{lcc}
\small
\tablewidth{0pt}
\tablecaption{\label{pah_ratio}Relationship between PAH Ratios and Ionization Field Hardness}
\tablehead{\colhead{Diagnostic} & \colhead{$\rho$\tablenotemark{1}} & \colhead{N\tablenotemark{2}}}

\startdata

\multicolumn{3}{l}{[NeV]/[NeII]\tablenotemark{3} vs.: }\\
\cline{1-3}\\

L$_{PAH 7.7\mu m}$/L$_{PAH 17\mu m}$   & -0.241 & 33 \\
L$_{PAH 11.3\mu m}$/L$_{PAH 17\mu m}$  & -0.564 & 37 \\
L$_{PAH 7.7\mu m}$/L$_{PAH 11.3\mu m}$ & 0.611  & 34 \\

\cline{1-3}\\
\multicolumn{3}{l}{D\tablenotemark{4} vs.: }\\
\cline{1-3}\\

L$_{PAH 7.7\mu m}$/L$_{PAH 17\mu m}$   & -0.193 & 266 \\
L$_{PAH 11.3\mu m}$/L$_{PAH 17\mu m}$  & -0.364 & 278 \\
L$_{PAH 7.7\mu m}$/L$_{PAH 11.3\mu m}$ & 0.179  & 340 \\

\cline{1-3}\\
\multicolumn{3}{l}{AGN sub-sample, D\tablenotemark{4} vs:}\\
\cline{1-3}\\
L$_{PAH 7.7\mu m}$/L$_{PAH 17\mu m}$   & 0.082  & 47 \\
L$_{PAH 11.3\mu m}$/L$_{PAH 17\mu m}$  & -0.247 & 58 \\
L$_{PAH 7.7\mu m}$/L$_{PAH 11.3\mu m}$ & 0.431  & 51 \\

\cline{1-3}\\
\multicolumn{3}{l}{Sources with z $\geq$ 0.03, D\tablenotemark{4} vs.: }\\
\cline{1-3}\\

L$_{PAH 7.7\mu m}$/L$_{PAH 17\mu m}$   & -0.283 & 240 \\
L$_{PAH 11.3\mu m}$/L$_{PAH 17\mu m}$  & -0.356 & 246 \\
L$_{PAH 7.7\mu m}$/L$_{PAH 11.3\mu m}$ & 0.084  & 314 \\

\cline{1-3}\\
\multicolumn{3}{l}{AGN with z$\geq$ 0.03, D\tablenotemark{4} vs:}\\
\cline{1-3}\\
L$_{PAH 7.7\mu m}$/L$_{PAH 17\mu m}$   & -0.208 & 22 \\
L$_{PAH 11.3\mu m}$/L$_{PAH 17\mu m}$  & -0.554 & 27 \\
L$_{PAH 7.7\mu m}$/L$_{PAH 11.3\mu m}$ & 0.590  & 26 \\

\enddata
\tablenotetext{1}{$\rho$ is the correlation coefficient. Higher $|\rho|$ values indicate more significant (anti-)correlations.}
\tablenotetext{2}{Number of sources.}
\tablenotetext{3}{$\rho$ calculated from survival analysis.}
\tablenotetext{4}{$\rho$ calculated from linear regression.}

\end{deluxetable}

\end{document}